\documentclass[12pt]{article}
\usepackage[T1]{fontenc}
\usepackage{amsmath}
\usepackage{amssymb} 
\usepackage{bbm} 
\usepackage[small]{caption2} 
\usepackage{fleqn} 
\usepackage{graphicx} 
\usepackage{mathrsfs}	
\usepackage[small,loose]{subfigure}  
\usepackage{cite} 
\usepackage{cancel} 
\usepackage{pifont} 
\usepackage{pstricks} 
\usepackage{longtable}
\usepackage{rotating}
\usepackage{nomencl} 
\makeglossary

\advance \headheight by 3.0truept       
\setcaptionwidth{0.75\textwidth}

\DeclareMathOperator{\re}{Re}

\DeclareMathOperator{\tr}{tr}

\newcommand{\CenterObject}[1]{\ensuremath{\vcenter{\hbox{#1}}}}

\newcommand{\I}{\mathrm{i}}
\newcommand{\BmL}{\ensuremath{B\!-\!L} }

\newcommand{\E}[1]{\ensuremath{\mathrm{E}_{#1}}} 
\newcommand{\G}[1]{\ensuremath{\mathrm{G}_{#1}}}
\newcommand{\SO}[1]{\ensuremath{\mathrm{SO}(#1)}}
\newcommand{\SU}[1]{\ensuremath{\mathrm{SU}(#1)}}
\newcommand{\U}[1]{\ensuremath{\mathrm{U}(#1)}}
\newcommand{\Z}[1]{\ensuremath{\mathbbm{Z}_{#1}}} 

\newcommand{\V}[0]{\ensuremath{\boldsymbol{V}}} 

\numberwithin{equation}{section}
\numberwithin{table}{section}
\allowdisplaybreaks

\begin{document}
\title{
\begin{flushleft}
{\normalsize DESY 06-059\hfill June 2006}
\end{flushleft}
\vspace{2cm}
{\bf Supersymmetric Standard Model\\ from the Heterotic String (II)}\\[0.8cm]}

\author{{\bf\normalsize 
Wilfried~Buchm\"uller$^1$, Koichi Hamaguchi$^{1,2}$, 
Oleg Lebedev$^3$, Michael Ratz$^{3}$}\\[1cm]
{\it\normalsize
${}^1$Deutsches Elektronen-Synchrotron DESY, 22603 Hamburg, Germany}\\
{\it\normalsize
${}^2$Department of Physics, University of Tokyo, Tokyo 113-0033, Japan}\\
{\it\normalsize
${}^3$Physikalisches Institut der Universit\"at Bonn, Nussallee 12, 53115 Bonn,
Germany}
}
\date{}
\maketitle \thispagestyle{empty} 

\abstract{
\noindent
We describe in detail a $\Z6$ orbifold compactification of the heterotic
$\E8\times\E8$ string which leads to the (supersymmetric) standard model gauge
group and matter content. The quarks and leptons appear as three ${\bf
16}$-plets of $\SO{10}$, two of which are localized at fixed points with local
$\SO{10}$ symmetry. The model has supersymmetric vacua without exotics at low
energies and is consistent with gauge coupling unification. Supersymmetry can be
broken via gaugino condensation in the hidden sector. The model has large vacuum
degeneracy. Certain vacua with approximate $B\!-\!L$ symmetry have attractive
phenomenological features. The top quark Yukawa coupling arises from gauge
interactions and is of the order of the gauge couplings. The other Yukawa
couplings are suppressed by powers of standard model singlet fields, similarly
to the Froggatt-Nielsen mechanism.}

\clearpage
\setcounter{tocdepth}{1}
\tableofcontents
\clearpage
\section{Introduction and Summary}
\label{sec:Intro}

The standard model is a remarkably successful theory of the structure of matter.
It is a chiral gauge theory with the gauge group
$G_\mathrm{SM}=\SU3_c\times\SU2_\mathrm{L}\times\U1_Y$ and three generations  of
quarks and leptons. All masses are generated by the Higgs mechanism which
involves an \SU2 doublet of scalar fields. Its unequivocal prediction is the
existence of the Higgs boson which still remains to be discovered. From a
theoretical perspective, the minimal supersymmetric extension of the standard
model, the MSSM, is particularly attractive. Apart from  stabilizing the
hierarchy between the electroweak and Planck scales and  providing a natural
explanation of the observed dark matter, it  predicts unification of the gauge
couplings at the unification scale $M_\mathrm{GUT}\simeq
2\cdot10^{16}\,\mathrm{GeV}$.

Even more than the unification of gauge couplings, the symmetries and the
particle content of the standard model point towards grand unified theories
(GUTs) \cite{Georgi:1974sy,Pati:1974yy}. Remarkably, one generation of matter,
including the right-handed neutrino, forms a single spinor representation of
\SO{10} \cite{Georgi:1975qb,Fritzsch:1974nn}. It therefore appears natural to
assume an underlying \SO{10} structure of the theory. The route of unification,
continuing via exceptional groups, terminates at $\E8$, which is
beautifully realized in the heterotic string \cite{Gross:1984dd,Gross:1985fr}.

An obstacle on the path towards unification are the Higgs fields, which are 
$\SU2_\mathrm{L}$ doublets, while the smallest \SO{10} representation containing
the Higgs doublets, the $\boldsymbol{10}$--plet, predicts additional $\SU3_c$
triplets. The fact that Higgs fields form incomplete `split' GUT representations
is particularly puzzling in supersymmetric theories where both matter and Higgs
fields are chiral multiplets. The triplets cannot have masses below 
$M_\mathrm{GUT}$ since otherwise proton decay would be too rapid. This then
raises the question why $\SU2_\mathrm{L}$ doublets are so much lighter than $\SU3_c$
triplets. This is the notorious doublet-triplet splitting problem of ordinary 
4D GUTs.

Higher-dimensional theories offer new possibilities for gauge symmetry breaking
connected with compactification to four dimensions. A simple and elegant scheme,
leading to chiral fermions in four dimensions, is the compactification on
orbifolds, first considered for the heterotic string \cite{Dixon:1985jw,
Dixon:1986jc,Ibanez:1986tp,Ibanez:1987xa,Ibanez:1987sn,Casas:1987us,Casas:1988hb},
and more recently  applied to GUT field theories
\cite{Kawamura:2000ev,Altarelli:2001qj,
Hall:2001pg,Hebecker:2001wq,Asaka:2001eh,Hall:2001xr}.  Such orbifold GUTs
appear as intermediate effective field theories in  compactifications of the
heterotic string when some of the compact dimensions are of order
$1/M_\mathrm{GUT}$ and therefore large compared to the string  length
\cite{Kobayashi:2004ud,Forste:2004ie,Kobayashi:2004ya,Buchmuller:2004hv}.

In orbifold compactifications, gauge symmetry of the 4D effective theory is an
intersection of larger symmetries at orbifold fixed points. Massless modes 
located at these fixed points all appear in the 4D theory and form 
representations of the larger local symmetry groups. Zero modes of bulk fields,
on the contrary, are only representations of the smaller 4D gauge symmetry  and
form in general `split multiplets'. When the local symmetry at some orbifold
fixed  points is a GUT symmetry, one obtains the picture of `local grand 
unification'. The SM gauge group can be thought of as an intersection of
different local GUT groups. Matter fields appear as complete GUT 
representations localized at the fixed points, whereas the Higgs doublets are 
associated with bulk fields, and therefore split multiplets. In this way the
structure of the standard model is naturally 
reproduced~\cite{Buchmuller:2004hv,Asaka:2003iy,Buchmuller:2005sh}. 

Recently, we have obtained the gauge group and matter content of the 
supersymmetric standard model from the heterotic string by using the picture of
local grand unification as the guiding principle \cite{Buchmuller:2005jr}.
Quarks and leptons appear as three $\boldsymbol{16}$--plets of \SO{10}, two of
which are localized at orbifold fixed points with local \SO{10} symmetry. For
generic vacua, no exotic states appear at low energies and the model is
consistent with gauge coupling unification. In this paper we describe our
construction in detail. 

It is well-known that the number of possible string vacua is huge. Early
estimates of the total number of different vacua of the heterotic string gave
numbers like $10^{1500}$ \cite{Lerche:1986cx}, which came as a complete
surprise. More recent studies, based on flux compactifications, give similarly
large numbers \cite{Bousso:2000xa}. Searches for standard model--like vacua have
been based on orbifold compactifications \cite{Ibanez:1987dw2,Bailin:1999nk}, the
free fermionic formulation \cite{Cleaver:1998sa,Faraggi:2004rq,Faraggi:2006bs},
intersecting D--brane models \cite{Blumenhagen:2005mu} and Gepner orientifolds
\cite{Anastasopoulos:2006da}.  Despite the huge number of vacua, it turned out
to be extremely difficult to  construct a consistent ultraviolet completion of
the (supersymmetric) standard model, and only recently several examples have
been obtained \cite{Buchmuller:2005jr,Braun:2005nv,Bouchard:2005ag}\footnote{
In intersecting brane constructions, a model containing  the spectrum 
of the MSSM plus vector--like exotics and additional U(1) factors was
obtained in \cite{Honecker:2004kb}.}. This
suggests that not all field theories can be embedded into string theory and that
a consistent ultraviolet completion of the standard model may eventually lead to
some testable low energy predictions.

In this paper, the model presented in \cite{Buchmuller:2005jr} is described in
detail. We hope that this will be useful for further phenomenological studies of
the model and also for the search for other embeddings of the standard model
into the heterotic string. In order to keep the paper self-contained, we recall
the basics of strings on orbifolds in
Secs.~\ref{sec:StringsOnOrbifold}--\ref{sec:CouplingsAndSelectionRules}. In
Sec.~\ref{sec:StringsOnOrbifold}, the boundary conditions for untwisted and
twisted strings, the mode expansion and the massless spectrum are discussed;
furthermore, a simple derivation of the projection conditions for physical
states is given. Our orbifold model is based on the 6D torus defined by the $\G2
\times \SU3 \times \SO4$ root lattice, which has a
$\Z{6-\mathrm{II}}=\Z3\times\Z2$ discrete symmetry. The geometry is described in
Sec.~\ref{sec:Z3xZ2geometry} with emphasis on the localization of twisted 
states. In Sec.~\ref{sec:CouplingsAndSelectionRules}, the string selection rules
for superpotential couplings of the $\Z{6-\mathrm{II}}$ orbifold are reviewed
and somewhat extended.

The main results of this paper are contained in
Secs.~\ref{sec:Model}--\ref{sec:Pheno} and in the appendices A--D. After
describing our search strategy for compactifications with local $\SO{10}$
symmetry, we study the unbroken gauge group $G$ and the massless spectrum of the
model in Sec.~\ref{sec:Model}. We also list the GUT representations at  various
fixed points and the 6D orbifold GUTs which one obtains for two compact
dimensions of size $1/M_\mathrm{GUT}$. The Fayet-Iliopoulos (FI) $D$-term of an
anomalous \U1 triggers further symmetry breaking \cite{Dine:1987xk}. In
particular, 
\begin{equation} G~\longrightarrow
~\SU{3}_c\times\SU{2}_\mathrm{L}\times \U1_Y \times G_\mathrm{hidden} \;,
\end{equation} with $G_\mathrm{hidden}= \SU4\times\SU2'$
is possible, in which case the model has a truly hidden sector admitting 
spontaneous SUSY breaking. We further show that, for generic vacua,
unwanted exotic states attain large masses and decouple. This is one of the
central results of our paper.

The decoupling of exotic states can be achieved without breaking supersymmetry.
In Sec.~\ref{sec:FandDflat}, we discuss $D$- and $F$-flat directions in the
field space as well as general supersymmetric field configurations, neglecting
supergravity corrections. The model naturally accommodates spontaneous
supersymmetry breaking via hidden sector gaugino condensation, which is
described in Sec.~\ref{sec:SUSYbreaking}.

In generic $F$-- and $D$--flat configurations,  our model yields the
MSSM spectrum, however  the analysis of proton decay and flavour becomes
intractable. In Sec.~\ref{sec:Pheno}, we therefore identify a simple and
phenomenologically  attractive $D$--flat  field  configuration, without
proving $F$--flatness, which preserves
\begin{equation}
G_\mathrm{SM}\times\U1_{\BmL}\times[\SU4 ] \;.
\end{equation}
Here we keep the hidden sector \SU4 unbroken which is needed for gaugino
condensation. We show that unwanted exotics can be decoupled in this case as well. 
Further, we 
identify two Higgs doublets and discuss the pattern of Yukawa couplings.
The top quark Yukawa coupling arises from gauge interactions and is of
the order of the gauge couplings. Other Yukawa couplings are suppressed
by powers of standard model singlet fields, similarly to the Froggatt--Nielsen
mechanism \cite{Froggatt:1978nt}.

Finally, in Sec.~\ref{sec:Outlook}, we conclude with a brief outlook on open
questions and further challenges for realistic compactifications of the
heterotic string.

\clearpage
\section{Strings on  orbifolds}
\label{sec:StringsOnOrbifold}

In the following subsections we collect the basic notions and formulae  which
are needed to describe  propagation of the $\E8\times\E8$  heterotic  string on 
orbifolds $\mathbbm{T}^6/\Z{N}$ \cite{Dixon:1985jw,Dixon:1986jc}. We follow the
definitions of Katsuki {\it et al.} \cite{Katsuki:1989bf}.

\subsection{Lattices and twists}
\label{sec:ConsitencyConditions}

The torus is obtained as the quotient  $\mathbbm{T}^6 =
\mathbbm{R}^6/2\pi\Lambda$, where $\Lambda$ is the lattice of a semi--simple Lie
algebra of rank 6 with a $\Z{N}$ discrete symmetry.  The 6 compact coordinates
of the torus $x^i$, $i=4\ldots 9$, are conveniently  combined into 3 complex
coordinates $z^i = {1\over \sqrt{2}} \left(x^{2i+2} + \I x^{2i+3}\right)$,
$i=1\ldots 3$.
\nomenclature[zi]{$z^i$}{complex coordinates of the torus\refpage}
Points in $\mathbbm{R}^6$  differing  by a lattice vector,
\begin{equation}\label{trans}
 z~\sim~z + 2\pi \ell\;, 
\end{equation}
\nomenclature[ea]{$e_a$}{lattice vectors\refeqpage}
with $\ell = m_a e_a$, $m_a \in \mathbbm{Z}$ ($a = 1 \ldots 6 $), are
identified. Here $e_a$ denote the basis vectors in the three planes of the
lattice.

The lattice has a $\Z{N}$ discrete symmetry which acts crystallographically,
i.e., it maps the lattice onto itself,
\begin{equation}\label{twist}
 z~\rightarrow~\theta z\;,\quad 
 \theta^i_j~=~e^{2\pi \I v_N^i}\delta^i_j\;, \quad i,j=1\ldots 3\;,
\end{equation}
with
\begin{equation}
 \theta^N~=~1\;,\quad  Nv_N^i = 0\mod1\;.
\end{equation}
\nomenclature[theta]{$\theta$}{twist\refeqpage}
Here we assume the factorization $\mathbbm{T}^6 = \mathbbm{T}^2 \otimes 
\mathbbm{T}^2 \otimes \mathbbm{T}^2$.
$N=1$ supersymmetry in 4D requires that the \Z{N} twist be contained in  the
\SU3 subgroup of \SO6, i.e.,
\begin{equation}
 \sum_i v_N^i = 0\mod 1 \;.
\end{equation}

\nomenclature[vN]{$v_N$}{twist vector\refeqpage}
Lattice translations and twists $\theta^k$ ($k=0,\dots,N-1$) form the space
group $\mathbbm{S}$ whose elements are denoted by $(\theta^k,\ell)$. The
orbifold $\mathbbm{T}^6/\Z{N}$ can also be defined as the quotient 
$\mathbbm{R}^6/\mathbbm{S}$, where
\begin{equation}
 z~\sim~(\theta^k,\ell)\, z~\equiv~\theta^k\, z + 2\pi\,\ell\;. 
\end{equation}
The multiplication rule in the space group is given by
\begin{equation}\label{eq:SpaceGroupMultiplication}
 (\theta^{k_1},\ell_{1})(\theta^{k_2},\ell_{2})~=~
 (\theta^{k_1}\theta^{k_2},\theta^{k_1}\ell_{2} + \ell_{1})\;. 
\end{equation}

An orbifold has fixed points $f$, which are invariant under the action
of a space group element $(\theta^k,\ell)$,
\begin{equation}\label{fixed}
 f~=~(\theta^k,\ell)\, f~=~\theta^k\, f + 2\pi\,\ell\;, 
 \quad \ell~=~m_a\, e_a\;, \quad  m_a \in \mathbbm{Z}\;.
\end{equation}
Here $k$ and  $\ell$ depend on the fixed point $f$.
\nomenclature[f]{$f$}{fixed point\refeqpage}
Since the position of the fixed point is defined only up to a lattice vector,
$\ell$ is defined up to a translation in the sublattice 
\begin{equation}
 \Lambda_k~\equiv~(\mathbbm{1}-\theta^k)\, \Lambda
 ~=~\left\{\lambda\in\Lambda|\ \lambda=(\mathbbm{1}-\theta^k)\,\mu,\;
  \mu\in\Lambda\right\}\;.
\end{equation}
Each fixed point $(\theta^k,\ell)$ is associated with a sublattice $\Lambda =
\ell + \Lambda_k$, and there are as many sublattices as fixed points. The
dimension of a sublattice $\Lambda_k$ can be smaller than $\dim \Lambda=6$ if
$(\mathbbm{1}-\theta^k)$ has eigenvectors with eigenvalue $0$. In this case the
element $(\theta^k,\ell)$ describes fixed planes.

\subsection{Untwisted and twisted strings}
\label{sec:UnTwisted}

\begin{figure}[h]
\centerline{\includegraphics{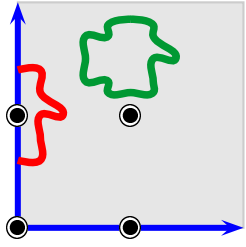}}
\caption{Twisted and untwisted strings. The dots denote orbifold fixed points.}
\label{Fig:untwistedstring}
\end{figure}

In the light-cone gauge the heterotic string can be described by the following 
world-sheet fields~\cite{Green:1987mn}: 8 string coordinates and 8 right-moving 
Neveu-Schwarz-Ramond fermions ($\sigma_{\pm} = \tau \pm \sigma$),
\begin{equation}
X^i(\tau,\sigma)~=~X^i_\mathrm{L}(\sigma_+) + X^i_\mathrm{R}(\sigma_-)\;, \quad
\psi^i(\sigma_-)\;,\quad i= 2 \ldots 9\;,
\end{equation}
\nomenclature[Xi]{$X_\mathrm{L,R}^i$}{string
coordinates\refeqpage}
\nomenclature[psii]{$\psi^i$}{right--moving fermions\refeqpage}
and 32 left-moving fermions $\lambda^I$, 
\begin{equation}
\lambda^I(\sigma_+)\;, \quad I = 1\ldots 32\;.
\end{equation} 
\nomenclature[lambdaI]{$\lambda^I$}{left--moving fermions\refeqpage}
Here $i$ is the space--time index, while index $I$ is associated with 
$\E8\times\E8$ gauge degrees of freedom.
It is convenient to combine the string coordinates in the compact dimensions
into 3 complex variables $Z^i$ and, similarly,  the right moving fermions
into  3 complex NSR fermions $\widetilde{\psi}^i$,
\begin{equation}
 Z^i~=~{1\over \sqrt{2}} \left(X^{2i+2} + \I X^{2i+3}\right)\;, \quad
 \widetilde{\psi}^i~=~{1\over \sqrt{2}} \left(\psi^{2i+2} + \I \psi^{2i+3}\right)\;,
\end{equation} 
\nomenclature[Zi]{$Z^i$}{complex string coordinates\refeqpage}
\nomenclature[psitildei]{$\widetilde{\psi}^i$}{complex NSR fermions\refeqpage}
where $i=1\ldots 3\;$.
The \Z{N} twist acts on these fields as
\begin{equation}\label{twistzf}
 Z~\rightarrow~\theta\, Z\;,\quad
 \widetilde{\psi}~\rightarrow~\theta\, \widetilde{\psi}\;.
\end{equation}
Closed strings on \Z{N} orbifolds can be untwisted or twisted. In the former
case the string is closed already on the torus and has the boundary conditions, 
\begin{eqnarray}
 Z(\sigma+2\pi) &=& Z(\sigma) + 2\pi m_a e_a\;, 
 \quad  m_a \in \mathbbm{Z}\;,\\ 
 \widetilde{\psi}(\sigma+2\pi) &=& \pm\ \widetilde{\psi}(\sigma)\;, \label{nsru}
\end{eqnarray}
whereas in the latter case the string is closed on the orbifold but not
on the torus and has the boundary conditions ($k=1\ldots N-1$),
\begin{eqnarray}
 Z(\sigma+2\pi) &=& \theta^k\, Z(\sigma) + 2\pi\, m_a\, e_a\;, 
 \label{twistZ}\\
 \widetilde{\psi}(\sigma+2\pi) &=& 
 \pm\, \theta^k\, \widetilde{\psi}(\sigma)\;, \label{nsrt}
\end{eqnarray}
where $k$ and $m_a$ depend on the fixed point $f$. The lattice translation in
Eq.~(\ref{twistZ}) enters the space group element associated with the fixed
point, Eq.~(\ref{fixed}). The plus and minus signs in Eqs.~(\ref{nsru}) and
(\ref{nsrt}) correspond to the Ramond and the Neveu-Schwarz sectors,
respectively. Twisted strings are localized at the orbifold fixed points,
whereas untwisted strings can propagate freely on the orbifold
(Fig.~\ref{Fig:untwistedstring}).

Modular invariance usually requires that the $\Z{N}\subset\SO{6}$ twist of the
space--time degrees of freedom  be accompanied by a $\Z{N}\subset\E8\times\E8$
twist of the fermions $\lambda^I$, representing the internal symmetry group. On
the complex fermions
\begin{equation}
\widetilde{\lambda}^I~=~{1\over \sqrt{2}}
\left(\lambda^{2I-1} + \I \lambda^{2I}\right)\;, \quad I=1\ldots 16\;,
\end{equation}
the \Z{N} twist acts as
\begin{equation}\label{twistg}
\widetilde{\lambda}~\rightarrow~\Theta\, \widetilde{\lambda}\;,\quad 
\Theta^I_J~=~e^{2\pi \I\, V^I_N}\,\delta^I_J\;,
\end{equation}
\nomenclature[VN]{$V_N$}{gauge shift vector\refeqpage}
where 
\begin{equation}
\Theta^N = 1\;, \quad 
N\sum_{I=1}^8 V_N^I = N\sum_{I=9}^{16} V_N^I = 0\mod 2\;,
\end{equation}
with integer $N V_N^I$. The fermions $\widetilde{\lambda}^I$ can have untwisted
($k=0$) or twisted ($k=1\ldots N-1$) boundary conditions,
\begin{equation}\label{bcferm}
 \widetilde{\lambda}(\sigma+2\pi)~=~\pm
 \Theta^k\, \widetilde{\lambda}(\sigma)\;.
\end{equation}
This makes the parallel between $\theta$ and $\Theta$ transparent. Extending
$v_N$ by a zero entry acting on the uncompactified dimensions, $v_N \rightarrow
(v_N^1, v_N^2, v_N^3; 0)$, we note that  vectors $N v_N$ and $N V_N$ lie on the
root lattices $\Lambda_{\SO8}$ and $\Lambda_{\SO{16}\times\SO{16}}$,
respectively. In an orthonormal basis, $\Lambda_{\SO{2N}}$ is defined by vectors
$(n_1,\dots,n_{N})$ with integer $n_i$ and $\sum_{i=1}^N n_i=0 \mod 2$. One can
show that the gauge symmetry of this theory is  $\E8\times\E8$ which contains $
\SO{16}\times\SO{16}$ as a subgroup \cite{Gross:1985fr}.

A convenient formulation of the heterotic string  is obtained by representing
fermionic degrees of freedom in terms of bosons. In this case one replaces the
8 right--moving and 32 left--moving fermions with 4 right--moving and 16
left--moving bosons, 
\begin{eqnarray}
 \widetilde{\psi}^i(\sigma_-) &=& e^{-2\I\, H^i(\sigma_-)}\;, 
 \quad  i=1\ldots 4\;,\label{bosonH}\\
 \widetilde{\lambda}^I(\sigma_+) &=& e^{2\I\, X^I(\sigma_+)}\;, 
 \quad  I=1\ldots 16\;. \label{bosonX}
\end{eqnarray}
The fields $X^I$ are compactified  on a 16--dimensional  torus represented
by  the $\E8\times\E8$ root lattice, 
\begin{equation}
 \Lambda_{\E8}:\quad  p = (n_1,...,n_8)\;\;{\rm or}\;\; 
 \left(n_1 +{1\over 2}, ..., n_8 +{1\over 2} \right)\;,
 \label{root}
\end{equation}
where $n_i$ integer with $\sum_{i=1}^8 n_i=0 \mod 2$, and similarly for the
second \E8. This gives rise to gauge multiplets of the $\E8\times\E8$ group in
10 dimensions, coupled to supergravity.

Compactifying the extra 6 dimensions on an orbifold amounts to modding the
string coordinates by the space group and its gauge counterpart. The latter is
obtained by  embedding the twists and lattice shifts into gauge degrees of
freedom  $X^I$ as
\begin{equation}
 (\theta^k, m_a e_a)~\longrightarrow~(\mathbbm{1}, k\, V_N^I + m_a\, W_{na}^I) \;,
\end{equation}
where $W_{na}^I$ denotes a Wilson line of order $n$.  Here $N V_N$ and $n W_{n}$
($n\leq N$) are required to lie on the  $\E8\times\E8$ root
lattice.\footnote{This generalizes $V_N$ of Eq.~(\ref{twistg}) in which case $N
V_N$ lies on the $\SO{16}\times\SO{16}$ root lattice.}
Thus, a twist of the space--time degrees of freedom is accompanied by a shift
$k V_N$ of the gauge coordinates, while a torus lattice translation is
accompanied by a gauge coordinate shift $m_a W_{na}$. This corresponds to
generalizing the boundary condition (\ref{bcferm}) for the left--moving
fermions to
\begin{equation}
 \widetilde{\lambda}^I(\sigma+2\pi)~=~\pm
 e^{2\pi \I\, (kV^I_N + m_a W^I_{na}) }\, \widetilde{\lambda}^I(\sigma)\;.
 \label{lambdaI}
\end{equation}
The bosonic field boundary conditions then read ($k = 0\ldots N-1$)
\begin{subequations}\label{eq:BosonicBoundaries}
\begin{eqnarray}
 H^i(\sigma+2\pi) &=& H^i(\sigma) -\pi\, k\, v_N^i \mod \pi\,
 \Lambda_{\SO8}^*\;, \label{Hi}\\
 X^I(\sigma+2\pi) &=& X^I(\sigma) + 
 \pi\, \left(k\,V_N^I + m_a\, W_{na}^I\right)\mod\pi\,\Lambda_{\E8\times\E8}\;.
 \label{XI}
\end{eqnarray}
\end{subequations}
Here $\Lambda_{\SO8}^*$ denotes the weight lattice of \SO8 given  in the
orthonormal basis by
\begin{equation}
 \Lambda_{\SO8}^*:\quad q = (n_1,n_2,n_3,n_4)\;,
\end{equation}
where $n_i$ integer with $\sum_i n_i$ odd or $n_i$ half-integer with  $\sum_i
n_i$ even. 

To summarize, the heterotic string can be described by  the left moving  bosonic
fields $Z^i_\mathrm{L}(\sigma_+)$, $Z^{i*}_\mathrm{L}(\sigma_+)$,
$X^I(\sigma_+)$ and the right moving bosonic fields $Z^i_\mathrm{R}(\sigma_-)$,
$Z^{i*}_\mathrm{R}(\sigma_-)$, $H^i(\sigma_-)$. They fall into untwisted or
twisted categories depending on whether they represent strings closed on a torus
or on an orbifold only.

\subsection{Modular invariance and local twists}

The gauge shift $V_N$ and the Wilson lines $W_n$ are subject to  consistency
conditions. First of all, $NV_N$ and $nW_n$ are vectors of the $\E8\times\E8$
root lattice,
\begin{equation}\label{eq:ConsistencyCondition1}
 N\, V_N~\in~\Lambda_{\E8\times\E8}\;,\quad
 n\, W_n~\in~\Lambda_{\E8\times\E8}\;.
\end{equation}
Second, modular invariance of the theory requires that they satisfy additional
constraints (see e.g., \cite{Forste:2004ie}):
\begin{subequations}\label{eq:WeakModularInvariance}
\begin{eqnarray}
 N\;\left(V_N^2-v_N^2\right) & = & 0\mod2\;,\\
 N\; V_N\cdot W_n & = & 0\mod 1\;,\\
 N\; W_n\cdot W_m & = & 0\mod 1\;,\quad ( W_n\ne W_m)\\
N\;W_n^2& = & 0\mod 2\;.
\end{eqnarray}
\end{subequations}
By adding $\E8\times\E8$ root lattice vectors to $V_N$ and $W_n$  satisfying
these conditions,  one can bring  $V_N$,$W_n$ to the form  which obeys a
stronger constraint\footnote{There are exceptions to this statement, for
instance, when $V=0$.},  
\begin{subequations}\label{eq:StrongModularInvariance}
\begin{eqnarray}
 \frac{1}{2}\,\left(V_N^2-v_N^2\right) & = & 0\mod 1\;,\\
 V_N\cdot W_n & = & 0\mod 1\;,\\
 W_n\cdot W_m & = & 0\mod 1\;,\quad (W_n\ne W_m)\\
 \frac{1}{2}\,W_n^2& = & 0\mod 1\;.
\end{eqnarray}
\end{subequations}
This form has the advantage that the analysis of physical states of the 
theory simplifies significantly. These equations can also be written as 
\begin{equation}\label{eq:StrongModularInvariance2}
 \frac{1}{2}\left[\left(rV_N+m_a\,W_{na}\right)^2-r^2v_N^2\right]
 ~=~0\mod 1\;,\quad r=0,1\;,
\end{equation}
where $0\le m_a\le n-1$ for a Wilson line $W_n$ of order $n$.

The twist can be thought of as a \emph{local} quantity, 
that is, depending on the fixed point and the twisted sector. Indeed, 
Eqs.~(\ref{lambdaI}) and (\ref{XI}) show that what matters
at a particular fixed point $f$  is the combination
\begin{equation}\label{localt}
 V_f^I~=~k\,V^I_N + m_a\, W^I_{na}\;,
\end{equation}
\nomenclature[Vf]{$V_f$}{local gauge shift\refeqpage}
which plays the role of the ``local'' gauge twist, as well as its right--moving
counterpart $k\,v_N$.  Each local twist $V_f$ can be expressed as the sum of the
twist  $k\,V_N$ for  vanishing Wilson lines and a linear combination of Wilson
lines determined by  the location of the fixed point $f$. The local twists
satisfy modular invariance conditions (\ref{eq:StrongModularInvariance2}) and
can be treated  on the same footing as $V_N$.  This observation will be
important for the concept of local GUTs.

\subsection{Mode expansion and massless spectrum}
\label{sec:Spectrum}

The boundary conditions discussed in Sec. 2.2 lead to the following mode 
expansion for the untwisted string ($i=1\ldots 3$),
\begin{subequations}\label{eq:modeZu}
\begin{eqnarray}
 Z^i(\tau,\sigma) &=& z^i + {1\over 2} p^i \tau + m_a e^i_a \sigma \nonumber\\
 &&{} + {\I\over 2} \sum_{n}
 {1\over n}\alpha^i_{n}e^{-\I n\sigma_-} +
 {\I\over 2} \sum_{n} 
 {1\over n}\widetilde{\alpha}^i_{n}e^{-\I n\sigma_+}\;,\label{modeZu}\\
 Z^{*i}(\tau,\sigma) &=& z^{*i} + {1\over 2} p^{*i} \tau 
 + m_a\, e^{*i}_a \sigma \nonumber\\
 && + {\I\over 2} \sum_{n}{1\over n}\alpha^{*i}_{n}e^{-\I n\sigma_-} +
 {\I\over 2} \sum_{n} 
 {1\over n}\widetilde{\alpha}^{*i}_{n}e^{-\I n\sigma_+}\;. \label{modeZ*u}
\end{eqnarray}
\end{subequations}
Twisted strings have the expansion (cf.\ Eq.~(\ref{twistZ})) 
\begin{eqnarray}
 Z^i(\tau,\sigma) &=& f^i  + {\I\over 2} \sum_{n\in \Z{} + kv_N^i}
 {1\over n}\alpha^i_{fn}e^{-\I n\sigma_-} + 
 {\I\over 2} \sum_{n\in \Z{} - kv_N^i} 
 {1\over n}\widetilde{\alpha}^i_{fn}e^{-\I n\sigma_+}\;,\label{modeZt}\\
 Z^{*i}(\tau,\sigma) &=& f^{*i} 
 + {\I\over 2} \sum_{n\in \Z{} - kv_N^i}
 {1\over n}\alpha^{*i}_{fn}e^{-\I n\sigma_-} +
 {\I\over 2} \sum_{n\in \Z{} + kv_N^i} 
 {1\over n}\widetilde{\alpha}^{*i}_{fn}e^{-\I n\sigma_+}\;. \label{modeZ*t}
\end{eqnarray}
In this case, there is no center--of--mass string motion,  i.e., $p^i = m_a =
0$. If there is a fixed plane, the boundary conditions for strings  in the fixed
plane are untwisted  and the expansion is given by
Eqs.~\eqref{eq:modeZu}.

The bosonized NSR fermions  have the expansion ($i=1\ldots 4$)
\begin{equation}
 H^i(\tau,\sigma)~=~h^i + {1\over 2}\left(q^i + kv_N^i\right) \sigma_-
 + {\I\over 2}\sum_{n\neq 0}{1\over n}\widetilde{\beta}^i_{n}e^{-\I n\sigma_-}\;,
\label{modeH}
\end{equation}
while the gauge coordinates are given by     ($I=1\ldots 16$)
\begin{equation}
 X^I(\tau,\sigma)~=~x^I 
 + {1\over 2}\left(p^I + \left(kV_N^I+m_aW^I_{na}\right)\right)\sigma_+ + 
 {\I\over 2}\sum_{n\neq 0}{1\over n}\widetilde{\alpha}^I_{fn}e^{-\I n\sigma_+}\;.
\label{modeX}
\end{equation}
\nomenclature[p]{$p$}{$p\in\Lambda_{\E8\times\E8}$: $\E8\times\E8$ root lattice
vector (`momentum')\refeqpage}
\nomenclature[q]{$q$}{$q\in\Lambda_{\SO8}^*$: \SO8 weight (`momentum') 
\refeqpage}
The momentum vectors $q^i$ and $p^I$ specify the Lorentz and gauge quantum 
numbers of the string states.  Note that the creation and annihilation operators
of the twisted string (\ref{modeZt}), (\ref{modeZ*t}) and the left-moving
string (\ref{modeX}) depend on the fixed point $f$.

States of the heterotic string are given by a direct product of the
right--moving and left--moving parts. A basis in the Hilbert space of the
quantised string is obtained by acting with the creation operators
$\alpha^i_{fn}$, $\widetilde{\alpha}^i_{fn}$, $\widetilde{\beta}^i_{fn}$,
$\widetilde{\alpha}^I_{fn}$ ($n < 0$) on the ground states of the untwisted sector
$U$ ($k=0$) and the twisted  sectors $T_k$ ($k=1\ldots N-1$). Massless states
in the untwisted sector as well as twisted states living on fixed planes have
$p^i=m_a=0$.  The ground states of the different sectors depend on the momentum
vectors $q^i$, $p^I$ and, for the twisted sectors, also on the fixed point $f$
(cf.~(\ref{localt})),
\begin{equation}
 |q,p\rangle~\equiv~|q\rangle \otimes |p\rangle\;, \quad
 |f;q,p\rangle~\equiv~|q + kv_N\rangle \otimes |p + V_f\rangle\;.
\end{equation}
It turns out that for the model discussed below only oscillator modes of the
left-moving strings $Z^i_\mathrm{L}(\sigma_+)$, $Z^{*i}_\mathrm{L}(\sigma_+)$ 
and $X^I(\sigma_+)$ are relevant. The corresponding twisted sector states are
($n_i, m_i < 0$) 
\begin{equation}
\widetilde{\alpha}^{i_1}_{fn_1}\widetilde{\alpha}^{i_2}_{fn_2}\ldots
\widetilde{\alpha}^{*{j_1}}_{fm_1}\widetilde{\alpha}^{*{j_2}}_{fm_2}\ldots
\widetilde{\alpha}^{*{I_1}}_{fl_1}\widetilde{\alpha}^{*{I_2}}_{fl_2}\ldots
|f;q,p\rangle\;.
\end{equation}
Massless states of the untwisted sector satisfy the 
following mass equations:
\begin{subequations}\label{eq:UntwMassEquations}
\begin{eqnarray}
 \frac{1}{8}m_\mathrm{R}^2 
 & = & 
 \frac{1}{2} q^2
 -\frac{1}{2}   + N
 + N^*
 = 0\;,\\
 \frac{1}{8}m_\mathrm{L}^2 
 & = & 
 \frac{1}{2} p^2 -1+ 
 \widetilde{N}
 +\widetilde{N}^* = 0\;,
\end{eqnarray}
\end{subequations}
where $N,N^*, \widetilde{N},\widetilde{N}^* $ are the integer oscillator
numbers. Twisted massless states obey 
\begin{subequations}\label{eq:MassEquations}
\begin{eqnarray}
 \frac{1}{8}m_\mathrm{R}^2 
 & = & 
 \frac{1}{2}(q+k\,v_N)^2
 -\frac{1}{2}+\delta c^{(k)}+\omega_i^{(k)}\,N_{fi}
 +\Bar{\omega}_i^{(k)}\,N_{fi}^*
 = 0\;,\\
 \frac{1}{8}m_\mathrm{L}^2 
 & = & 
 \frac{1}{2}(p+V_f)^2 -1+\delta c^{(k)} + 
 \omega_i^{(k)}\,\widetilde{N}_{fi}
   +\Bar{\omega}_i^{(k)}\,\widetilde{N}_{fi}^* = 0\;,
\end{eqnarray}
\end{subequations}
where 
\begin{equation}
 \delta c^{(k)}~=~\frac{1}{2}\sum_i \omega^{(k)}_i\,(1-\omega^{(k)}_i)\;,
\end{equation}
with $\omega^{(k)}_i\,=\,(k\,v_N)_i \mod 1$, so that $0<\omega^{(k)}_i\le1$, 
and $\Bar{\omega}^{(k)}_i\,=\,(-k\,v_N)_i \mod 1$ so that
$0<\Bar{\omega}^{(k)}_i\le1$. This implies that
$\omega^{(k)}_i=\Bar{\omega}^{(k)}_i=1$ for $(k\,v_N)_i$ integer. 
In  Eq.~(\ref{eq:MassEquations}), $N_{fi}$,  $N_{fi}^*$, $\widetilde{N}_{fi}$
and $\widetilde{N}_{fi}^*$ $\in \mathbb{N}$ represent the oscillator numbers of 
the right- and left-movers in $z_i$ and $\Bar{z}_i$ directions, respectively. 
Note that $N_{fi}$ and $N_{fi}^*$, as well as $\widetilde{N}_{fi}$ and 
$\widetilde{N}_{fi}^*$, denote independent quantities.  They are the eigenvalues
of the corresponding number operators $\hat{N}_{fi}$,
\begin{equation}
 \hat{N}_{fi}~=~{1\over \omega^{(k)}_i}~
\sum_{n>0} \alpha^i_{f-n}\alpha^i_{fn} \;, 
\end{equation}
and analogously for $N_{fi}^*$,  $\widetilde{N}_{fi}$,  $\widetilde{N}_{fi}^*$.
The sum  $\sum_i(\omega^{(k)}_{ki}\,\widetilde{N}_{fi} + 
\Bar{\omega}^{(k)}_{ki}\,\widetilde{N}_{fi}^*)$ is often referred to as
$\widetilde{N}$ in the literature.

\subsection{Projection conditions for physical states}

As discussed in Sec.~\ref{sec:UnTwisted}, an orbifold is obtained by identifying
points in flat space which transform into each other under the action of the
space group,
\begin{equation}
 x~\sim~g x\;, \quad x \in \mathbbm{R}^6\;, \quad g \in \mathbbm{S}\;.
\end{equation}
Quantized strings whose boundary conditions are related by a symmetry transformation
must lead to the same Hilbert space of physical states.
In particular, strings with the boundary conditions
\begin{equation}
 \phi(\sigma + 2\pi)~\sim~g\phi(\sigma)\quad {\rm and}\quad
 \phi(\sigma + 2\pi)~\sim~hgh^{-1} \phi(\sigma) 
\end{equation}
produce the same Hilbert space  for any  $h\in \mathbbm{S}$ \cite{Dixon:1986jc}.
Here $\phi$ stands for $Z^i$,  $Z^{*i}$, $H^i$ and $X^I$. For each
\emph{conjugacy class} consisting of  elements   $h\,g\,h^{-1}$ one therefore
has a separate Hilbert space.

Space group elements $\bar{h}$ which commute with $g$,   i.e.\
$\bar{h}\,g\,\bar{h}^{-1}=g$, leave the string boundary conditions invariant. 
Hence, their representation in the Hilbert space must act as the identity on
physical states,
\begin{equation}\label{phys}
 \bar{h}\,|\mathrm{phys}\rangle~=~|\mathrm{phys}\rangle\;.
\end{equation}
This is the invariance or `projection' condition for 
 physical states.

A space group element $\bar{h}=(\theta^{\bar{k}},\bar{\ell})$ acts as a
translation on the  center--of--mass coordinates of the bosonic fields $H^i$ and
$X^I$  (cf.~\eqref{eq:BosonicBoundaries}),
\begin{equation}
 h^i~\rightarrow~h^i - \pi\, \bar{k}\,v_N^i\;, \quad
 x^I~\rightarrow~x^I + \pi\, \left(\bar{k}\,V_N^I + \bar{m}_a\, W^I_{na}\right) \;.
\end{equation}
Hence, the momentum eigenstates in twisted sectors transform as
\begin{eqnarray}
 |f;q,p\rangle & \rightarrow &
 e^{2\pi\I\, (-\bar{k}\,v_N\cdot (q + kv_N) + 
 (\bar{k}\,V_N + \bar{m}_a\,W_{na}) \cdot (p + V_f))}\, |f;q,p\rangle \;,
 \label{eq:Transform}
\end{eqnarray}
and similarly for untwisted states.
From Eqs.~(\ref{modeZt}) and (\ref{modeZ*t}) one reads off the transformation
properties of the creation operators,
\begin{equation}
 \widetilde{\alpha}^i_{fn}~\rightarrow~
 e^{2\pi\I\, \bar{k}\,v_N^i}\, \widetilde{\alpha}^i_{fn}\;, \quad
 \widetilde{\alpha}^{*i}_{fn}~\rightarrow~
 e^{-2\pi\I\, \bar{k}\,v_N^i}\, \widetilde{\alpha}^{*i}_{fn}\;.
\end{equation}
A state with non--vanishing oscillator numbers then transforms as 
\begin{eqnarray}
 \lefteqn{\widetilde{\alpha}^i_{fn}\ldots\widetilde{\alpha}^{*i}_{fm}\ldots
 |f;q,p\rangle\rightarrow} 
 \nonumber \\
 & &{}e^{2\pi\I\, \left(\bar{k}\,v_N\cdot \left(\widetilde{N} - \widetilde{N}^*\right)
 -\bar{k}\,v_N\cdot (q+kv_N) + 
 (\bar{k}\,V_N + \bar{m}_a\,W_{na})\cdot (p+V_f)\right)} 
 \,\widetilde{\alpha}^i_{fn}\ldots\widetilde{\alpha}^{*i}_{fm}\ldots
 |f;q,p\rangle\;.
 \nonumber\\
 \label{transf}
\end{eqnarray}
Physical states have to satisfy Eq.~(\ref{phys}), which yields the projection
conditions
\begin{eqnarray}
 \lefteqn{\bar{k}v_N\cdot \left(\widetilde{N}_f - \widetilde{N}_f^*\right) -
 \bar{k}v_N\cdot (q+kv_N)}
 \nonumber\\
 & &\hspace*{3cm} {} + (\bar{k}V_N + \bar{m}_a\,W_{na})\cdot (p+V_f)
 ~=~
 0\mod 1\;,
 \label{phase}
\end{eqnarray}
for values of $\bar{k}$ and $\bar{m}^a$ which depend on the conjugacy class. As
we will discuss in Sec.~\ref{sec:ProjectionTwistedSector}, in non--prime
orbifolds Eq.~\eqref{phase} gets modified for higher twisted sector states.
Below we analyze in detail the projection conditions for the untwisted and twisted sectors.

\subsubsection{Untwisted sector}

The untwisted sector ($k=0$) is associated with the space group element 
$g=(\mathbbm{1},0)$, and Eq.~(\ref{phase}) has to be satisfied  for the full
space group, i.e., for all values $\bar{k}$ and  $\bar{m}_a$. This yields the
projection conditions
\begin{equation}
 v_N\cdot q - V_N\cdot p~=~0~{\rm mod}~1\;, \quad W_n\cdot p~=~0~{\rm mod}~1\;,
\end{equation}
where $p$ is the $\E8\times\E8$ root lattice momentum ($p^2=2$)   and 
$q$ is the \SO8  weight lattice momentum ($q^2=1$). 
The \E8 momenta lie on the same lattice as the \E8 coordinates because of self-duality. 

The untwisted sector contains gauge and matter supermultiplets of the 4D effective theory.
For the former $v_N\cdot q=0$ mod 1 yielding gauge bosons with $q=(0^3;\pm1)$
and gauginos with $q=\pm\left(\tfrac{1}{2},\tfrac{1}{2},\tfrac{1}{2};\tfrac{1}{2}\right)$.
For the matter multiplets, $v_N\cdot q=n/N$ mod 1 with $n=1,..,N-1$ leading to the
bosonic \SO8 momenta $({\underline {\pm1,0,0}};0)$ where the underline denotes permutations,
and their fermionic partners.

Since gauge multiplets satisfy $v_N\cdot q=0$ mod 1, the conditions
\begin{equation}\label{4dgauge}
 V_N\cdot p~=~0~{\rm mod}~1\;, \quad  W_n\cdot p~=~0~{\rm mod}~1\;,
\end{equation}
determine the roots $p$ of the unbroken 4D gauge group. 
It is instructive to rewrite this 
set of equations as 
\begin{equation}
 V_f\cdot p~=~0~{\rm mod}~1\;,\quad \text{for all fixed points}\: f\;,
\end{equation}  
where $V_{f}$ is the \emph{local} shift (\ref{localt}) associated with the 
fixed point.
At each fixed point the gauge group is  broken  locally   to a 
subgroup of $\E8\times\E8$. The states surviving all local projection 
conditions, i.e., those corresponding to the intersection of all local gauge groups, yield the gauge 
fields of the low-energy gauge group.

Matter multiplets ($v_N\cdot q \not=0$ mod 1) originate from the 10D gauge fields
polarized in the compact directions and their fermionic partners. They form
chiral superfields transforming as the coset of $\E8\times\E8$ and the unbroken 4D gauge group.
All untwisted states are bulk fields in the compactified dimensions.

\subsubsection{Twisted sectors}
\label{sec:ProjectionTwistedSector}

For the twisted sectors $T_k$  ($k=1\ldots N-1$), the projection conditions depend
on $k$. 
Consider  $k=1$  and a fixed point $f$ with the space group
element $g=(\theta, \ell)$. The space group elements commuting with $g$ are
$\bar{h}=(\theta^{\bar{k}}, \bar{\ell}) = (\theta, \ell)^n$,  $n\in
\mathbbm{N}$. The resulting  projection condition is 
\begin{equation}\label{cond}
 v_N\cdot \left(\widetilde{N}_f - \widetilde{N}_f^*\right)
 - v_N\cdot (q+v_N) + V_f \cdot (p+V_f)~=~0\mod 1\;,
\end{equation}
where $V_f = V_N + m_aW_{na}$. 
Using `strong' modular invariance (\ref{eq:StrongModularInvariance}), one 
can show that all massless states (cf.~(\ref{eq:MassEquations})) satisfy
this condition. Therefore all massless modes in the first twisted
sector correspond to physical states. In the case of prime orbifolds, Eq.~(\ref{cond}) also holds for
higher twisted sectors with $V_f = k\,V_N + m_a\,W_{na}$.

For non--prime orbifolds the situation is more complicated. Some of the higher
twisted sectors $T_k$, $k>1$, are  related to lower order twists
$\mathbb{Z}_{N/k}$  which leave one of the $\mathbb{T}^2$ tori invariant. This
results in additional projection conditions.  Furthermore, fixed points of the
lower order twists  are not necessarily fixed points of the original twist
$\mathbb{Z}_{N}$. The $\mathbb{Z}_{N}$ twist transforms these fixed points into
each other such that they are mapped into the same singular point in the 
fundamental domain of the orbifold. Physical states correspond to linear
combinations of the states appearing at the fixed points of the
$\mathbb{Z}_{N/k}$ twist.

The conjugacy classes of higher twisted sectors $T_k$   are given by 
$h\,g\,h^{-1}$ where both $g$ and $ h$ have the form $(\theta^k, \ell)$. The
number of the conjugacy classes  is the number of the fixed points of the lower
order twist $\mathbb{Z}_{N/k}$. In general,  twists of other orders  
$\mathbb{Z}_{N/k'}$   transform these classes into each other. In particular,
the $\mathbb{Z}_{N}$ twist acts on the  $\mathbb{Z}_{N/k}$ conjugacy classes
$g_i$  as 
\begin{equation}\label{mixed}
 \bar{h}\,g_1\,\bar{h}^{-1}~=~g_2\;,\quad
 \bar{h}\,g_2\,\bar{h}^{-1}~=~g_3\;,\quad\dots\quad
 \bar{h}\,g_n\,\bar{h}^{-1}~=~g_1\;,
\end{equation}
with  $\bar{h}$ of the form $(\theta, \ell)$ and $n\ge 1$.
In this case, the higher twisted states transform as 
\begin{eqnarray}
 \bar{h}\,|1\rangle~=~|2\rangle\;, \quad
 \bar{h}\,|2\rangle~=~|3\rangle\;,\quad
 \dots\quad
 \bar{h}\,|n\rangle~=~|1\rangle\;.
 \label{physmixed}
\end{eqnarray}
From linear combinations of these localized states one obtains a basis of
physical states which are $\mathbb{Z}_{N}$ twist and  $\bar{h}$--eigenstates 
\cite{Dixon:1986jc,Dixon:1986qv,Kobayashi:1991rp},
\begin{equation}\label{eq:qgammaStates}
 |\mathrm{phys},q_\gamma\rangle
 ~=~\frac{1}{\sqrt{n}}\sum\limits_{s=1}^n
 e^{-2\pi\I\,s\,q_\gamma}|s\rangle\;,
\end{equation} 
where $q_\gamma=0,1/n,2/n,\dots, 1$. 
\nomenclature[qgamma]{$q_\gamma$}{additional quantum number in $T_{k>1}$ twisted
sectors of non--prime orbifolds\refeqpage}
As a consequence,
\begin{eqnarray}
 \bar h\,|\mathrm{phys},q_\gamma\rangle
 &=&
 e^{2\pi\I\,q_\gamma}\,e^{ 2\pi\I
 \left(\bar k v_N\cdot \left(\widetilde{N}_f - \widetilde{N}_f^*\right)
 - \bar k v_N\cdot (q+kv_N) + (\bar k V_N +\bar m_a W_a)\cdot (p+V_f)\right)}\,
\nonumber\\ 
&\times& |\mathrm{phys},q_\gamma\rangle\;,
\end{eqnarray}
where we have  used Eq.~\eqref{transf} and 
$\bar h=(\theta^{\bar k},\bar m_a\,e_a)$ is assumed to mix the conjugacy classes of $T_k$ as above.
This
leads to the modified projection conditions for the  superpositions
\eqref{eq:qgammaStates}:
\begin{eqnarray}
 \lefteqn{\bar{k}v_N\cdot \left(\widetilde{N}_f - \widetilde{N}_f^*\right) -
 \bar{k}v_N\cdot (q+kv_N)}
 \nonumber\\
 & &\hspace*{2cm}{} + (\bar{k}V_N + \bar{m}_a\,W_{na})\cdot (p+V_f)+q_\gamma
 ~=~
 0\mod 1\;.
 \label{phaseqgamma}
\end{eqnarray}

In this paper we are especially interested in a $\Z{6-\mathrm{II}}$ orbifold 
which has \Z3 and \Z2 subtwists with invariant tori. 
The corresponding twist vector is $v_6= (-1/6,-1/3,1/2)$.
As we shall discuss in
detail in Sec.~\ref{sec:Z3xZ2geometry}, two different fixed points in the
$T_{2,4}$ twisted sectors are related by Eq.~(\ref{mixed}) with
$\bar{h}=(\theta^3,0)$. 
The eigenstates of $(\theta^3,0)$ are 
\begin{equation}
 |\mathrm{phys},\pm\rangle~=~
 {1\over \sqrt{2}} \left(|1\rangle \pm |2\rangle\right) \;,
\end{equation}
where the states $|1\rangle, |2\rangle$  correspond to the two fixed points
of $\theta^2$ away from the origin.
The projection condition  (\ref{phaseqgamma}) becomes 
\begin{equation}\label{eq:ProjectionT2T4}
 3v_6\cdot \left(\widetilde{N}_f - \widetilde{N}_f^*\right)
 - 3v_6\cdot (q+k\,v_6) + 3V_6 \cdot (p+V_f) +  q_{\gamma}~=~0\mod 1\;,
\end{equation} 
with $q_{\gamma}=1/2,1$ for $k=2,4$. Here $V_f = k\,V_6 + m_3W_3$ is the local
\Z3 gauge shift.  Physical states of $T_{2,4}$ must also satisfy additional
projection conditions which stem from invariance of the third  $\mathbb{T}^2$ 
torus (`the \SO4 torus')  under $\theta^2$. Clearly, translations $\ell_3$ in
this torus commute with $\theta^2$. Thus invariance under space group
transformations $ (\mathbbm{1},\ell_3 )$ requires     
\begin{subequations}
\begin{eqnarray}
 W_2\cdot (p+V_f) &=&  0\mod 1\;, \\*
 W'_2\cdot (p+V_f) &=&  0\mod 1\;, 
\end{eqnarray}
\end{subequations}
where $W_2$ and $W_2'$ are two discrete Wilson lines in  the \SO4 torus.

The $\Z{6-\mathrm{II}}$ orbifold also has a \Z2 subtwist. 
The fixed points of the $T_3$ twisted
sector are mapped into each other by the space group element 
$\bar{h}=(\theta^2,0)$. 
Invariance under $\theta^2$ leads to 
the projection condition 
\begin{equation}\label{eq:ProjectionT3}
 2v_6\cdot \left(\widetilde{N}_f - \widetilde{N}_f^*\right)
 - 2v_6\cdot (q+3v_6) + 2V_6 \cdot (p+V_f) +  q_{\gamma}~=~0\mod 1\;,
\end{equation} 
with $q_{\gamma}=1/3,2/3,1$ and 
the local \Z2 gauge shift   $V_f = 3V_6 + n_2W_2 + n_2'W_2'$.
The \Z2 twist leaves the second torus (`the \SU3 torus') invariant.
Invariance of the $T_3$ states  under translations in this torus requires
\begin{equation}
 W_3\cdot (p+V_f)~=~ 0\mod 1\;.
\end{equation}
Here $W_3$ is a discrete Wilson line in the \SU3 torus. 

$T_5$ twisted sector contains  anti--particles of the $T_1$ sector, and will not
be treated separately in the following.

The above projection conditions are relevant to our model. 
A sample calculation of the physical spectrum is given
in App.~\ref{sec:SampleCalculation}.

\subsection{Local GUTs}

Consider a fixed point $f$ which is associated with the  local gauge shift
$V_f^I = k\,V^I_N + m_a\, W^I_{na}$. A local GUT can be defined by the
$\E8\times\E8$ roots $p$ ($p^2=2$) satisfying
\begin{equation}
p \cdot V_f~=~0 \mod 1 \;. 
\label{localGUT}
\end{equation}
These roots represent a local gauge symmetry supported at the fixed point. 
Twisted matter appears in a representation of the local GUT. Each representation
is characterized by the square of the shifted momentum, $(\widetilde{p}+V_f)^2$,
which is the same for members of the same multiplet.   

The concept of local GUTs is important for construction of realistic models. In
particular, all massless states of the $T_1$ sector survive the $\Z{N}$ 
projection and represent physical states. They form complete multiplets of the
corresponding local GUT, although this GUT does not appear in 4D. As discussed
in Sec.~\ref{sec:Intro}, this may naturally explain  why the SM gauge (and
Higgs) bosons do not form complete GUT multiplets, while the matter fields do.

Let us illustrate how a local SO(10) structure arises. Consider a
$\Z{6-\mathrm{II}}$ heterotic orbifold based on the Lie lattice  $\G2 \times
\SU3 \times \SO4$ with $v_6=(-1/6,-1/3,1/2;0)$, the gauge shift 
\begin{equation}
 V_6 ~=~
 \left(\frac{1}{2},\frac{1}{2},\frac{1}{3},0,0,0,0,0\right) \, 
 \left(\frac{1}{3},0,0,0,0,0,0,0\right)\;,
\end{equation}
and arbitrary Wilson lines.

The local gauge shift at the origin in the $T_1$ sector is $V_f= V_6$. The local
GUT roots are found from
\begin{equation}
p \cdot V_6 ~=~0 \mod 1 \;. 
\end{equation}
This corresponds to $\SO{10}\times\SU{2}^2$ symmetry in the observable sector. 
The $\SO{10}$ roots are given by
\begin{equation}
p~=~(0,0,0,\underline{\pm 1,\pm 1,0,0,0}) \;,
\end{equation}
where the underline denotes all possible permutations of the corresponding 
entries.

Relevant twisted matter fields of the $T_1$ sector satisfy the masslessness 
condition\footnote{The invariance conditions \eqref{cond} are satisfied 
automatically once $V_6$ is brought to the `strong' modular invariant form
\eqref{eq:StrongShift}.}
\begin{equation}
(p+V_6)^2_\mathrm{obs}~=~ {23\over 18}
\end{equation}
for the  $p+V_6$ components in the first \E8.
The solution is
\begin{equation}
(p+V_6)_\mathrm{obs}~=~
 \left(0, 0, -\tfrac{1}{6}, 
 \text{odd}\:(\pm\tfrac{1}{2})^5\right) \;,
\end{equation}
where ``odd $(\pm 1/2)^5$'' denotes all combinations containing an odd
number of minus signs. 
This is a $\boldsymbol{16}$--plet of \SO{10}. The $\Z6$ invariant  states  have
the right--mover shifted momentum $q+v_6=(-1/6,-1/3,-1/2;0)$ for space--time
bosons and analogously for space--time fermions. All of these states appear in
the physical spectrum of the model.

The Wilson lines can be chosen such that the gauge group in 4D is
that of the standard model (times extra factors). This does 
not affect the above considerations and the local \SO{10} GUT structure
remains intact.

\section{Geometry of the  $\boldsymbol{\Z{6-\mathrm{II}}}$ orbifold}
\label{sec:Z3xZ2geometry}

In this section we describe geometrical features of the
$\Z{6-\mathrm{II}}=\Z3\times\Z2$ orbifold based on the $\G2 \times \SU3 \times
\SO4$ Lie algebra lattice, which is required for construction of our model.

\subsection{Fixed points and fundamental region}

\begin{figure}[!h]
\begin{center}
\CenterObject{\includegraphics{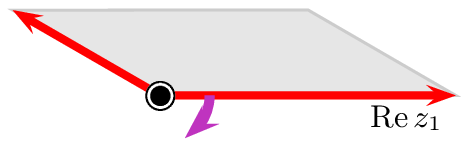}}
$\times$
\CenterObject{\includegraphics{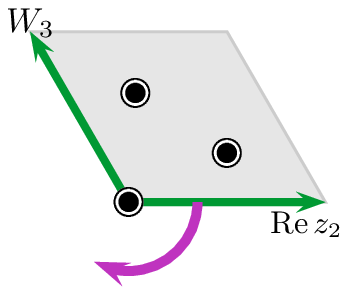}}
$\times$
\CenterObject{\includegraphics{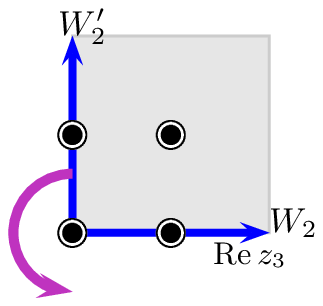}}
\end{center}
\caption{   $\G2 \times \SU3 \times \SO4$ torus lattice of the $\Z{6-\mathrm{II}}$ orbifold.
Possible Wilson lines  are denoted by $W_3$, $W_2$ and $W_2'$.} 
\label{fig:G2SU3SO4torus}
\end{figure}

\begin{figure}[!h]
\begin{center}
\begin{tabular}{ccc}
\CenterObject{\includegraphics{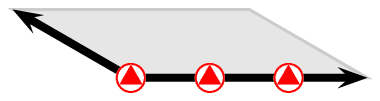}}
&
\CenterObject{\includegraphics{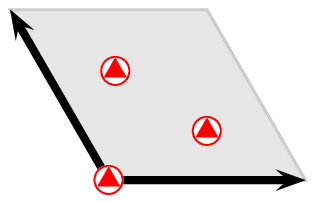}}
&
\CenterObject{\includegraphics{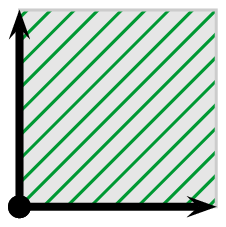}}
\end{tabular}
\end{center}
\caption{\Z3 fixed points.}
\label{fig:Z3fp}
\end{figure}

\begin{figure}[!h]
\begin{center}
\begin{tabular}{ccc}
\CenterObject{\includegraphics{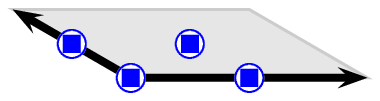}}
&
\CenterObject{\includegraphics{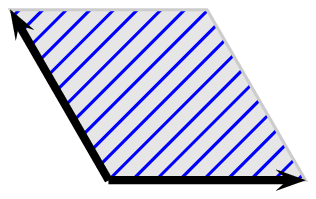}}
&
\CenterObject{\includegraphics{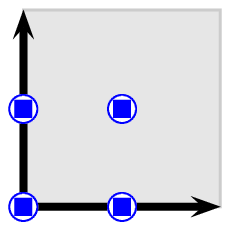}}
\end{tabular}
\end{center}
\caption{\Z2 fixed points.}
\label{fig:Z2fp}
\end{figure}

The $\Z{6-\mathrm{II}}$ orbifold with the $\G2 \times \SU3 \times \SO4$ lattice
is based on the twist vector\footnote{The overall sign of $v_6$ is chosen such
that one obtains left--chiral states ($q_4=-1/2$ for fermions) in the first
twisted sector. This convention differs from that of  our earlier work
\cite{Buchmuller:2005jr}.}
\begin{equation}
 v_6~=~\frac{1}{6}(-1,-2,3;0)\;.
\end{equation}
This orbifold  allows  for one discrete Wilson line of degree 3 in the \SU3 plane   
and two Wilson lines   of degree 2 in the \SO4 plane. 
The $\mathbbm{Z}_6$ action on the torus coordinates $z^i$,
\begin{equation}\label{eq:Rotationz}
z^i \rightarrow e^{2\pi\I {v_6}^i} z^i\;,
\end{equation}
is illustrated in Fig.~\ref{fig:G2SU3SO4torus}. This orbifold  has $\mathbbm{Z}_6$,
$\mathbbm{Z}_3$ and $\mathbbm{Z}_2$ fixed points  defined by
\begin{equation}
f^i - e^{2\pi \I \frac{6}{\kappa}{v_6}^i} f^i \in
\Lambda_{\G2 \times \SU3 \times \SO4}\;, \quad \kappa = 6, 3, 2\;,
\end{equation}
where $\Lambda_{\G2 \times \SU3 \times \SO4}$ is the torus lattice. 
The 12 \Z6 fixed points are shown in Fig.~\ref{fig:G2SU3SO4torus},
the 9 \Z3 fixed points -- in Fig.~\ref{fig:Z3fp} and the 16 \Z2 fixed points
-- in Fig.~\ref{fig:Z2fp}. 
It is a characteristic feature of 
non-prime orbifolds that the \Z3 and \Z2 fixed points are generally  different 
from the \Z6 fixed points. The  \Z3  subtwist  leaves the \SO4 plane
invariant, whereas  under  the \Z2  subtwist the  \SU3  plane is fixed. 


The orbifold is flat apart from the singular points (`conical singularities')
corresponding to the $\mathbbm{Z}_6$,
$\mathbbm{Z}_3$ and $\mathbbm{Z}_2$ fixed points.
Twisted states are localized at these singularities. In what follows, we
detail their localization properties in each $\mathbbm{T}^2$ torus.

\subsection[Twisted states location]
{Twisted states location}

\subsubsection{$\boldsymbol{\G2}$ plane}

\begin{figure}[!t]
 \subfigure[Modding out to the $\mathbbm{Z}_6$ `pillow'.]{%
 	\CenterObject{\includegraphics{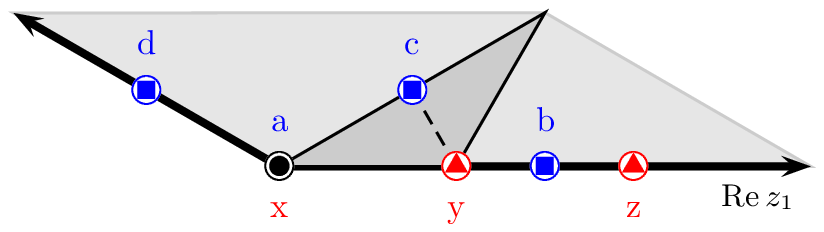}}}
 \hfil
   \subfigure[$\mathbbm{Z}_6$ `pillow'.]{%
   \CenterObject{\includegraphics{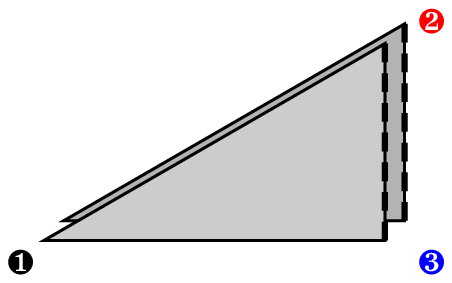}}}
\caption{The $\G2$ plane. The two simple roots of $\G2$ are given by the arrows
in (a) with  the shaded area spanned by them being the fundamental   region of
the torus. The fundamental region  of the orbifold is one sixth of this region
(darker area) and can be  represented by the `pillow' in (b).  The latter
corresponds to folding the fundamental region along the dashed edge and gluing
the other edges together (cf.\ \cite{Quevedo:1996sv2,Hebecker:2003jt}). }
 \label{fig:Z6triangle}
\end{figure}

In the $\mathrm{G}_2$ plane, there is one point fixed under \Z6 
located at the origin, 3  points x,y,z fixed under  \Z3, and
4 points a,b,c,d fixed under  \Z2 (Fig.~\ref{fig:Z6triangle}).
Some of them transform into each other under \Z6 twisting and
correspond to the same points in the fundamental domain of the orbifold.
For the three \Z3 fixed points
\begin{equation} 
  \text{\red x}\equiv\text{\ding{"0B6}}\;,\quad 
  \text{\red y}\equiv\text{\red\ding{"0B7}}\;,\quad 
  \text{\red z}\;,
\end{equation}
one has
\begin{equation}
\text{\red x}\to\text{\red x}\;, \quad 
\text{\red y}\to \text{\red z}\;, \quad
\text{\red z}\to\text{\red y}\;, 
\end{equation}
under the \Z2 twist $\theta^3$, and the four \Z2 fixed points 
\begin{equation}
  \text{\blue a}\equiv\text{\ding{"0B6}}\;,\quad 
  \text{\blue b}\;,\quad 
  \text{\blue c}\equiv\text{\blue\ding{"0B8}}\;,\quad
  \text{\blue d}\;, 
\end{equation}
transform under the \Z3 twist $\theta^2$     as 
\begin{equation}
\text{\blue a}\to\text{\blue a}\;, \quad 
\text{\blue b}\to\text{\blue c}\;, \quad
\text{\blue c}\to\text{\blue d}\;, \quad 
\text{\blue d}\to \text{\blue b}\;.
\end{equation}
Thus we have the following mapping from the fundamental domain of the torus  to
the fundamental domain of the orbifold:
\begin{equation}
 \left.\begin{array}{c}
  \text{\red x}\\
  {\blue\text{a}}
 \end{array}\right\} \to \text{\ding{"0B6}}\;,\quad
 \left.\begin{array}{c}
  \text{\red y}\\
  \text{\red z}
 \end{array}\right\} \to \text{\red\ding{"0B7}}\;,\quad
 \left.\begin{array}{c}
  {\blue\text{b}}\\
  {\blue\text{c}}\\
  {\blue\text{d}}
 \end{array}\right\} \to \text{\blue\ding{"0B8}}\;.
\end{equation}
Consequently, $T_{2,4}$ twisted matter lives at \ding{"0B6} or 
{\red\ding{"0B7}} points of the orbifold `pillow', whereas $T_3$ twisted matter lives at \ding{"0B6} or
{\blue\ding{"0B8}}.

As explained in Sec.~\ref{sec:ProjectionTwistedSector}, the fact that the \Z3
and \Z2 fixed points are not fixed under \Z6  introduces a new quantum number
for physical states, a phase $\gamma=e^{2\pi\I q_\gamma}$ with fractional
$q_\gamma$. Consider the  $T_{2,4}$ twisted sectors. Among the states localized
at {\red\ding{"0B7}}, there are two linear combinations 
\begin{equation}
|\text{\red\ding{"0B7}}; +1\rangle ~=~ {1\over \sqrt{2}}
\left(|{\red\text{y}}\rangle+|{\red\text{z}}\rangle\right)\;, \quad
|\text{\red\ding{"0B7}}; -1\rangle ~=~ {1\over \sqrt{2}}
\left(|{\red\text{y}}\rangle-|{\red\text{z}}\rangle\right)\;, 
\end{equation}
which are \Z2 (and \Z6) eigenstates 
with  eigenvalues $\gamma=\pm 1$,
\begin{equation}
\theta^3\, |\text{\red\ding{"0B7}}; +1\rangle ~=~ 
|\text{\red\ding{"0B7}}; +1\rangle\;, \quad
\theta^3\, |\text{\red\ding{"0B7}}; -1\rangle ~=~ 
- |\text{\red\ding{"0B7}}; -1\rangle\;.
\end{equation}
These  eigenstates  can  be labelled by the  order of the twist $k=2,4$
and the parameter $q_\gamma$,
\begin{equation}
|\text{\red\ding{"0B7}}; +1\rangle ~=~ |k=2,4;q_{\gamma} = 1\rangle\;, \quad
|\text{\red\ding{"0B7}}; -1\rangle ~=~ |k=2,4;q_{\gamma} = 1/2\rangle\;.
\end{equation}
The state at the origin has $\gamma=1$ and can be labelled as 
\begin{equation}
|\text{\red x}\rangle ~=~ |\text{\ding{"0B6}}; +1\rangle 
~=~ |k=2,4;q_{\gamma} = 0\rangle\;.
\end{equation}
To distinguish $\gamma=1$ states at {\ding{"0B6}} from those at
{\red\ding{"0B7}}, we assign  $q_{\gamma} = 0$ to the former and $q_{\gamma} =
1$ to the latter.

The $T_3$ states are  treated analogously.  There are  three  linear
combinations of states located at {\blue\ding{"0B8}}, with \Z3 eigenvalues  $1$,
$\omega \equiv e^{2\pi\I/3} $, and $\omega^{-1}$,
\begin{subequations}
\begin{eqnarray}
|\text{\blue\ding{"0B8}};1\rangle &=& 
{1\over \sqrt{3}}\left(|{\blue\text{b}}\rangle+
|{\blue\text{c}}\rangle+|{\blue\text{d}}\rangle\right)\;, \\
|\text{\blue\ding{"0B8}};\omega\rangle &=& 
{1\over\sqrt{3}}\left(|{\blue\text{b}}\rangle+
\omega^{-1}|{\blue\text{c}}\rangle+\omega^{-2}|{\blue\text{d}}\rangle\right)\;,\\
|\text{\blue\ding{"0B8}};\omega^{-1}\rangle &=& 
{1\over\sqrt{3}}\left(|{\blue\text{b}}\rangle+
\omega\,|{\blue\text{c}}\rangle+\omega^2\,|{\blue\text{d}}\rangle\right)\;.
\end{eqnarray}
\end{subequations}
The \Z3  (and \Z6)   eigenstates can again be characterized by the order of the twist 
and $q_{\gamma}$,
\begin{equation}
|\text{\blue\ding{"0B8}}; 1\rangle ~=~ |k=3;q_{\gamma} = 1\rangle\;, \quad
|\text{\blue\ding{"0B8}}; \omega^{\pm 1}\rangle~=~
|k=3;q_{\gamma}= \pm 1/3\rangle\;.
\end{equation}
The state at the origin is now labelled as 
\begin{equation}
|\text{\blue a}\rangle ~=~ |\text{\ding{"0B6}}; 1\rangle 
 ~=~ |k=3;q_{\gamma} = 0\rangle\;.
\end{equation}

\begin{figure}[h]
\centerline{\CenterObject{\includegraphics{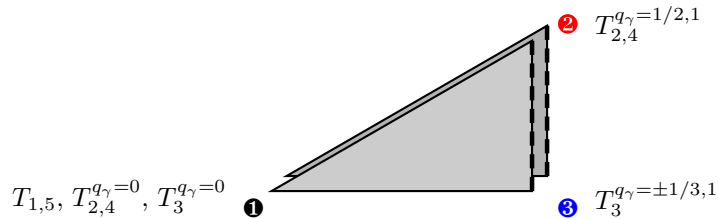}}}
 \caption{Location of the twisted states in the \G2 plane. } 
 \label{fig:LocationTk}
\end{figure}

The $T_{1,5}$  twisted sector states are  localized at the origin, which corresponds
to a \Z6 eigenstate with eigenvalue $\gamma=1$, i.e.,
\begin{equation}
|\text{\ding{"0B6}}; 1\rangle 
 ~=~ |k=1,5;q_{\gamma} = 0\rangle\;.
\end{equation}
The location of all $T_k$ twisted states is illustrated in
Fig.~\ref{fig:LocationTk}.

The effect of the quantum number $q_{\gamma}$ on the projection conditions
for physical states has been discussed in Sec.~\ref{sec:StringsOnOrbifold}.

\subsubsection{$\boldsymbol{\SU3}$ plane}

\begin{figure}[t]
\centerline{
\subfigure[\SU3 plane.\label{fig:n3}]{%
\CenterObject{\includegraphics{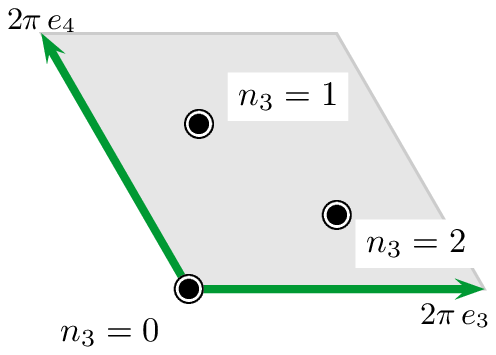}}}
\qquad
\subfigure[\SO4 plane.\label{fig:n2}]{%
\CenterObject{\includegraphics{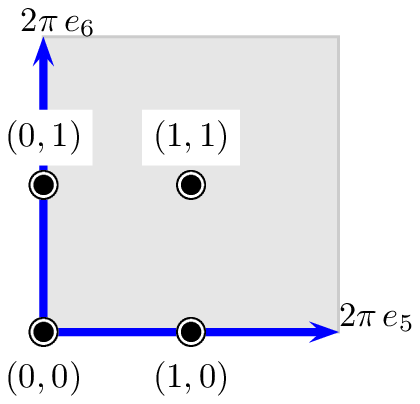}}}
}
\caption{Localization quantum numbers $n_3$, $n_2$ and
$n_2'$.}
\label{fig:n3n2}
\end{figure}

States twisted by $\theta^k$ with $k=1,2,4,5$ are localized at the three fixed
points in the \SU3 plane, whereas $T_3$ and untwisted states live in the bulk.
The localization is specified by the 
quantum number $n_3$ (cf.\ Fig.~\ref{fig:n3}). Tab.~\ref{tab:n3} lists the coordinates
of the fixed points in the \SU3 torus as well as the corresponding space group
elements. 
\nomenclature[n3]{$n_3$}{localization quantum number in the \SU3 plane\refpage}
The coordinates are defined up to translations in the
sublattice    $ 2\pi [ n\,e_3+(-n-3m)\,e_4 ]$ with $n,m\in\mathbbm{Z}$. 

\begin{table}[h]
\centerline{
$
\begin{array}{|c|c|cc|cc|cc|}
 \hline
  & \text{location} & \multicolumn{6}{|c|}{\text{space group element}}\\
 \cline{3-8}
 n_3 & \text{(in $2\pi$ units)}  & k=1 & m_3 & k=2 & m_3 & k=4 & m_3 \\
 \hline
 0 & 0 & (\theta,0) & 0 & (\theta^2,0) & 0 & (\theta^4,0) & 0\\
 1 & \frac{1}{3}e_3+\frac{2}{3}e_4 &
 (\theta,e_3) & 1 & (\theta^2,-e_3) & 2 & (\theta^4,e_3) & 1\\
 2 & \frac{2}{3}e_3+\frac{1}{3}e_4 & 
 (\theta,-e_3)& 2 & (\theta^2,e_3)& 1 & (\theta^4,-e_3)& 2\\
 \hline
 \end{array}
$}
\caption{ Localization quantum numbers and space group elements.}
\label{tab:n3}
\end{table}

As discussed in Sec.~\ref{sec:StringsOnOrbifold}, a fixed point or plane  with the 
space group element $(\theta^k,a\,e_3+b\,e_4)$ corresponds to the local gauge shift 
\begin{equation}\label{eq:Vf1}
 V_f~=~ k\,V_6 + m_3\, W_3\;, \quad m_3~=~a+b  ~ {\rm mod} ~3\;,
\end{equation} 
up to terms involving $W_2$ and $W_2'$. Note that $m_3$ depends not only on the
location ($n_3$) but also on the order of the twist $k$ (Tab.~\ref{tab:n3}).
The above local shift is equivalent to 
\begin{equation}
 V_f~=~k (V_6+n_3\,W_3) \;.
\end{equation}

\subsubsection{$\boldsymbol{\SO4}$ plane}

Twisted states from $T_{1,5}$ and $T_3$ are localized at the four fixed points in
the \SO4 plane whereas $T_2$, $T_4$ and untwisted states correspond to bulk
fields. The fixed points are labelled by $n_2$ and $n_2'$ (Fig.~\ref{fig:n2}). 
Tab.~\ref{tab:n2} lists the coordinates of the fixed points
and the corresponding space group elements. 
\nomenclature[n2]{$n_2$}{localization quantum number in the \SO4 plane\refpage}
\nomenclature[n2p]{$n_2'$}{localization quantum number in the \SO4 plane\refpage}
The coordinates are defined up to translations in the sublattice
$ 2\pi [  2n\,e_5+2m\,e_6 ] $ where $n,m\in\mathbbm{Z}$. The local shift for the
$\theta^k$ sectors ($k=1,3,5$) reads
\begin{equation}
 V_f~=~k\,(V_6+n_2\,W_2+n_2'\,W_2')
\end{equation}
up to terms involving $W_3$.

\begin{table}[!h]
\centerline{
$
\begin{array}{|c|c|c|c|}
 \hline
  & \text{location}& \multicolumn{2}{|c|}{\text{space group element}}\\
 \cline{3-4}
 (n_2,n_2') & \text{(in $2\pi$ units)} & k=1 & k=3 \\
 \hline
 (0,0) & 0 & (\theta,0) & (\theta^3,0) \\
 (0,1) & \frac{1}{2}e_6 &
 (\theta,e_6) & (\theta^3,e_6) \\
 (1,0) & \frac{1}{2}e_5 &
 (\theta,e_5)& (\theta^3,e_5) \\
 (1,1) & \frac{1}{2}(e_5+e_6) &
 (\theta,e_5+e_6)& (\theta^3,e_5+e_6) \\
 \hline
 \end{array}
$}
\caption{Localization quantum numbers and space group elements.}
\label{tab:n2}
\end{table}


\section{Superpotential}
\label{sec:CouplingsAndSelectionRules}

In this section, we discuss the superpotential couplings in heterotic orbifolds.
Interactions on orbifolds are calculated using superconformal field theories
\cite{Hamidi:1986vh,Dixon:1986qv}. This leads to a set of selection rules
dictating which couplings are allowed. For our purposes, it suffices to
identify the allowed couplings without knowing their precise strength. The
following discussion is closely related to the analysis of Kobayashi {\it et
al.} \cite{Kobayashi:2004ya}, with some extensions.\footnote{Certain
corrections to the selection rules of \cite{Kobayashi:2004ya} will
be discussed in detail in Ref.\cite{futurepaper}.  }

\subsection{Vertex operators and correlation functions}

In orbifold conformal field theory, couplings are obtained from correlation
functions of vertex operators for the corresponding physical states. The vertex
operators for bosons in the ($-1$)--ghost picture read (cf.\
\cite{Kobayashi:2004ya}):
\begin{equation}\label{eq:V-1}
 \V^{(f)}_{\!\!-1}~=~e^{-\phi}\, e^{2\I(q+kv_N)\cdot H}\,
 e^{2\I(p+V_f)\cdot X}
 \prod_{i=1}^3 \left(\overline{\partial} Z^i\right)^{\widetilde{N}_{fi}}
 \left(\overline{\partial} Z^{*\,i}\right)^{\widetilde{N}^*_{fi}}\sigma_f\;.
\end{equation}
\nomenclature[V-1]{$\V_{\!\!-1}$}{Bosonic vertex operator\refeqpage}
Here $q$, $k$, $p$, $f$ and $\widetilde{N}_{fi}$, $\widetilde{N}_{fi}^*$ are the
quantum numbers described in Sec.~\ref{sec:StringsOnOrbifold}, and $\sigma_f$ is
the twist field which creates the vacuum of the twisted sector at the fixed
point $f$ from the untwisted vacuum (cf.\
\cite{Hamidi:1986vh,Dixon:1986qv,Font:1988tp,Font:1988mm,Font:1989aj}); $\phi$
is the bosonized superconformal ghost (cf.\ \cite{Lust:1989tj}). Vertex
operators for untwisted states correspond to $k = V_f = 0$, $\sigma_f = 1$. 

In the 0--ghost picture, \eqref{eq:V-1} is replaced with 
\begin{eqnarray}
 \V^{(f)}_{\!\!0} & = & e^{2\I(q+kv_N)\cdot H}\, e^{2\I(p+V_f)\cdot X}\,
 \prod_{i=1}^3 \left(\overline{\partial} Z^i\right)^{\widetilde{N}_{fi}}
 \left(\overline{\partial} Z^{*\,i}\right)^{\widetilde{N}^*_{fi}} \nonumber
 \\
&&\hspace{3cm} 
\times\sum_{j=1}^3\left(e^{2\I H^j} \partial Z^j 
 + e^{-2\I H^j} \partial Z^{*\,j} \right)\sigma_f\;.\label{eq:V0}
\end{eqnarray}
\nomenclature[V0]{$\V_{\!\!0}$}{Bosonic vertex operator\refeqpage}
The vertex operator for fermions is given by
\begin{equation}
 \V^{(f)}_{\!\!-1/2}~=~e^{-{\phi\over 2}}\, e^{2\I(q+kv_N)\cdot H} \,
 e^{2\I(p+V_f)\cdot X}\,
 \prod_{i=1}^3 \left(\overline{\partial} Z^i\right)^{\widetilde{N}_{fi}}
 \left(\overline{\partial} Z^{*\,i}\right)^{\widetilde{N}^*_{fi}}\sigma_f\;.
\end{equation}
\nomenclature[V-1/2]{$\V_{\!\!-1/2}$}{Fermionic vertex operator\refeqpage}
In what follows, we will mainly be interested in the superpotential couplings.
These are extracted from couplings between 2 fermions and $n-2$ bosons given by
the correlation functions
\begin{equation}\label{eq:cor}
 \left\langle \V^{(f_1)}_{\!\!-1/2}\,\V^{(f_2)}_{\!\!-1/2}\,
 \V^{(f_3)}_{\!\!-1}\,\V^{(f_4)}_{\!\!0}
 \ldots \V^{(f_{n})}_{\!\!0}\right\rangle \;.
\end{equation}
The correlation function \eqref{eq:cor} factorizes into correlators involving
separately the fields $\phi$, $H$, $X^I$, $Z^i$ and the twist fields
\cite{Hamidi:1986vh,Dixon:1986qv,Font:1988tp,Font:1988mm,Font:1989aj}. $\Z{6}$
invariance of each correlator leads to various selection rules which we 
discuss in the following.

\subsection{Selection rules}

\subsubsection{Gauge invariance}

Consider a coupling of $n$ massless physical states labelled by index $r$. As
expected, the coupling has to obey gauge invariance. The gauge quantum numbers
are specified by the shifted momenta $p + V_f$ which play the role of the
weight vectors w.r.t.\ the unbroken subgroup of $\E8\times\E8$. For the 
correlation function to be non--zero, the states have to form a gauge singlet, 
\begin{equation}\label{eq:GaugeInvariance}
 \sum_{r=1}^n (p + V_f)_{(r)}~=~0\;.
\end{equation}

It is instructive to interpret a coupling among twisted fields in terms of local
gauge groups. Suppose that the twisted states form representations
$\boldsymbol{R}$, $\boldsymbol{R'}$, etc. under the local non--Abelian gauge
groups $G_\mathrm{local}$, $G_\mathrm{local}'$, etc. Then the coupling among
these states is invariant under the intersection of these groups,
\begin{equation}\label{eq:intersectingG}
 G_\mathrm{intersection}~=~G_\mathrm{local}\cap G_\mathrm{local}'\cap G_\mathrm{local}''\cap\dots
 ~\subset~\E8\times\E8\;,
\end{equation}
which is given by the $\E8\times\E8$ roots common to all of the local groups.
The remaining gauge invariance conditions concern \U1 charges.
$\boldsymbol{R}$, $\boldsymbol{R'}$, etc. can be decomposed into representations
of $G_\mathrm{intersection}$ such that the invariant couplings involve the
latter. This implies, for instance, that a coupling between localized
$\boldsymbol{16}$--plets of $\SO{10}$ and other twisted states need not be
invariant under the full $\SO{10}$. As a result, a mass term for the SM
singlet in the $\boldsymbol{16}$--plet can be written without invoking large 
$\SO{10}$ representations such as $\boldsymbol{126}$--plets, which are necessary
in 4D GUTs.

\subsubsection{$\boldsymbol{H}$--momentum rules}

Twist invariance of the compact 6D space requires that the superpotential be
a scalar with respect to discrete rotations in the compact space. In other
words, the $H$--momenta must add up to zero (up to a discrete ambiguity). The
$H$--momenta invariant under the ghost picture changing are defined by 
\cite{Kobayashi:2004ya}
\begin{equation}
 R^i_{(r)}~=~(q^i+kv^i_6)_{(r)} - (\widetilde{N}_{fi}-\widetilde{N}^*_{fi})_{(r)}
\end{equation}
\nomenclature[Ri]{$R_i$}{invariant $H$--momenta\refeqpage}
and can be thought of as discrete $R$--charges \cite{Font:1988tp,Font:1988nc}.
They lie on the $\SO8$ weight lattice.

For an allowed coupling between 2 fermions and $n-2$ bosons, the sum of the
$H$--momenta must vanish. This rule can be reformulated in terms of bosonic
$H$--momenta only. Specifically,
 \begin{subequations}\label{eq:H-momentumRules}
\begin{eqnarray}
\sum_{r=1}^n R^1_{(r)} &=& -1\mod6\;, \\*
\sum_{r=1}^n R^2_{(r)} &=& -1\mod3\;, \\*
\sum_{r=1}^n R^3_{(r)} &=& -1\mod2\;,\label{eq:H-momentumRule3}
\end{eqnarray}
\end{subequations}
where $R^i_{(r)}$ are the $H$--momenta of the bosonic components of chiral
superfields. For the $\Z{6-\mathrm{II}}$ orbifold these are listed in
Tab.~\ref{tab:Example4s}.\footnote{Our sign convention is opposite to that of 
\cite{Kobayashi:2004ya}.}
\begin{table}[t]
\centerline{
\begin{tabular}{|l|l|}
\hline
$k$ & $H$--momentum\\
\hline
1 & $\frac{1}{6}(-1,-2,-3)$ \\
2 & $\frac{1}{3}(-1,-2,0)$ \\
3 & $\frac{1}{2}(-3,0,-3)$ \\
4 & $\frac{1}{3}(-2,-1,0)$ \\
5 & $\frac{1}{6}\left(-5 ,-4 ,-3\right)$\\
\hline
\end{tabular}}
\caption{$\Z{6-\mathrm{II}}$ orbifold: $H$--momenta for bosons 
containing no oscillators.}
\label{tab:Example4s}
\end{table}
We note that gauge invariance requires strict vanishing of the sum of 
$\E8\times\E8$ momenta, whereas the sum of $H$--momenta must vanish up to a 
discrete shift as given above. The difference between the two rules stems from
the fact that the gauge 16D torus possesses continuous symmetries, while in the
case of the 6D orbifold they are only discrete.

\subsubsection{Space group selection rules}

\begin{figure}[h]
\centerline{%
\subfigure[Allowed.]{\CenterObject{\includegraphics{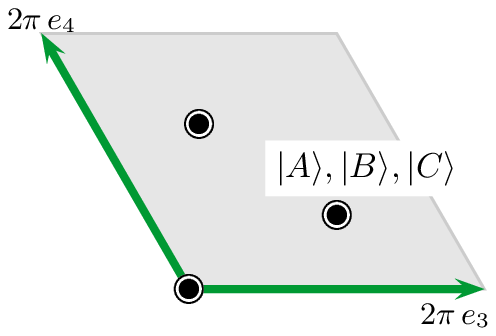}}}
\quad
\subfigure[Allowed.]{\CenterObject{\includegraphics{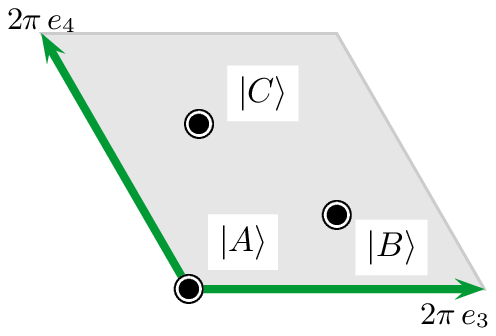}}}
\subfigure[Forbidden.]{\CenterObject{\includegraphics{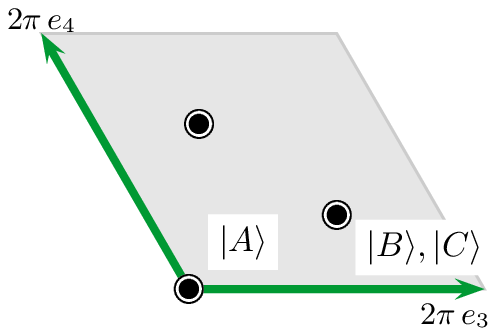}}}
}	
\caption{Allowed (a,b) and forbidden (c) 3--point couplings between localized
states $|A\rangle,|B\rangle,|C\rangle$ in the \SU3 plane.}
\label{fig:CouplingsInSU3}
\end{figure}

The space group selection rule \cite{Hamidi:1986vh,Dixon:1986qv} states that
the string boundary conditions have to match in order for the coupling to be
allowed. Consider twisted states living at the fixed points $f_1$, $f_2$,\dots
$f_n$ corresponding to the space group elements $(\theta^{k_{(1)}},\ell_{(1)})$,
$(\theta^{k_{(2)}},\ell_{(2)})$, \dots , $(\theta^{k_{(n)}},\ell_{(n)})$. A coupling of
these states is allowed if (cf.\ \cite{Font:1988tp})
\begin{equation}\label{sgr}
 (\theta^{k_{(1)}},\ell_{(1)})\,(\theta^{k_{(2)}},\ell_{(2)}) \dots 
 (\theta^{k_{(n)}},\ell_{(n)})~=~(\mathbbm{1}, 0 )
\end{equation}
up to a torus lattice vector $ \sum_{r=1}^n\Lambda_{k_{(r)}}$, where
$\Lambda_{k_{(r)}}=(\mathbbm{1}-\theta^{k_{(r)}})\Lambda$. The untwisted sector
corresponds to the space group element $(\mathbbm{1},0)$. The above condition
is equivalent to (cf.\ App.~\ref{sec:AESR}) 
\begin{subequations}\label{eq:SpaceGroupSelectionRules}
\begin{eqnarray}
 \sum\limits_{r=1}^n k_{(r)} & = & 0\mod 6\;,\\*
 \sum\limits_{r=1}^n 
 \ell_{(r)} & = & 0\mod {\sum_{r=1}^n \Lambda_{k_{(r)}}}
 \;.
\end{eqnarray}
\end{subequations}
The first equation restricts the twisted sectors that can couple and states 
that the total twist of the coupling must be $0\mod 6$. The second condition
puts a restriction on the fixed points. In terms of the localization quantum
numbers, it reads
\begin{subequations}\label{eq:SpaceGroupRules}
\begin{eqnarray}
 \label{eq:gammaRule}
  \text{\SU3\ plane} & : & 
 \sum\limits_{r=1}^n k_{(r)}\, n_{3(r)} ~=~ 0\mod3\;,\\*
 \text{\SO4\ plane}& : & 
 \sum\limits_{r=1}^n n_{2(r)}~=~ 0\mod2\;,\\*
 & & \sum\limits_{r=1}^n n^\prime_{2(r)}~= ~0\mod2\;,
 \label{eq:n2primeRule}
\end{eqnarray}
\end{subequations}
plus an additional condition to be discussed below. The quantum numbers
 $n_{3(r)}$, $n_{2(r)}$ and $n^\prime_{2(r)}$ have been defined
in Sec.~\ref{sec:Z3xZ2geometry}.\footnote{Note that the sum rule for the \SU3
plane differs from the corresponding rule in \cite{Kobayashi:2004ya} by the
factors $k_{(r)}$.} The space group selection rule for the \SU3 plane is
illustrated in Fig.~\ref{fig:CouplingsInSU3}.

For the \G2 plane, there is a non--trivial selection rule  if only $T_2$ and $T_4$,
or only $T_3$ states are involved in the coupling. As we show in 
App.~\ref{sec:AESR},  the coupling
must satisfy
\begin{equation}\label{eq:qgammarule2}
 \{q_{\gamma(1)},\dots q_{\gamma(n)}\}
 ~\not\in~\text{permutations}\{x,0,\dots0\}\;
\end{equation} 
for $x\not= 0$.

To summarize, we have presented the string selection rules which determine
whether a given superpotential coupling is allowed. Apart from gauge
invariance, such couplings enjoy certain discrete symmetries related to the
localization properties of the states involved.


\section{The MSSM from the heterotic string}
\label{sec:Model}

In this section, we present an orbifold compactification of the $\E8
\times \E8$ heterotic string which yields the MSSM spectrum and gauge
group at low energies. Apart from the MSSM sector, the model contains
a hidden sector which can account for low--energy
supersymmetry breakdown.  In this section we present  basic
features of the model, whereas other important aspects  such as vacuum
configurations, SUSY breaking, and phenomenology  will be
discussed in Secs.~\ref{sec:FandDflat}--\ref{sec:Pheno}.

\subsection{Search Strategy}

It is well known that with an appropriate choice of the gauge shift $V$ and
Wilson lines, it is not difficult to get the standard model gauge group times
extra group factors.  The real challenge  however is to get three generations of
the SM matter. 

We base our search on the concept of local GUTs. Since one complete matter
generation (plus a right--handed neutrino) is a $\boldsymbol{16}$--plet of
$\SO{10}$, we use the  gauge shifts which admit  local $\SO{10}$ symmetry and
$\boldsymbol{16}$--plets at the fixed points. There are only two such shifts in
a $\Z{6-\mathrm{II}}$ orbifold \cite{Katsuki:1989cs,Katsuki:1989qz},
\begin{eqnarray}
 V_6 & = &
 \left(\frac{1}{2},\frac{1}{2},\frac{1}{3},0,0,0,0,0\right) \, 
 \left(\frac{1}{3},0,0,0,0,0,0,0\right) 
 \;,\nonumber\\
 V_6' & = & 
 \left(\frac{1}{3},\frac{1}{3},\frac{1}{3},0,0,0,0,0\right) \, 
 \left(\frac{1}{6},\frac{1}{6} ,0,0,0,0,0,0\right) \;. 
\end{eqnarray}
Each of them ensures that there are $\boldsymbol{16}$--plets in the
$T_1$ sector, which remain in the massless spectrum regardless of the
Wilson lines.  Further, one adjusts the Wilson lines such that the
gauge group in 4D is that of the standard model times additional
factors.

\begin{figure}[h]
\centerline{\hfill%
\subfigure[{}\label{fig:3SequentialFamilies}]{%
\CenterObject{\includegraphics{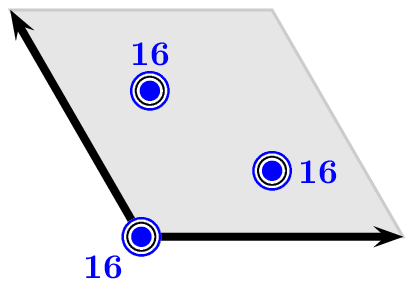}}}
\hfill\subfigure[{}\label{fig:2SequentialFamilies}]{%
\CenterObject{\includegraphics{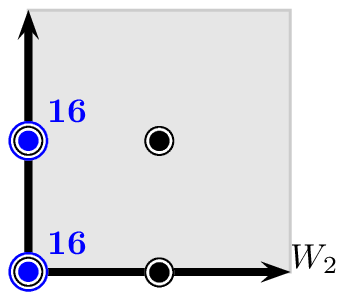}}}
\hfill
}
\caption{3 vs.\ $2$ equivalent families.}
\label{fig:SequentialFamilies}
\end{figure}

To obtain three matter generations, the simplest option is to use  three
equivalent fixed points with local $\SO{10}$ symmetry \cite{Buchmuller:2004hv},
Fig.~\ref{fig:3SequentialFamilies}.   This  would provide an intuitive
explanation for triplication of fermion families. However, our scan over such
models shows that in this case there are always chiral exotic states in the
spectrum (cf.\ App.~\ref{app:NoGo}).\footnote{We find that some models have
exotic matter which is vector--like with respect to $\SU{3}_c \times
\SU{2}_\mathrm{L}$ but chiral with respect to correctly normalized $\U1_Y$. In
particular, our earlier model \cite{Buchmuller:2004hv} suffers from this
problem.}  Such states get masses due to electroweak symmetry breaking and
generally are inconsistent with experiment. A similar statement applies to
other  $\Z{n}$  orbifolds with $n \leq 6$.

This result implies that the three families of
$\boldsymbol{16}$--plets are \emph{not} all equivalent, at least in
the context of $\Z{\leq 6}$ orbifold models. We are thus led to
consider the next--to--simplest possibility: 2 equivalent families and
one different family, Fig.~\ref{fig:2SequentialFamilies}. The
equivalent $\boldsymbol{16}$--plets can appear due to 2 equivalent
fixed points in the SO(4) plane with one Wilson line $W_2$. The
remaining family then has to come from other sectors of the
model. We find that this procedure is successful and, in many cases,
the exotic matter is vector--like with respect to the standard
model. Furthermore, we find that the vector--like matter can be
consistently decoupled at least in one case.

\subsection{The model}

Our model is a $\Z{6-\mathrm{II}}$ heterotic orbifold based on the Lie lattice 
$\G2 \times \SU3 \times \SO4$. It involves two Wilson lines: one of order 2,
$W_2$, and another of order 3, $W_3$, and has the gauge shift $V_6$  consistent
with the local SO(10) structure.  Specifically, the gauge shift and the Wilson
lines  are given by \cite{Buchmuller:2005jr}
\begin{eqnarray}
 V_6 & = &
 \left(\frac{1}{2},\frac{1}{2},\frac{1}{3},0,0,0,0,0\right) \, 
 \left(\frac{1}{3},0,0,0,0,0,0,0\right) 
 \;,\nonumber\\
 W_2 & = & 
 \left(\frac{1}{2},0,\frac{1}{2},\frac{1}{2},\frac{1}{2},0,0,0\right) 
 \,\left(-\frac{3}{4},\frac{1}{4},\frac{1}{4},-\frac{1}{4},\frac{1}{4},\frac{1}{4},\frac{1}{4},-\frac{1}{4}\right) 
 \;,\nonumber\\
 W_3 & = &
 \left(\frac{1}{3},0,0,\frac{1}{3},\frac{1}{3},\frac{1}{3},\frac{1}{3},\frac{1}{3}\right) \, 
 \left(1,\frac{1}{3},\frac{1}{3},\frac{1}{3},0,0,0,0\right)
 \;.
\end{eqnarray}
By adding elements of the root lattice $\Lambda_{\E8\times\E8}$ to the shift and
Wilson lines, one can transform this set to
\begin{eqnarray}
 V_6' & = &
 \left(-\frac{1}{2},-\frac{1}{2},\frac{1}{3},0,0,0,0,0\right) 
 \,\left(\frac{17}{6},-\frac{5}{2},-\frac{5}{2},-\frac{5}{2},-\frac{5}{2},-\frac{5}{2},-\frac{5}{2},\frac{5}{2}\right) 
 \;,\nonumber \\
 W_2' & = &
 \left(-\frac{1}{2},0,-\frac{1}{2},\frac{1}{2},\frac{1}{2},0,0,0\right) 
 \,\left(\frac{23}{4},-\frac{25}{4},-\frac{21}{4},-\frac{19}{4},-\frac{25}{4},-\frac{21}{4},-\frac{17}{4},\frac{17}{4}\right) 
 \;,\nonumber\\
 W_3' & = &
 \left(-\frac{1}{6},\frac{1}{2},\frac{1}{2},-\frac{1}{6},-\frac{1}{6},-\frac{1}{6},-\frac{1}{6},-\frac{1}{6}\right)
 \,\left(0,-\frac{2}{3},\frac{1}{3},\frac{4}{3},-1,0,0,0\right)
 \;,\label{eq:StrongShift}
\end{eqnarray}
which fulfills the `strong' modular invariance conditions
\eqref{eq:StrongModularInvariance}.

The gauge group after compactification is 
\begin{equation}
 G~=~\SU{3}\times\SU{2}\times[\SU{4}\times\SU2']\times\U1^9\;.
\label{G}
\end{equation}
Here the brackets $[\dots]$ indicate a subgroup of the second \E8 factor.
The generators of the \U1 factors can be chosen as
\begin{eqnarray}\label{eq:U1generatorsSaturdayAfternoon}
 \mathsf{t}_1 & = & \mathsf{t}_Y ~=~
 \left(0,0,0,\frac{1}{2},\frac{1}{2},-\frac{1}{3},-\frac{1}{3},-\frac{1}{3}\right) 
  \, (0,0,0,0,0,0,0,0)\;,\nonumber\\ 
 \mathsf{t}_2 
 & = & 
 (1,0,0,0,0,0,0,0) \, (0,0,0,0,0,0,0,0) 
 \;,\nonumber\\
 \mathsf{t}_3 & = &
 (0,1,0,0,0,0,0,0) \, (0,0,0,0,0,0,0,0) 
 \;,\nonumber\\
 \mathsf{t}_4 & = & 
 (0,0,1,0,0,0,0,0) \, (0,0,0,0,0,0,0,0) 
 \;,\nonumber\\
 \mathsf{t}_5 & = &
 (0,0,0,1,1,1,1,1) \, (0,0,0,0,0,0,0,0) 
 \;,\nonumber\\
 \mathsf{t}_6 & = & 
 (0,0,0,0,0,0,0,0) \, (1,0,0,0,0,0,0,0) 
 \;,\nonumber\\
 \mathsf{t}_7 & = & 
 (0,0,0,0,0,0,0,0) \, (0,1,1,0,0,0,0,0) 
 \;,\nonumber\\
 \mathsf{t}_8 & = & 
 (0,0,0,0,0,0,0,0) \, (0,0,0,1,0,0,0,0) 
 \;,\nonumber\\
 \mathsf{t}_9 & = & 
 (0,0,0,0,0,0,0,0) \, (0,0,0,0,-1,-1,-1,1)
 \;.\label{eq:U1generators}
\end{eqnarray}
One of the \U1 factors is `anomalous'. It is generated by
\begin{equation}
 \mathsf{t}_\mathrm{anom}~=~\sum\limits_{i=1}^8c_i\,\mathsf{t}_i
 \quad\text{with}\quad
 c_i=
 \left(0,\frac{11}{6},\frac{1}{2},-\frac{3}{2},-\frac{1}{6},1,\frac{1}{3},
 \frac{1}{3},0\right)
 \;.
\end{equation} 
The sum of the anomalous \U1 charges  is
\begin{equation}
 \tr \mathsf{t}_\mathrm{anom}~=~88\;,
\end{equation}
which is relevant to the calculation of the Fayet--Iliopoulos term.

The factors \SU3 and \SU2 in $G$ are identified with the  color $\SU3_c$ and the
weak $\SU2_\mathrm{L}$ of the standard model.  The hypercharge generator is
given by $\mathsf{t}_Y$.  It is embedded in SO(10) just like in usual 4D GUTs,
\begin{equation}
 \SU3_c \times \SU2_\mathrm{L} \times \U1_Y ~\subset~ \SO{10} \;.
\end{equation}
 Thus it automatically has the correct normalization and is consistent with
gauge coupling unification. It is also important that this  hypercharge is
non--anomalous, $ \mathsf{t}_Y\cdot\mathsf{t}_\mathrm{anom}=0$.

The massless matter states are listed in
Tab.~\ref{tab:FieldNamingSaturdayAfternoon}. They appear in both the untwisted
and twisted sectors, apart from $T_5$ which has no left--chiral  superfields.
The spectrum can be summarized as follows:
\begin{equation}  
 \text{matter:}~~~~3\times\boldsymbol{16} ~~+~~ \text{vector-like} \;.
\end{equation}
Two generations are localized in the compactified space and come from the first
twisted sector $T_1$, whereas the third generation is partially twisted and
partially untwisted:
\begin{equation}
2\times\boldsymbol{16} \in T_1 ~~,~~ \boldsymbol{16} \in U,T_2,T_4 \;.
\end{equation}
In particular, the up--quark and the quark doublet of the third
generation are untwisted, which results in a large Yukawa coupling,
whereas the down--quark is twisted and its Yukawa coupling is
suppressed. The $\boldsymbol{16}$--plet quantum numbers of the third
generation are not enforced by local GUTs, but are related to the
standard model anomaly cancellation.
 
Apart from the 3 matter families, the model contains extra states which are
vector--like with respect to the standard model gauge group. These include a
pair of Higgs doublets and additional exotic matter which, as we show in the
subsequent sections, can be consistently decoupled. A complete list of quantum
numbers of the massless states is given in
Tabs.~\ref{tab:LongTableNonSingletsSaturdayAfternoonModel} and
\ref{tab:LongTableSingletsSaturdayAfternoonModel}.

\begin{center}
\begin{table}[!t]
\begin{center}
 \begin{tabular}{|c|c|c|c|c|c|c|}
 \hline
 name & irrep & count & &
 name & irrep & count\\
 \hline
 $q_i$ & $(\boldsymbol{3},\boldsymbol{2};\boldsymbol{1},\boldsymbol{1})_{1/6}$ & 3 & &
 $\bar u_i$ & $(\overline{\boldsymbol{3}},\boldsymbol{1};\boldsymbol{1},\boldsymbol{1})_{-2/3}$ 
 	& 3\\
 $\bar d_i$ & $(\overline{\boldsymbol{3}},\boldsymbol{1};\boldsymbol{1},\boldsymbol{1})_{1/3}$ 
 	& 7 & & 
 $d_i$ & $(\boldsymbol{3},\boldsymbol{1};\boldsymbol{1},\boldsymbol{1})_{-1/3}$ & 4\\
 $\bar\ell_i$ & $(\boldsymbol{1},\boldsymbol{2};\boldsymbol{1},\boldsymbol{1})_{1/2}$ & 5 & &
 $\ell_i$ & $(\boldsymbol{1},\boldsymbol{2};\boldsymbol{1},\boldsymbol{1})_{-1/2}$ & 8\\
 $m_i$ & $(\boldsymbol{1},\boldsymbol{2};\boldsymbol{1},\boldsymbol{1})_{0}$ & 8 & &
 $\bar e_i$ & $(\boldsymbol{1},\boldsymbol{1};\boldsymbol{1},\boldsymbol{1})_{1}$ & 3 \\
 $s^-_i$ & $(\boldsymbol{1},\boldsymbol{1};\boldsymbol{1},\boldsymbol{1})_{-1/2}$ & 16 & & 
 $s^+_i$ & $(\boldsymbol{1},\boldsymbol{1};\boldsymbol{1},\boldsymbol{1})_{1/2}$ & 16\\
 ${s}_i$ & $(\boldsymbol{1},\boldsymbol{1};\boldsymbol{1},\boldsymbol{1})_{0}$ &
 69
 & & $h_i$ &
 $(\boldsymbol{1},\boldsymbol{1};\boldsymbol{1},\boldsymbol{2})_{0}$& 14\\
 $f_i$ & $(\boldsymbol{1},\boldsymbol{1};\boldsymbol{4},\boldsymbol{1})_{0}$ & 4 & &
 $\bar f_i$ &
 $(\boldsymbol{1},\boldsymbol{1};\overline{\boldsymbol{4}},\boldsymbol{1})_{0}$ & 4 \\
 $w_i$ & $(\boldsymbol{1},\boldsymbol{1};\boldsymbol{6},\boldsymbol{1})_{0}$ &
 5 & & & &\\
 \hline
 \end{tabular}	
\end{center}
\caption{Quantum numbers  of the massless states 
w.r.t.\ $G_\mathrm{SM}\times[\SU{4}\times\SU2]$
 and a field naming convention.}
\label{tab:FieldNamingSaturdayAfternoon} 
\end{table} 
\end{center}

\subsection{Local GUT  representations}

The matter states of the model can be viewed as originating from
representations of local GUTs supported at certain fixed points or
planes.  States from the first twisted sector correspond to `brane'
fields living at the orbifold fixed points.  As discussed in
Sec.~\ref{sec:StringsOnOrbifold}, such states are invariant under the
orbifold action. Thus they all survive in 4D and furnish complete
representations of the local GUTs. On the other hand, states from
higher twisted (as well as untwisted) sectors are not automatically
invariant under the orbifold action. Part of the GUT multiplet is
projected out such that the surviving states produce incomplete
(`split') multiplets in 4D. In particular, the gauge multiplets of
$\E8$ reduce to those of the standard model (and extra group
factors). The latter can be viewed as an intersection of local GUTs at
various orbifold fixed points (see e.g.\ \cite{Buchmuller:2005sh}). We
survey the local GUTs and their representations in
Tab.~\ref{tab:LocalGUTs}.

\subsection{Spontaneous gauge symmetry breaking}

The effective low energy theory of our orbifold model has, in general, smaller
gauge symmetry and fewer massless states than those in Eq.~\eqref{G} and 
Tab.~\ref{tab:FieldNamingSaturdayAfternoon}. One of the reasons is that there is
an anomalous \U1 which induces a FI $D$--term,
\begin{equation}\label{eq:FI}
 D_\mathrm{anom}
 ~=~
 \sum q_\mathrm{anom}^{(i)}\,|\phi_i|^2+
 \frac{g\,M_\mathrm{P}^2}{192\pi^2}
 ~{\tr \mathsf{t}_\mathrm{anom}}
 \;,
\end{equation}
where the sum runs over all scalars $\phi_i$ with anomalous charges 
$q_\mathrm{anom}^{(i)}$. This $D$--term must be zero in a supersymmetric vacuum,
so at least some of the scalars are forced to attain  large vacuum expectation
values, typically not far  below the string scale.  As a result, the anomalous
\U1 gets broken. Generically, this  also triggers  breakdown of other gauge
symmetries, under which the above mentioned scalars are charged. The resulting
gauge group and matter fields at low energies are  therefore a subset of those
in Eq.~\eqref{G} and  Tab.~\ref{tab:FieldNamingSaturdayAfternoon}.

More generally, some of the scalars can attain VEVs as long as it is
consistent with supersymmetry, $F_i=D_a=0$. In the simplest case, such
scalars are associated with flat directions in the field space. In
general, supersymmetric configurations are described by non--trivial
solutions of $F_i=D_a=0$, which correspond to points or
low--dimensional manifolds in the field space. In either case, this
breaks part of the gauge symmetry,
\begin{equation}
 G~\xrightarrow{\mathrm{VEVs}}~G_\mathrm{low-energy}\;.
\end{equation} 
Furthermore, such VEVs provide mass terms for some of the matter states. In
particular, if the superpotential coupling 
\begin{equation}
 \Delta W = x_i\,\bar x_j\times \langle s_{\alpha_1} ...  s_{\alpha_n}   \rangle  
\end{equation}
exists, with  $x_i$ and $\bar x_j$ being  vector-like  states w.r.t.\
$G_\mathrm{low-energy}$ and  $s_{\alpha_k}$ being the scalars attaining VEVs, 
then $x_i$ and $\bar x_j$  become massive and decouple from the low energy
theory.

It is common that orbifold models contain states which are charged under
both $G_\mathrm{SM}$ and other gauge factors  originating from the second
$\E8$. As long as such gauge factors are unbroken, there is no hidden sector in
the model, which is usually required for spontaneous SUSY breaking. The
separation between the ``visible'' and the ``hidden'' comes about when some of the
scalars attain VEVs thereby breaking  the unwanted gauge factors. In our model,
this occurs, in particular, when some of the 69 $s_i$ states break $\U1^8$.

An interesting property of the model is that none of the oscillator states is
charged under $G_\mathrm{SM}$ (cf.\
Tabs.~\ref{tab:LongTableNonSingletsSaturdayAfternoonModel},
\ref{tab:LongTableSingletsSaturdayAfternoonModel}). 
If all the oscillators develop VEVs, the unbroken gauge group is
$G_\mathrm{SM}\times[\SU4\times\U1]$, while the SM matter is neutral 
under the  additional \U1.
This might be important as it has been
argued that giving VEVs to oscillator modes corresponds to resolving the conical
singularities associated with the fixed points~\cite{Font:1988tp}. This means that
the phenomenologically relevant gauge group survives  the  naive `blowing--up'
procedure. 

Orbifold models with the same gauge shifts and Wilson lines but  different
scalar VEVs lead to distinct low--energy theories. For example, in some of them,
the standard model gauge group is broken. To obtain realistic models, one has
to make sure that, first of all, 
\begin{itemize}
 \item $G_\mathrm{SM}$ is unbroken,
 \item exotic matter is heavy.
\end{itemize} 
There are also further phenomenological constraints which we discuss in the
subsequent sections.

\subsection{Decoupling the exotic states}

A necessary condition for the decoupling of  vector--like exotic
states, without  breaking the  standard model gauge group, is
the existence of the superpotential couplings
\begin{equation}
 x_i\,\bar x_j\,\times (\text{SM singlets})\;.
\end{equation}
Furthermore, the rank of the $x_i,\;\bar x_j$ mass matrix must be maximal such
that no massless vector--like states survive. 
We find  that in our model the required mass terms are allowed and the exotic
states can be decoupled.

The exotic states charged under $G_\mathrm{SM}$ are pairs of $d_i$ and $\bar
d_i$, $\ell_i$ and $\bar \ell_i$, $s_i^-$ and $s_i^+$, and $m_i$. The  mass
terms for these states have the form
\begin{equation}
 W_\mathrm{mass}~=~
 d_i\,\mathcal{M}_d^{ij}(s)\,\bar d_j
 +\bar\ell_i\,\mathcal{M}_\ell^{ij}(s)\,\ell_j
 +m_i\,\mathcal{M}_m^{ij}(s)\,m_j
 +s^+_i\,\mathcal{M}_s^{ij}(s)\,s^-_j\;,
\end{equation}
where $s$ denotes some SM singlets. Taking $s=\{s_i\}$, we find
\begin{eqnarray}
 \mathcal{M}_d^{ij}(s) & = &
 \left(\begin{array}{ccccccc}
 s^5 & s^5 & s^5 & s^5 & s^5 & s^3 & s^3 \\
 s^1 & s^1 & s^3 & s^3 & s^3 & s^3 & s^3 \\
 s^1 & s^1 & s^3 & s^3 & s^3 & s^3 & s^3 \\
 s^6 & s^6 & s^6 & s^3 & s^3 & s^6 & s^6
 \end{array}\right)\;,\\
 \mathcal{M}_\ell^{ij}(s) & = &
 \left(\begin{array}{cccccccc}
 s^3 & s^4 & s^4 & s^1 & s^1 & s^1 & s^1 & s^1 \\
 s^1 &s^2 & s^2 & s^5 & s^5 & s^3 & s^3 & s^3 \\
 s^1 &s^2 & s^2 & s^5 & s^5 & s^3 & s^3 & s^3 \\
 s^1 &s^2 & s^2 & s^5 & s^5 & s^6 & s^3 & s^3 \\
 s^1 &s^6 & s^6 & s^3 & s^3 & s^6 & s^3 & s^3
 \end{array}\right)\;.
\end{eqnarray}
$\mathcal{M}_m^{ij}(s)$ and $\mathcal{M}_s^{ij}(s)$ are given in
Eqs.~\eqref{eq:Mm(s)} and \eqref{eq:Ms(s)} in App.~\ref{sec:Tables},
respectively. Here, an entry $s^N$  indicates the existence of a coupling  which
involves $N$ singlets. For instance, the $(1,1)$ entry of the $d$-$\bar{d}$ mass
term includes
\begin{equation}
 W_{d_1\bar d_1}~=~d_1 \bar d_1 (
 s_3 s_{20} s_{39} s_{44} s_{65}
 +
 s_7 s_{34} s_{35} s_{40} s_{41}
 +
 \cdots)\;,
\end{equation}
where the coefficients are omitted. Different entries generally involve
different  combinations of the singlets as well as different couplings, such
that the rank of each mass matrix is maximal. We note that higher $N$ does not
necessarily imply significant suppression of the coupling \cite{Cvetic:1998gv}: $s$ can be
close to the string scale and, furthermore, the coefficient in front of the
coupling grows with $N$. We find that all mass matrices have maximal rank at
order 8. A zero in the mass matrices \eqref{eq:Mm(s)}--\eqref{eq:Mf(s)} of
App.~\ref{sec:MassMatrices} indicates that up to order 8 no coupling appears.

This result implies that all of the exotic states can be decoupled below  the
GUT scale or so.  In particular, the rank of  $\mathcal{M}_d$ is 4 such that
only  3 down--type quarks survive. $\mathcal{M}_\ell$ has, in general, rank 5
resulting in 3 massless  doublets of hypercharge $-1/2$. In order to get an
extra pair of  (``Higgs'')   doublets with hypercharge $-1/2$ and $1/2$,  one
has to adjust the singlet VEVs such that the rank reduces to 4. This large
finetuning   constitutes the  well known   supersymmetric $\mu$--problem and
will be  discussed in subsequent sections.    A further constraint on the above
texture comes from the top Yukawa coupling: it is order one if the up--type
Higgs doublet has a significant component of  $\bar \ell_1$.

In the above mass matrices, $s$ are chosen to be singlets under
$\SU4\times\SU2'$  such that their VEVs break
\begin{equation}
 G~\longrightarrow ~\SU{3}_c\times\SU{2}_\mathrm{L}\times \U1_Y   
 \times  G_\mathrm{hidden} \;,
\end{equation}
with $G_\mathrm{hidden}= \SU4\times\SU2'$. Now the model has a truly hidden
sector which can be responsible for spontaneous SUSY breaking.

In the next section we show that the required configurations of the singlet VEVs
are in general consistent with supersymmetry, e.g.\ $F_i=D_a=0$. The
$D$--flatness is ensured by constructing gauge invariant monomials out of the
singlets \cite{Buccella:1982nx,Gatto:1986bt} involved in the mass terms for the
exotic states. We further show  that generally there exist non--trivial
solutions to $F_i=D_a=0$ in the form of low--dimensional manifolds in the field
space.

Not all vacuum configurations consistent with supersymmetry and the decoupling
are phenomenologically viable. Further important constraints are due to 
\begin{itemize}
 \item absence of rapid proton decay,
 \item realistic flavour structures,
 \item small $\mu$--term.
\end{itemize} 
This strongly restricts allowed VEVs for the singlets. As we show in
Sec.~\ref{sec:Pheno}, these constraints motivate certain patterns of the VEVs,
in particular those which preserve \BmL symmetry at the GUT scale.

Finally, let us remark on gauge invariance of the couplings in the framework of local GUTs. 
As stated in  Eq.~\eqref{eq:intersectingG}, a coupling among
twisted states is invariant under the intersection of local
gauge groups supported at the corresponding fixed points, but not necessarily under each
of the groups. To give an example, consider  an allowed  coupling $s_4\,s_{26}\,s_{57}$.
Each of these singlets originates from a larger representation of the local group.
The above coupling arises from the coupling of states contained in
$(\boldsymbol{16},\boldsymbol{1},\boldsymbol{1};\boldsymbol{1})$ of 
$\SO{10}\times\SU2\times\SU2\times\SO{14}$,
$(\boldsymbol{1};\boldsymbol{1},\overline{\boldsymbol{4}})$ of
$\SU7\times\left[\SO8\times\SU4\right]$, and $(\boldsymbol{14};\boldsymbol{1})$ of
$\SO{14}\times[\SO{14}]$. Clearly, it is not $\SO{10}$ invariant.
This is a special feature of local GUTs.

To summarize, we have shown that our model reproduces the exact MSSM
spectrum and the gauge group at low energies. The matter multiplets
appear as 3 $\boldsymbol{16}$--plets of \SO{10}. Since $\mathcal{M}_d$
is a $4\times 7$ matrix and $\mathcal{M}_\ell$ is a $5\times 8$
matrix, there exists one pair of $\SU{2}$ `Higgs' doublets which do not
form complete GUT representations. The model also has a hidden sector.

\subsection{Orbifold GUT limits }

\begin{figure}[t]
\centerline{\subfigure[\SO4 plane.\label{fig:OGL1}]{%
\CenterObject{\includegraphics{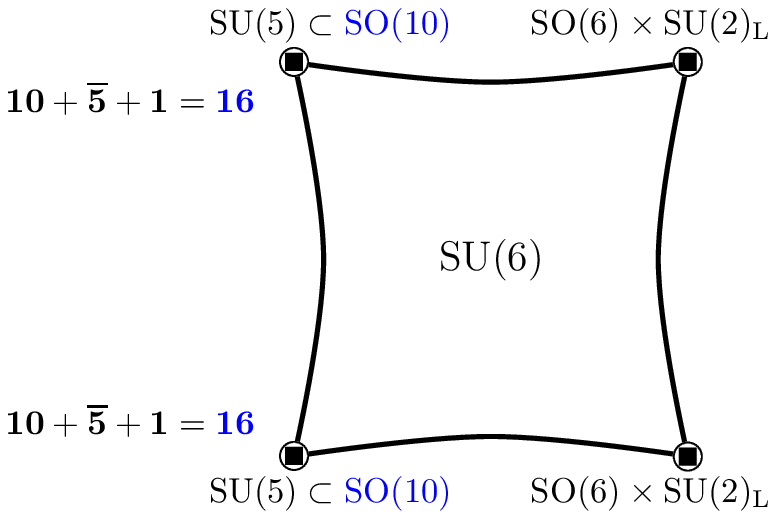}}}
\quad
\subfigure[\SU3 plane.\label{fig:OGL2}]{%
\CenterObject{%
\includegraphics{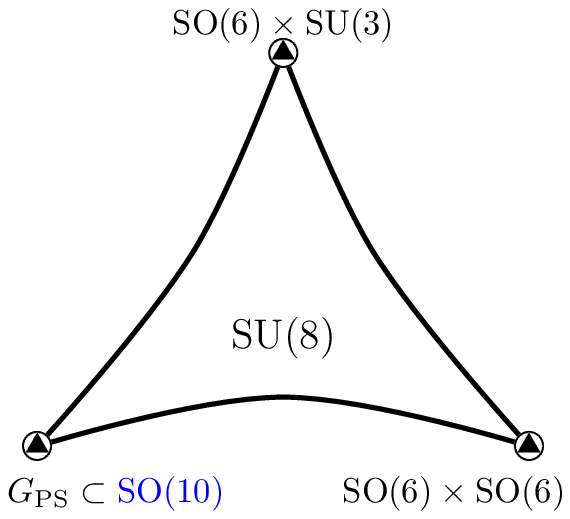}}}
}
\centerline{\subfigure[\G2 plane.\label{fig:OGL3}]{%
\CenterObject{%
\includegraphics{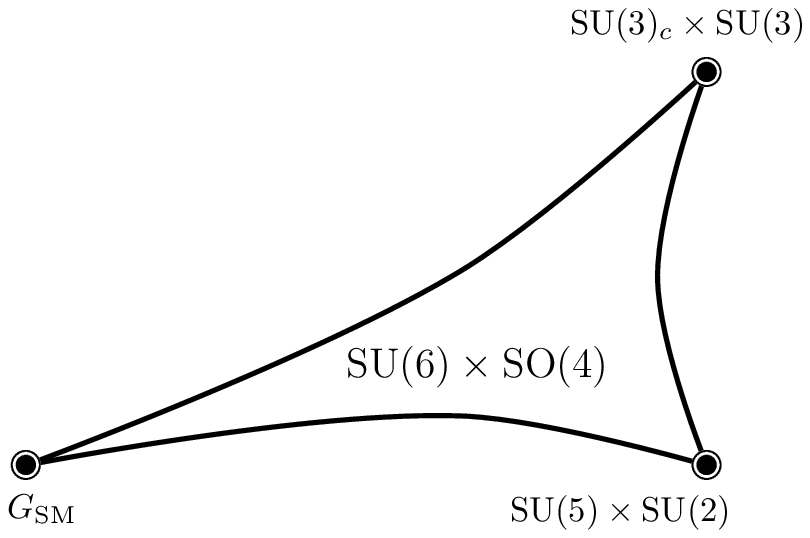}}}}
\caption{Orbifold GUT limits. In each case, a plane with a large
compactification radius is displayed. Only subgroups of the first \E8 are shown
and \U1 factors are omitted.}
\label{fig:OGL}
\end{figure}

One of the motivations for revisiting orbifold compactifications of the
heterotic string is the phenomenological success of orbifold GUTs
\cite{Kawamura:2000ev,Altarelli:2001qj,Hall:2001pg,%
Hebecker:2001wq,Asaka:2001eh,Hall:2001xr}. In our model, the hypercharge is
correctly normalized and the  spectrum is that of the MSSM, which leads to gauge
coupling unification at about $2 \times 10^{16}$ GeV. It is therefore
interesting to study orbifold GUT limits of the model, which  correspond to
anisotropic compactifications where some radii are significantly larger than the
others.  Such  anisotropy may mitigate the discrepancy  between the GUT and the
string scales  and can be consistent with perturbativity for one or two large
radii of order $(2\times 10^{16} ~ {\rm GeV})^{-1}$ 
\cite{Witten:1996mzF3,Hebecker:2004ce}. In the energy range between
the compactification scale and the string scale  one  obtains an effective
higher-dimensional field theory.

In $\Z{6-\mathrm{II}}$ orbifolds, there are four independent radii: two are
associated with the $\G2$ and $\SU3$ planes, respectively, and the other two are
associated with the two independent directions in the $\SO4$-plane. Any of these
radii can in principle be large leading to a distinct GUT model.

The bulk gauge group and the amount of supersymmetry are found via a subset of
the invariance conditions \eqref{4dgauge} with $N<6$. Consider a subspace ${\cal
S}$  of the 6D compact space with large compactification radii. This subspace is
left invariant under the action of some elements  of the orbifold space group,
i.e.\ a subset of twists and translations ${\cal G }$. The bulk gauge multiplet 
in ${\cal S}$ is part of the $N=4$ $\E8 \times \E8$ gauge multiplet which is
invariant under the action of ${\cal G }$, i.e.\ a subset of conditions
\eqref{phys} restricted to ${\cal G }$ \cite{Buchmuller:2005sh}.

In our model, the intermediate orbifold picture  can have any dimensionality
between 5 and 10. For example, the 6D orbifold GUT limits are (up to $\U1$
factors):
\begin{eqnarray}   
 \SO4 ~\text{plane}& : &~\text{bulk GUT}=\SU6~, ~N=2 \;, \nonumber\\
 \SU3 ~\text{plane}& : &~\text{bulk GUT}=\SU8~,~N=2 \;, \nonumber\\
 \G2  ~\text{plane}& : &~\text{bulk GUT}=\SU6\times \SO4 ~,~N=4 \;, \nonumber
\end{eqnarray} 
where the plane with a ``large'' compactification radius is indicated  and $N$
denotes the amount of supersymmetry. In all of these cases, the bulk
$\beta$--functions of the SM gauge couplings coincide. This is because either
$G_\mathrm{SM}$ is contained in a simple gauge group or there is $N=4$
supersymmetry. It is remarkable that regardless of which radii are large,
unification in the bulk occurs.  
This observation may indeed be relevant to  the  discrepancy  between
the GUT and the string scales. 
However, to check whether this is really the case, 
logarithmic corrections from localized 
fields, contributions from vector--like heavy fields and string thresholds
have to be taken into account.

The 6D orbifold GUT limits of the model are displayed in Fig.~\ref{fig:OGL}. We
note that the standard model gauge group is obtained as an intersection of the
local gauge groups at the orbifold fixed points. For completeness, in
Tab.~\ref{tab:OGL} we survey all possible orbifold GUT limits. For $D\ge6$, we
find that they are all consistent with gauge coupling unification in the bulk.
The different geometries differ, however, in the values of the gauge couplings
at the unification scale as well as the Yukawa couplings.


\section{Supersymmetric vacuum configurations}
\label{sec:FandDflat}

In this section we discuss supersymmetric vacuum configurations of our model. In
globally supersymmetric theories, these require vanishing of the  $D$- and
$F$--terms. We start with the discussion of the $D$--terms.

\subsection{$\boldsymbol{D}$--flatness}

In a supersymmetric gauge theory with an anomalous \U1, vanishing of the
$D$--terms requires 
\begin{eqnarray}
 D_a & = &\sum_i \phi_i^* T_a \phi_i=0 \;, \nonumber\\
 D_\mathrm{anom} & = & \sum_i q_\mathrm{anom}^{(i)}\,|\phi_i|^2+
 \frac{g\,M_\mathrm{P}^2}{192\pi^2}
 ~{\tr \mathsf{t}_\mathrm{anom}}=0 \;,
\end{eqnarray}
where $T_a$ are generators of the gauge group and $\mathsf{t}_\mathrm{anom}$ is 
the generator of an anomalous \U1. In particular, to have a vanishing FI
$D$--term, there must exist at least one field whose anomalous charge is
opposite in sign to that of ${\tr \mathsf{t}_\mathrm{anom}}$.

In theories without an anomalous \U1,  these conditions are satisfied if there
exists a gauge invariant monomial $I(\phi_i)$ \cite{Buccella:1982nx}. The
$D$--flat field configurations are found from
\begin{equation}
 \left\langle{\partial I \over \partial \phi_i }\right\rangle
 ~=~c \,\langle\phi_i^*\rangle \;,
\end{equation}
where $c$ is a constant and $\langle x\rangle$ denotes the VEV of 
$x$. On solutions of this equation, gauge invariance of $I$ simply
means $D_a=0$ \cite{Buccella:1982nx}. If an anomalous \U1 is present,
$D$--flatness requires the existence of $I(\phi_i)$ which is gauge invariant
with respect to all symmetries apart from $\U1_\mathrm{anom}$ and which carries
a net anomalous charge whose sign is opposite to that of ${\tr
\mathsf{t}_\mathrm{anom}}$ (see, e.g., \cite{Casas:1987us,Cleaver:1997jb,Cleaver:1998im}).

Therefore, searching for a $D$--flat configuration amounts to finding gauge
invariant monomials with the above properties. Clearly, such monomials can be
multiplied together while preserving the required properties. We are
particularly interested in the $\U1^N$ gauge theory which is relevant when the
non--Abelian singlets $s_i$ get VEVs. In this case, the gauge invariant  
monomial 
\begin{equation} 
(s_1)^{n_1}\, (s_2)^{n_2}\dots(s_k)^{n_k}\;,\quad n_i \in\mathbbm{N}\;,
\end{equation}
represents the $D$--flat direction 
\begin{equation}
\vert s_1 \vert / \sqrt{n_1}= \vert s_2 \vert / \sqrt{n_2}
~=~\dots~=~\vert s_k \vert / \sqrt{n_k}
\end{equation}
in the field space. The overall scale of these VEVs is fixed by the FI
$D$--term.

Starting with the anomalous $\U1$ $D$--term, an example of a gauge  invariant
monomial with a negative anomalous charge is  given by
\begin{equation}
 I_1~=~ (s_{12})^2\,s_{39}\,s_{55}\,s_{56} \;.
\end{equation}
Clearly, it is not unique
(cf.~Tab.~\ref{tab:GaugeInvariantMonomialsGen}).
In particular, it can be multiplied by a monomial
with zero anomalous charge. We find that every $s_i$ that enters the mass
matrices for the exotic states also enters a gauge invariant monomial (see
Tab.~\ref{tab:GaugeInvariantMonomials}).  This shows that $s_i$ can be given
large VEVs while having vanishing $D$--terms.

There is also another algorithm to check the $D$--flatness for the required
singlet configuration. Mass terms for vector--like exotic matter $x_i$ are
generated by
\begin{equation} 
 W~=~ \sum_{ij} x_i \bar x_j \mathcal{M}_{ij}(s)
\end{equation}
with 
\begin{equation}
 \mathcal{M}_{ij}(s)~=~ \sum_n \sum_{k_1,\dots,k_n}   c_{ij} ({k_1},\dots,{k_n}) ~
 s_{k_1}\dots s_{k_n}\;,
\end{equation}
where $s_{k_a}$ are the singlets and $c_{ij} ({k_1},\dots,{k_n})$ are some
coefficients. Any monomial of fields in the superpotential is gauge invariant
and represents a $D$--flat direction. Thus multiplying all of the monomials
together, we again get a flat direction. 
However, we do not want to give VEVs to the
exotic matter fields $x_i$ since this would break the  standard model gauge group.
So, one needs to replace those with some SM--singlets which have the same 
total $\U1$--charges:  
\begin{equation}
\U1\text{-charges} \left( \prod_{M_{ij}\not= 0} x_i \bar x_j   \right)
~=~\U1\text{-charges} \; \bigl(  s_{l_1}\dots s_{l_n}      \bigr) \;.
\end{equation}
This is just one equation. In our case, there are many singlet monomials
satisfying this equation and one of them is 
\begin{equation}
 s_7 (s_{19})^4 s_{26} s_{36} s_{39} (s_{40})^5
 (s_{48})^{18} s_{55}  (s_{56})^3 (s_{57})^7  (s_{64})^2
 (s_{68})^{42}   (s_{69})^{27} \;. 
\end{equation}
To cancel the FI term, one has to multiply it with the monomial $I_1$ which has
a negative anomalous charge.
This shows that one can give VEVs to all singlets involved in the decoupling of
extra matter consistently with the $D$--flatness.

\subsection{Some of the $\boldsymbol{F}$--flat directions}

The requirement  $F_i=0$ for a singlet $s_i$ is most easily  satisfied if
$s_i$ is an $F$--flat direction, i.e.,
\begin{equation}
 { \partial W \over \partial  s_i }~=~0
\end{equation}
for arbitrary values of $s_i$. (When $s_i$ is not a flat direction,
$F_i=0$ is satisfied only at special values of $s_i$.)
The existence of such flat directions usually  requires that the VEVs
of some other singlets appearing in the superpotential be zero.

Many exactly $F$--flat directions can be obtained from the selection rule 
\eqref{eq:H-momentumRule3} for the superpotential couplings,
\begin{equation} 
\sum R^3~=~-1 \mod2 \;.
\label{R3rule}
\end{equation}
   As seen from
Tabs.~\ref{tab:LongTableNonSingletsSaturdayAfternoonModel} and
\ref{tab:LongTableSingletsSaturdayAfternoonModel}, all of the non--Abelian
singlets in $U$, $T_2$, $T_4$ sectors have  $R^3=0$. Thus they cannot couple
among themselves  consistently with the rule (\ref{R3rule}).  Furthermore, they
cannot couple to a single state in $T_{1,3}$ since the latter  have $R^3=-1/2$
and at least two of such states are  needed to have an allowed coupling. That
means that the $F$--terms are proportional to a VEV of  some state in 
$T_{1,3}$:
\begin{equation}
 F_i~\sim ~ {\partial W \over \partial s_i}~\sim~
 \langle  \text{singlet from}~ T_{1,3}     \rangle~=~0 \;,
\end{equation}
as long as all singlets in $T_{1,3}$ have zero VEVs. Thus one immediately
gets 39 exactly $F$--flat directions associated with $s_i$ from the
\begin{equation}
 U,~ T_2, ~T_4  
\end{equation}
sectors. By this  we mean that  the 39 fields are allowed to attain non--zero VEVs
simultaneously, without referring to the number of real variables 
parametrizing such VEVs.

One can also show that these directions are $D$--flat\footnote{ The 
requirement of the $D$--terms cancellation  fixes some  of the  
field VEVs.}. In particular,
each non--Abelian singlet from  $U, T_2, T_4$   enters a gauge invariant
monomial which involves only  $U, T_2, T_4$ singlet states. Furthermore,
it is possible to construct a monomial with a negative net anomalous charge. 
An example is (see also Tab.~\ref{tab:GaugeInvariantMonomialsFflat})
\begin{equation}
 I~=~
 s_{34}\,s_{35}\,s_{40}\,s_{39}\,s_{67}\;.
\end{equation}
That means one can give non--zero VEVs to the 
$U, T_2, T_4$ singlets while preserving supersymmetry.
Some of such states presumably correspond  to the ``blowing--up'' modes
of the orbifold which allow one to interpolate between a smooth Calabi--Yau
manifold and an orbifold.

These flat directions allow us to decouple many exotic states but not all.   
One can perhaps increase the dimensionality of the $F$-- and $D$--flat space by
including non--Abelian flat directions or by other considerations. We also note
that, for practical purposes, flatness is only required up to a certain order
in superpotential couplings and one may exploit approximately flat
directions.

In any case, flat directions are not necessary for the decoupling. As we 
discuss   below, supersymmetric field configurations are in general more
complicated and allow for the decoupling of the exotic states.

\subsection{General supersymmetric field configurations}

Given a set of 69 states $s_i$,  supersymmetric field configurations are given
by the sets of VEVs  $\langle s_i \rangle $ which satisfy $F_i=D_a=0$.
Naively, it appears  that the number of constraints, that is 69 plus 
the number of the gauge group generators, 
is larger than the number of variables, 69.  The system seems
to be overconstrained. However, this is  not the case. 
As well known,
complexified gauge transformations allow us to eliminate the $D$--term
constraints (\cite{Ovrut:1981wa},\cite[Chapter VIII]{Wess:1992cp}), 
such that the number of variables
equals the number of equations. In what follows, we demonstrate this for 
Abelian and non--Abelian cases.

\subsubsection{Abelian case}

Consider a supersymmetric $\U1^N$ gauge theory  with $n$ charged fields $z_i$.
The superpotential can be
written as 
\begin{equation}
 W~=~\sum_{(a)} I^{(a)}(z_1,\dots,z_n) \;.
\label{1}
\end{equation}
Here $I^{(a)}$ are gauge invariant monomials (some of which may be reducible,
i.e.\ a product of lower order monomials),
\begin{equation}
I^{}(z_1,\dots,z_n)~=~  c\; z_1^{k_1}\dots z_n^{k_n}
\end{equation}
with $c$ being a constant  and
\begin{equation}
 k_1 \boldsymbol{Q_1} +\dots+ k_n \boldsymbol{Q_n}~=~0 \;,
\end{equation}
where $\boldsymbol{Q_i}=(q^1_i,\dots,q_i^N)$ is an $N$--vector of
$\U1$ charges of the fields $z_i$.

Supersymmetry is preserved in the vacuum if 
\begin{equation}
 F_i~=~0 ~~,~~D_a~=~0\;,\quad i=1,\dots,n~;~a=1,\dots,N \;.
\end{equation}
Start with the $F$--terms.  $F_i=0$ can be written as\footnote{Here we define the $F$ component such that
it has the quantum numbers of $z^*$.}
\begin{equation}
  F_i(z)~\equiv~{ \partial W \over \partial z_i  }~=~0 
\end{equation}
for all $i$. Since there are $n$ such equations and $n$ variables, there are
solutions. In general, there are solutions with $z_i\not=0$ (for example, when
$W$ is a non--trivial polynomial).\footnote{%
In particular, this is generally the case  in string orbifold   models.  The
reason is that  if a superpotential  $W_0$ is allowed by string selection rules,
$W_0^N$ is also allowed for some integers $N$.  For example, in the $\Z6$ case,
one has $W \sim W_0 + W_0^7+\dots$. Such superpotentials allow for non--trivial
solutions to the $F$--term equations (consider, e.g., $W_0=z_1 z_2 z_3$).}   

Consider a solution with $z_i \not =0$.
Note that $F_i(z)$ is not gauge invariant, but  transforms as $z_i^{-1}$.
As a consequence, if $\{z_k^0\}$ is a  solution to
$F_i(z)=0$, then the transformation 
\begin{equation}
  z_k^0~\rightarrow~z_k'~=~z_k^0 
  ~(\alpha_1)^{q_k^1} (\alpha_2)^{q_k^2}\cdots (\alpha_N)^{q_k^N}\;,
\end{equation}
leaves the $F$--terms vanishing,
\begin{equation}
  F_i(z^0)~\rightarrow~F_i(z')~=~F_i(z^0) 
  ~(\alpha_1)^{-q_i^1} (\alpha_2)^{-q_i^2}\cdots (\alpha_N)^{-q_i^N}~=~0\;,
\end{equation}
where $\alpha_i$ are arbitrary complex numbers and ${q_i^a}$ is
the $a$-th $\U1$ charge of $z_i$. Therefore, given a solution $z_i^0$
to the $F$--term equations,  it  can be rescaled as above to give a family
of solutions. In fact, it can be rescaled in such a way that all the
$D$--terms vanish:
\begin{eqnarray}
 D_a(z')~=~\sum_i  q_i^a \vert z'_i  \vert^2~=~\sum_i  q_i^a 
 \vert z^0_i \vert^2 
 \vert\alpha_1\vert^{2q_i^1} 
 \vert\alpha_2\vert^{2q_i^2}
 \dots
 \vert \alpha_N \vert^{2q_i^N}~=~0 
 \label{3}
\end{eqnarray}
for    $a=1,\dots,N$.
The $N$
rescaling parameters $\vert \alpha_i \vert $ are found from the above
$N$ equations.  In terms of the rescaled variables $z'_i$, these
solutions are encoded in the gauge invariant monomials $z_1^{\prime\, k_1}
\dots z_n^{\prime\, k_n}$ such that
\begin{equation}
 \vert z_1' \vert /\sqrt{k_1}~=~  \vert z_2' \vert /\sqrt{k_2} ~=~\dots~=~ \vert
z_3'  \vert/\sqrt{k_n} 
\end{equation}
is a $D$--flat direction. This latter equation allows to find
$\alpha_i$ most easily and also shows that sensible solutions to
Eq.~\eqref{3} exist, i.e.\ $\vert \alpha_i \vert^2 >0$.
\footnote{Note that if  2 fields $z_{1,2}$
with identical charges are present, one has to be cautious. As far as
the $D$--term equations go, these two fields can be treated as one,
i.e.  $z_1^{k_1} z_2^{k_2} \rightarrow z_2^{k_1+k_2}$ and $q_1^a
\vert z_1 \vert^2 + q_2^a \vert z_2 \vert^2 \rightarrow q_2^a \vert
z_2 \vert^2 $.}

Let us now turn to the $D$--term of an anomalous $\U1$. The 
complexified gauge transformation     which leaves  $F_i=0$ intact  is
\begin{eqnarray}
  z_i^0~\to~z'_i &=& \alpha^{q_\mathrm{anom}^{(i)}} z_i^0 \;,
  \nonumber\\
 D_\mathrm{anom}(z') & = & \sum_i q_\mathrm{anom}^{(i)}\,|z_i^0|^2
 \vert\alpha\vert^{2q_\mathrm{anom}^{(i)}}
 +
 \frac{g\,M_\mathrm{P}^2}{192\pi^2}
 ~{\tr \mathsf{t}_\mathrm{anom}} \;.
\end{eqnarray}
As long as there is a field whose anomalous charge is opposite in sign to that
of ${\tr \mathsf{t}_\mathrm{anom}}$, the $D$-term can be cancelled.
Suppose ${\tr \mathsf{t}_\mathrm{anom}}<0 $, then   $D_\mathrm{anom}<0$ for 
$\alpha^{q_\mathrm{anom}^{(i)}} \rightarrow 0$ and $D_\mathrm{anom}>0$ for
$\alpha^{q_\mathrm{anom}^{(i)}} \rightarrow \infty  $. 
Therefore, there is a solution to
$D_\mathrm{anom}=0$ for finite $\alpha^{q_\mathrm{anom}^{(i)}}$.

It is now clear that the $D$--term constraints can be satisfied by an
appropriate choice of complexified gauge transformations. This means that the
number of SUSY conditions $F_i=0$ equals the number of variables $z_i$, such
that (non--trivial) solutions generally exist. Such solutions can be points (up
to gauge transformations) or low dimensional manifolds in the field space.

\subsubsection{Non--Abelian case}

Let us now consider the case of a non--Abelian gauge theory following 
Ref.~\cite[Chapter~VIII]{Wess:1992cp}. This situation arises in our construction
when one assigns VEVs to the doublets of the hidden sector \SU2. As in
the Abelian case, if $\{z_k^0\}$ is a  solution to the $F$--term
equations $F_i(z)=0$, then
\begin{equation}
  z'~=~\exp\left(\I\sum_a \lambda_a T_a\right)z^0
\end{equation}
is also a solution, 
where $T_a$ are the group generators and $\lambda_a$ are complex parameters.
This is 
because $F_i$ transforms as $z_i^{-1}$, i.e.,
$F(z') = \exp(-\I\sum\lambda_a T_a)F(z^0)$.  

The $D$--terms,
$D_a(z) =  \sum_i  z_i^\dagger T_a z_i$, transform in the adjoint representation
under this transformation. 
There is always a group element which
transforms  vector $D_a$ into $(x,0,..,0)$
corresponding to the direction of 
one of the Cartan generators  $T_{\hat{a}}$,
i.e.\ $D_a \rightarrow D_{\hat {a}}= \sum_i  z_i^\dagger T_{\hat{a}} z_i $.
Writing  $(T_{\hat{a}})_{ij} = \mu_i \delta_{ij}$ with real $\mu_i$,
the only non--vanishing $D$--term is
\begin{equation}
  D_{\hat{a}}(z')~=~\sum_i  \mu_i  \vert z'_i \vert^2 \;.
\end{equation}
The complexified gauge transformation along this direction,
\begin{equation}
  z'_i~\to~z^\eta_i~=~\exp(\mu_i \eta)\, z'_i\;,
\end{equation}
with real $\eta$, leaves $F_i(z^\eta)=0$ and
transforms the $D$--term into 
\begin{equation}
  D_{\hat{a}}(z^\eta)~=~  \sum_i  \mu_i e^{2\mu_i\eta}    \vert z'_i \vert^2\;.
\end{equation}
In the non--degenerate case, 
$D_{\hat{a}}(z^\eta)\to \pm\infty$ for $\eta\to\pm\infty$.\footnote{An example of the degenerate case
is  an $\SU{2}$ theory with 2 fundamental multiplets
$h_{1,2}$ and $W = (h_1 h_2)^n$. The solution to the $F$--term
equations is $h_2 = c h_1$ such that all gauge invariant monomials vanish.
The $D$--terms vanish only for  $h_{1,2} = 0$ corresponding to
$|\eta|\to \infty$.} Therefore, there is a solution to $D_a=0$ for 
finite $\eta$ and hence finite $z_i^\eta$.

\subsubsection{Summary and applications}

Employing complexified gauge symmetry, we have shown that the system
of equations $F_i=D_a=0$ in globally supersymmetric models is not
overconstraining. In particular, solutions to $F_i=0$ exist since the
number of equations equals the number of complex variables and in
general some of these solutions are non-trivial.  Once a non--trivial
solution to $F_i=0$ is found, it can be transformed using complexified
gauge symmetry to satisfy $D_a=0$.  This conclusion is based on the
observation that the $F$--term equations constrain gauge invariant
monomials, while such monomials are also associated with $D$--flat
directions.
 
Consequently, supersymmetric field configurations in orbifold models form low
dimensional manifolds or points (up to gauge transformations). In such
configurations, SM  singlets generally attain non--zero VEVs, typically not far
below the string scale. As a result,  when such VEVs play a role of the mass
terms for vector--like exotic states, the decoupling of the latter  can be made
consistently with supersymmetry.

The above considerations apply to globally supersymmetric models at the
perturbative  level.  In practice, we expect supergravity as well as
non--perturbative effects to play a role in selecting vacua. However, it would
be  very difficult to   quantify such effects at this stage. We note that, in
existing literature, it is rather common to amend the global SUSY conditions 
$F_i=D_a=0$ by $\langle W \rangle =0$  (see e.g.\ \cite{Font:1988mm}), which
implies a vanishing cosmological constant in supergravity. Such a condition
should however be imposed on the $total$ superpotential which includes, in
particular, non--perturbative potentials for moduli.  Thus, requiring $\langle W
\rangle =0$ does not set any immediate constraint on the charged matter VEVs. 
At this stage, we include only the most important supergravity effect, that is
gaugino condensation in the hidden sector, which we discuss in the next
section.


\section{Spontaneous supersymmetry breaking}
\label{sec:SUSYbreaking}

\subsection{Hidden sector gaugino condensation }

As supersymmetry is broken in nature, realistic models should admit spontaneous
supersymmetry breakdown. An attractive scheme for that is  hidden sector gaugino
condensation \cite{Nilles:1982ik,Ferrara:1982qs,Derendinger:1985kk,Dine:1985rz}.
In this case, a  hierarchically small supersymmetry breaking scale, which is
favoured by phenomenology, is explained by dimensional transmutation.

The basic idea is that one or more gauge groups in the hidden sector become
strongly coupled at an intermediate scale. This leads to confinement and gaugino
condensation.  Under certain circumstances, that is if the dilaton is
stabilized,  gaugino condensation translates into supersymmetry breaking. In
particular,
\begin{equation}
 \langle \lambda \lambda \rangle^{1/3} ~\sim~10^{13} \,\mathrm{GeV} \;,
\end{equation}
leads to the gravitino mass in the TeV range, $m_{3/2} \sim \langle
\lambda \lambda \rangle / M_\mathrm{P}^2$.  The condensation scale $
\Lambda \sim \langle \lambda \lambda \rangle^{1/3}$ is given by the
Landau pole of the condensing gauge group,
\begin{equation}
 \Lambda~=~M_\mathrm{GUT}\,
 	\exp\left(-\frac{1}{2\beta}\frac{1}{g^2(M_\mathrm{GUT})}\right)\;.
\end{equation}
For certain gauge groups and matter content, $ \Lambda$ can be in the right 
range. 

Gaugino condensation leads to supersymmetry breaking only if the dilaton is
stabilized at a realistic value. Models with a single gaugino condensate and a
classical K\"ahler potential suffer from the notorious dilaton run--away
problem. That is, gaugino condensation creates a  non--perturbative
superpotential for the dilaton $W \sim \exp (- a\,S ) \sim  \langle \lambda
\lambda \rangle $ (where $a=3/2\beta$) which leads to $S\rightarrow \infty$ at
the minimum of the scalar potential. There are two common options to avoid this
problem: employ multiple gaugino condensates or use non--perturbative
corrections to the K\"ahler potential. The first option is not available in our
model as the  hidden sector \SU2 either does not condense or  its condensation
scale is too low, and we are left with a single \SU4. Thus, we use the second
option. In this case, the  classical K\"ahler potential for the dilaton is
amended by non--perturbative corrections,
\begin{equation}
 K~=~-\ln (S + \overline{S}) + \Delta K_\mathrm{np} \;. 
\end{equation}
The functional form of $\Delta K_\mathrm{np}$ has been studied in the
literature \cite{Shenker:1990uf2,Banks:1994sg,Casas:1996zi,Binetruy:1996xj,
Binetruy:1996nx,Binetruy:1997vr}. For a
favourable choice of the  parameters, this correction allows one to stabilize
the dilaton at a realistic value, $\re S\simeq 2$, while breaking supersymmetry 
\cite{Casas:1996zi,Binetruy:1996nx,Binetruy:1997vr,Barreiro:1997rp,Buchmuller:2004xr}.
Supersymmetry is broken spontaneously by the dilaton $F$--term,
\begin{equation}
 F_S~\sim~{ \langle \lambda \lambda \rangle \over M_\mathrm{P}  } ~,~ F_T \sim 0 ~,
\end{equation}
where $T$ is the heterotic $T$--modulus. In what follows, we  will estimate the
gaugino condensation scale in our model without going into details of the
dilaton stabilization mechanism.

\begin{figure}[h]
\centerline{\includegraphics[scale=0.75]{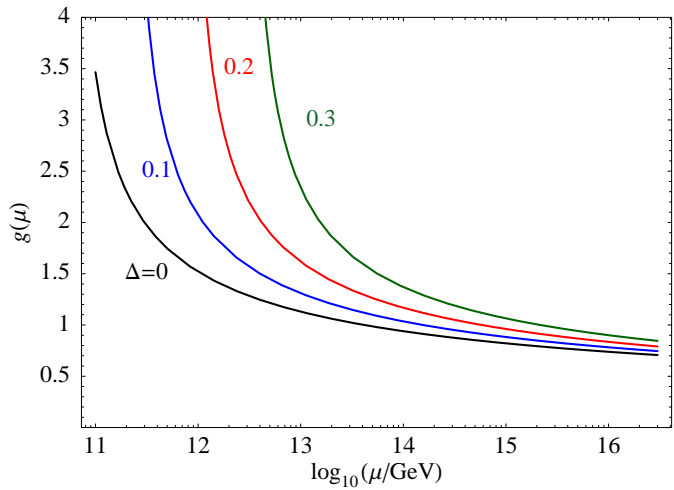}}
\caption{Scale dependence   of the 
hidden sector \SU4   gauge coupling    
for  different threshold corrections  $\Delta$.}
\label{fig:StrongSU4}
\end{figure}

The condensing gauge group in our case is \SU4. The condensation scale depends
on the matter content. If all the singlets have zero VEVs, there are 5 
$\boldsymbol{6}$--plets and 4 pairs of $\boldsymbol{4} +
\overline{\boldsymbol{4}}$.  The corresponding beta function is 
\begin{equation}
 \beta_{\SU4}
 ~=~ 
 \frac{1}{16\pi^2}\left\{12
 -\#(\boldsymbol{6})
 -\#(\boldsymbol{4}  + \overline{\boldsymbol{4}}  )  \right\}
 \;,
\end{equation}
where $\#(\boldsymbol{R})$ counts the number of representations
$\boldsymbol{R}$. With the above matter content, \SU4 is asymptotically free but
the condensation scale is too low. In a general field configuration, the
$\boldsymbol{6}$--plets and the pairs $\boldsymbol{4} +
\overline{\boldsymbol{4}}$ receive large masses (see Eqs.~\eqref{eq:Mf(s)} and
\eqref{eq:Mw(s)}) and are all decoupled. In this case,  the beta function
becomes
\begin{equation}
 \beta_{\SU4}~=~\frac{3}{4\pi^2}\;.
\end{equation}
The condensation scale is then $10^{10}-10^{11}$ GeV. There are many factors
that can affect it.  In particular, there are string threshold corrections
\cite{Ibanez:1986xy,Dixon:1990pc,Mayr:1993mq,Nilles:1997vk,Stieberger:1998yi}
which lead to different gauge couplings in the visible and hidden \E8. The
corresponding gauge kinetic functions are given by
\cite{Ibanez:1986xy,Nilles:1997vk,Stieberger:1998yi}
\begin{equation}\label{eq:fvishid}
 f_{\mathrm{vis}/\mathrm{hid}}~=~S\pm\epsilon\,T\;,
\end{equation}
where $\epsilon$ is a small parameter and,   
for simplicity, we have taken a large
$T$ limit. The gauge couplings are found from
\begin{equation}
 \re f~=~g^{-2} \;.
\end{equation} 
In the visible sector, the apparent gauge coupling unification requires
$g^{-2}_\mathrm{GUT}\simeq2$, whereas the hidden sector gauge coupling is
\begin{equation}
 g^{-2}_\mathrm{hid}(M_\mathrm{GUT})
 ~=~
 \re f_\mathrm{hid}
 ~\simeq~
 2\,(1-\Delta)\;,
\end{equation}
where $\Delta$ parametrizes string threshold corrections.  The corresponding
condensation scale is
\begin{equation}
 \Lambda~\simeq~M_\mathrm{GUT}\,
 	\exp\left[-\frac{1}{\beta}(1-\Delta)\right]\;.
\end{equation}
For $\Delta$ between 0 and 0.3, the condensation scale  ranges between $5\times
10^{10}$ and $10^{13}\,\mathrm{GeV}$ (cf.\ Fig.~\ref{fig:StrongSU4}). Thus a TeV
scale gravitino mass can in principle be obtained.

There are of course other factors that can affect the above estimate.  For
example, the SUSY breaking scale depends on coefficients entering  a particular
dilaton stabilization mechanism. Also the identification of the Landau pole with
$\langle \lambda \lambda \rangle^{1/3}$ is not precise. The main point, however,
is that the model contains the necessary ingredients for gaugino condensation
and SUSY breaking in the phenomenologically interesting range.

\subsection{Soft SUSY breaking terms}

The K\"ahler stabilization mechanism leads to a specific pattern of the  soft
terms, the so called ``dilaton dominated scenario'' \cite{Kaplunovsky:1993rd}. 
The resulting soft terms are universal and given by
\begin{eqnarray}
 \mathscr{L}_{\rm soft} & =  &{1\over 2} (M  \lambda^a \lambda^a + \text{h.c.}) 
 - m^2 \phi^{*\alpha} \phi^{\alpha} - \left( 
 {1\over 6 } A\; Y_{\alpha \beta \gamma} \phi^{\alpha} \phi^{\beta} \phi^{\gamma} + {\rm h.c.}
 \right)\;,
\end{eqnarray}
where  $\lambda^a$ are the gauginos, $\phi^{\alpha}$ are the scalars and
$Y_{\alpha \beta \gamma}$ are the Yukawa couplings.  Dilaton dominated SUSY
breaking implies the following relations  among the  soft breaking  parameters
(see e.g.\ \cite{Brignole:1997dp}):
\begin{eqnarray} 
 M ~=~ \pm \sqrt{3} m_{3/2} \;,\quad
 m ~=~ m_{3/2} \;,\quad
 A ~=~ - M \;.
\label{softterms}
\end{eqnarray}
This is a restricted version of mSUGRA with 
the only independent parameter being  the gravitino mass $m_{3/2}$.
Here we do not discuss the $\mu$ and $B\mu$ terms  which depend on
further details of the model. 

The dilaton dominated scenario has a number of phenomenologically attractive
features. In particular, due to flavour universality in the soft breaking
sector,  it avoids the SUSY FCNC problem. Also, most of the physical CP phases,
e.g.\ $\arg(A^* M)$, vanish which ameliorates the SUSY CP problem.  Other
phenomenological aspects have been discussed in Ref.~\cite{Abel:2000bj}.

The above considerations are based on
the assumption that the dilaton is stabilized via non--perturbative corrections
to the K\"ahler potential. Dilaton and other moduli stabilization is a difficult
issue and there may exist other possibilities which could lead to other
patterns of the soft terms.


\section{ $\boldsymbol{\BmL}$  symmetry and phenomenology}
\label{sec:Pheno}

Realistic string vacua must satisfy  a number of  phenomenological constraints, 
in addition to those imposed by the spectrum and the gauge group of the MSSM.  
In particular, the proton  should be sufficiently stable as well as flavour
structures should be realistic. This constrains  vacuum configurations 
for  the SM singlets. In generic vacua, there are baryon  number violating 
operators already at the renormalizable level, so in order to avoid rapid proton
decay one must be able to tune the VEVs and suppress  such operators. This
appears rather artificial and one may ask whether there is a deeper reason
behind it.

In this section, we explore vacua preserving the \BmL symmetry at the
high energy (GUT) scale, which appear phenomenology attractive. In
this case, the renormalizable $R$--parity violating couplings
\begin{equation}
\label{eq:RparityViolation}
 W_{\cancel{R}}~=~
 \mu_i\,\ell_i\,\phi_u
 +\lambda_{ijk}\ell_i\,\ell_j\,\overline{e}_k
 +\lambda'_{ijk}\ell_i\,q_j\,\overline{d}_k
 +\lambda''_{ijk}\overline{u}_i\,\overline{d}_j \,\overline{d}_k
\end{equation}
are prohibited, leading to suppression of proton decay. The \BmL symmetry fits 
naturally into the concept of local GUTs:  it is related to  the \SO{10}  \BmL
generator, although there are differences. Finally, \BmL can be broken at an
intermediate scale which might induce small $R$--parity violating couplings and
could be related to the smallness of  the neutrino masses.

Having suppressed \BmL violation, we  further  study whether the 
required singlet VEV configurations allow for the decoupling of the  
exotic matter, realistic flavour structures and a small $\mu$--term.

In this section, we study  a particular singlet VEV configuration
for which we are  able to prove the  $D$--flatness, but not the  $F$--flatness.
Since the phenomenological analysis is intractable for the general case,
we hope that the considerations below will provide some guidance to the
further search for realistic vacua.

\subsection{Vacuum configurations with unbroken $\boldsymbol{\BmL}$}
\label{sec:BmL}

The first step is to obtain singlet VEV configurations which 
preserve 
\[
G_\mathrm{SM}\times\U1_{\BmL}\times[\SU4] \;.
\]
Here we keep  the hidden sector  \SU4 unbroken which is  needed for gaugino
condensation.

Let us now identify a \BmL generator.  An obvious option would be to
use the \BmL of SO(10).  This however leads to anomalous \BmL
symmetry, $\mathsf{t}_{\BmL}^{\SO{10}}\cdot\mathsf{t}_\mathrm{anom}\ne
0$. It is possible to modify this generator such that the resulting
\BmL is non--anomalous and the \BmL charges for the members of the
$\boldsymbol{16}$--plets are the standard ones. Requiring further that
the hidden sector $\SU2$ doublets $h_i$ be neutral under $\U1_{\BmL}$,
fixes\footnote{This is the only phenomenologically viable $\U1_{\BmL}$
generator, up to an irrelevant component in the $\mathsf{t}_9$ direction.}
\begin{equation}
 \mathsf{t}_{\BmL}~=~
 \left(
 0,1,1,0,0,-\frac{2}{3},-\frac{2}{3},-\frac{2}{3}
 \right)
 \,
 \left(
 \frac{1}{2},\frac{1}{2},\frac{1}{2},-\frac{1}{2},0,0,0,0
 \right)
 \;.
\end{equation}

The \BmL charges of matter fields are shown in Tabs.~\ref{tab:BmL},
\ref{tab:BmLs}-\ref{tab:BmLm}.  The $q_i$ and
$\overline u_i$ states have the standard charges, while only four out
of seven $\overline d_i$ have the right charge ($-1/3$) to be
identified with the down type anti--quarks.  The $\overline d_i$
states with exotic \BmL charges as well as one linear combination of
the $\overline d_i$'s with charge $-1/3$ pair up with four $d_i$'s and
decouple from the low energy spectrum.  Similar considerations apply
to the lepton sector. The lepton doublets carry charge $-1$, while the
Higgs doublets are neutral. One pair of $\ell_i$ and $\overline
\ell_i$ with $q_{\BmL}=0$ must remain in the massless spectrum and is
identified with the physical Higgs bosons.

Among the 69 SM singlets $s_i$,  30 are neutral under  \BmL, 21 have charge
$+1$ and 18 have charge $-1$ (cf.\ Tab.~\ref{tab:BmLs}). This excess of
positively charged $s_i$  leads to a net number of three `right--handed'
neutrinos.

\begin{table}[!t]
\centerline{
\begin{tabular}{l|l}
field & $\BmL$ charges \\
\hline
$q_i$ & 
$\left\{\frac{1}{3},\frac{1}{3},\frac{1}{3}\right\}$\\
$\bar u_i$ &
$\left\{-\frac{1}{3},-\frac{1}{3},-\frac{1}{3}\right\}$\\
$\bar d_i$ &
$\left\{-\frac{1}{3},-\frac{1}{3},\frac{2}{3},-\frac{1}{3},-\frac{1}{3},\frac{2}{3},\frac{2}{3}\right\}$\\
$d_i$ &
$\left\{-\frac{2}{3},-\frac{2}{3},-\frac{2}{3},\frac{1}{3}\right\}$\\
$\ell_i$ &
$\{0,-1,-1,0,0,0,-1,-1\}$\\
$\bar\ell_i$ &
$\{0,0,0,1,0\}$\\
$\bar e_i$ &
$\{1,1,1\}$
\end{tabular}
}
\caption{$\BmL$ charges of the relevant matter fields. }
\label{tab:BmL}
\end{table}

Let us now consider configurations in which only states neutral under 
$G_\mathrm{SM}\times\U1_{\BmL}\times[\SU4]$ are allowed to develop VEVs. 
Such states include $s_i$ with zero $q_{\BmL}$ and  the $\SU2'$ doublets
$h_i$. For our purposes, it suffices to restrict ourselves to a certain subset
of these fields. In particular, we assume that  
\begin{eqnarray}
 \langle 
 s_{2}, s_{5}, s_{7}, s_{9}, s_{20}, s_{23}, s_{34}, s_{41}, s_{48}, 
 s_{58}, s_{59}, s_{62}, s_{65}, s_{66}, \;&&
 \nonumber\\
 h_{1}, h_{3}, h_{6}, h_{8},  h_{9}, h_{10}, h_{11}, h_{12}, h_{13}
 \rangle &= & 0 \;,
\end{eqnarray}
while 
\begin{eqnarray}
 \{\widetilde{s}_{i}\} & = &
 \left\{
 s_{1},s_{3}, s_{12}, s_{14}, s_{16}, s_{18}, s_{19}, s_{22}, s_{24},  s_{39},
 s_{40}, s_{53}, s_{54}, s_{57}, s_{60}, s_{61},
\right.
 \nonumber\\
 & & \left.{}{\hphantom{\left\{\right.}}
 h_{2}, h_{4}, h_{5}, h_{7}, h_{14}
 \right\} 
\label{eq:stilde}
\end{eqnarray}
develop non--zero VEVs. We find that such a configuration is $D$--flat 
since every field  from $\{\widetilde{s}_{i}\}$
 enters a gauge invariant monomial consisting
exclusively of $\{\widetilde{s}_{i}\}$ states 
(Tab.~\ref{tab:GaugeInvariantMonomials2}). Also, it is
possible to construct a monomial out of these states  which has  
a negative net anomalous charge (Tab.~\ref{tab:GaugeInvariantMonomialsNC}).
The set  $\{\widetilde{s}_{i}\}$    does not represent an 
$F$-flat direction. To preserve supersymmetry, we assume that
there exist non--trivial field configurations in $\{\widetilde{s}_{i}\}$
with vanishing $F$--terms.
Then, as described in Sec.~\ref{sec:FandDflat},
complexified gauge transformations allow us to satisfy $D_a=0$ at the
same time. The set $\{\widetilde{s}_{i}\}$ breaks all extra $\U1$'s but
$\U1_{\BmL}$.

\subsection{Decoupling  the exotic states}

The first question is whether it is possible to decouple all of the
exotic states by giving VEVs to the set $\{\widetilde{s}_{i}\}$ only. To answer
this question, we recalculate the mass matrices for the  exotic matter.  The
relevant  superpotential couplings are of the form
\begin{equation}\label{eq:xbarxU1B-L}
 W~=~
 x_i\,\bar x_j\,\mathcal{M}_x^{ij}(\widetilde{s})
 \quad\text{with}\quad
 \mathcal{M}_x^{ij}(\widetilde{s})~=~\sum
 \widetilde{s}_{i_1}\cdots\widetilde{s}_{i_n}
 \;,
\end{equation}
and  $x_i , \bar x_j$ being the vector--like pairs.
Including the couplings up to order 10,
the resulting mass matrices are 
\begin{eqnarray}
 \mathcal{M}_d^{ij}(\widetilde{s}) & = &
 \left(\begin{array}{ccccccc}
 0 & 0 & \widetilde{s}^6 & 0 & 0 & \widetilde{s}^6 & \widetilde{s}^6 \\
 0 & 0 & \widetilde{s}^6 & 0 & 0 & \widetilde{s}^7 & \widetilde{s}^7 \\
 0 & 0 & \widetilde{s}^6 & 0 & 0 & \widetilde{s}^7 & \widetilde{s}^7 \\
 \widetilde{s}^8 & 0 & 0 & \widetilde{s}^6 & \widetilde{s}^6 & 0 & 0
 \end{array}\right)\;,\label{eq:Mdstilde}\\
 \mathcal{M}_\ell^{ij}(\widetilde{s}) & = &
 \left(\begin{array}{cccccccc}
 \widetilde{s}^3 & 0 & 0 & 0 & 0 & \widetilde{s}^8 & 0 & 0 \\
 \widetilde{s} & 0 & 0 & 0 & 0 & \widetilde{s}^6 & 0 & 0 \\
 \widetilde{s} & 0 & 0 & 0 & 0 & \widetilde{s}^6 & 0 & 0 \\
 0 & \widetilde{s}^8 & \widetilde{s}^8 & 0 & 0 & 0 & \widetilde{s}^6 & \widetilde{s}^6 \\
 \widetilde{s} & 0 & 0 & \widetilde{s}^6 & \widetilde{s}^6 & 0 & 0 & 0
 \end{array}\right)\;.\label{eq:Mlstilde}
\end{eqnarray}
Here the columns in $\mathcal{M}_d^{ij}$ correspond to $\bar d_j$ and the rows
to $d_i$; in $\mathcal{M}_\ell^{ij}$, the columns  correspond to $\ell_j$ and
the rows to $\bar \ell_i$.   $\mathcal{M}_m^{ij}(\widetilde{s})$,
$\mathcal{M}_s^{ij}(\widetilde{s})$, $\mathcal{M}_f^{ij}(\widetilde{s})$, and
$\mathcal{M}_w^{ij}(\widetilde{s})$ are given by
Eqs.~\eqref{eq:Mmstilde}--\eqref{eq:Mwstilde} in App.~\ref{sec:MassMatrices}. 
As before, an entry $\widetilde{s}^N$ implies that there is an allowed coupling
involving $N$ states $\widetilde{s}_i$. For instance, the $d_1,\bar{d}_3$ mass
term includes
\begin{equation}
 W_{d_1 \bar d_3}~=~d_1 \bar d_3 (
 s_{16}\,s_{40}\,h_4\,h_{14}\,h_5\,h_{14}
 +\cdots )\;.
\end{equation}

Although the form of the mass matrices is quite restricted, all of
them have the maximal rank, apart from the $\bar \ell_i,\ell_j$ matrix
whose rank is 4. This means that all of the exotic states are
decoupled and one Higgs pair $\bar \ell,\ell$ is massless, as
required.

Clearly, some of the zeros of the mass matrices are dictated by the
\BmL symmetry (see Tabs.~\ref{tab:BmL}, \ref{tab:BmLspsm},
\ref{tab:BmLm}).  The massless down type anti--quarks are 3 linear
combinations of the 4 $\bar d_i$ states with $q_{\BmL}=-1/3$, namely
$\bar d_1$, $\bar d_2$, $\bar d_4$ and $\bar d_5$. The remaining
linear combination couples to $d_4$ and becomes heavy.  Likewise, the
physical lepton doublets are the 3 linear combinations of $\ell_2$,
$\ell_3$, $\ell_7$ and $\ell_8$ which do not couple to $\bar \ell_4$.
Interestingly, this type of structure has recently been explored in
the context of orbifold GUTs \cite{Asaka:2003iy}. It was shown that a
mixing between chiral and vector--like states can lead to realistic
flavour patterns.

Not all texture zeros can be understood from the $\BmL$ symmetry. For
instance, \BmL does not forbid the $\bar\ell_1,\ell_{j>1}$
couplings such that additional input is needed.  As we shall see,
these zeros are crucial for  identification of the Higgs doublets.

\subsection{Higgs doublets and flavour structure}

In our model, the only renormalizable \BmL conserving Yukawa coupling  which
involves SM matter is
\begin{equation}
W = g ~q_1\,\bar u_1\,\bar\ell_1 \;.
\label{topmass}
\end{equation}
This is a superpotential of the type 
\begin{equation}
U_1\, U_2\, U_3
\end{equation} 
with the untwisted superfields $U_i$ formed out of the compactified
components of the \E8 gauge multiplets in 10D. For example, for the
scalar components we have $\tilde U_1 \propto A_{4} + \I A_{5}$, etc.,
where $A_{i}$ are the gauge field components in the compact
directions. This can be understood by recalling that the gauge
supermultiplet in 10D decomposes into 1 vector and 3 chiral $N=1$
multiplets in 4D. The above superpotential results from the kinetic
term of the gauge supermultiplet in 10D with the corresponding Yukawa
coupling being the gauge coupling at the string scale.

\begin{figure}[h]
\centerline{\CenterObject{\includegraphics{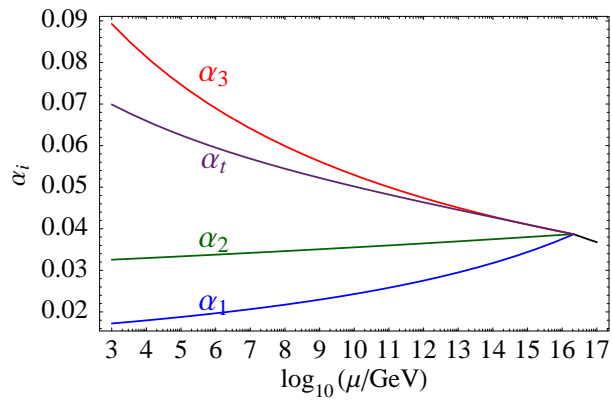}}}
\caption{An idealized picture of  gauge--top Yukawa unification. Here $\alpha_i
= g_i^2/(4 \pi)$, $\alpha_t = Y_t^2/(4 \pi)$ and we have assumed that $Y_t=g_i$
at the GUT scale.}
\label{fig:GaugeYukawaUnification}
\end{figure}

As long as $\bar\ell_1$ has a significant component in the physical 
up--type Higgs doublet, the superpotential (\ref{topmass}) 
naturally leads to a heavy top quark. The top Yukawa coupling is then 
of the order of the gauge coupling at the string scale,
\begin{equation}
Y_t ~\sim~ g \;.
\end{equation} 
This remarkable  top Yukawa--gauge unification
(Fig.~\ref{fig:GaugeYukawaUnification}) stems from the fact that the top quark
is a gaugino in 10D. The other quark Yukawa couplings vanish  at the
renormalizable level. We note that a large top Yukawa coupling has also been obtained 
in earlier analyses of the fermion masses
\cite{Ibanez:1986ka} - 
\cite{Cleaver:1998gc}, 
\cite{Kobayashi:2004ya}.

This attractive mechanism is at work when $\bar\ell_1\simeq
\phi_u$. Inspection of the $\ell_i \bar\ell_j$ mass matrix
\eqref{eq:Mlstilde} shows that the first 3 rows are linearly dependent
and the $\bar\ell_1\,\ell_1$ coupling appears only at order 5 while
the $\bar\ell_i\,\ell_1$ couplings with $i=2,3$ occur already at order
3. Thus, if the relevant $\tilde s_i$ VEVs are below the string scale,
one expects at least mild suppression of the $(1,1)$ entry.  Then,
the massless up-type Higgs is dominated by $\bar\ell_1$,
\begin{equation}
 \phi_u~\simeq~\bar\ell_1+\sum\limits_{i=2,3}\varepsilon_i\bar\ell_i\;,
 \quad |\varepsilon_i|\ll1\;,
\end{equation}
whereas the down-type Higgs is a linear combination of $\ell_4$ and $\ell_5$, 
\begin{equation}
 \phi_d~=~\ell_4\,a+\ell_5\,b\;,\quad
\end{equation}
with $|a|,\,|b|$ of order one.
Our choice of the  vacuum configuration  $\widetilde{s}_i$ 
(Eq.~\eqref{eq:stilde}) was in part  motivated by these  considerations. 

At this stage, the Higgs doublets $\phi_{u,d}$ are massless.  In contrast to
the  conventional  4D GUTs, no finetuning is required to keep the doublets light
while decoupling the color triplets. 
If the \BmL symmetry gets broken at an intermediate scale, a small
$\mu$-term will be generated.

Having identified the Higgs doublets and the top quark, we turn to the
discussion of the remaining Yukawa couplings. The relevant superpotential is
\begin{equation}
 W_\mathrm{Yukawa}
 ~=~
 Y_u^{ij}(\widetilde{s})\,\phi_u\,q_i\,\bar u_j
 +Y_d^{ia}(\widetilde{s})\,\phi_d\,q_i\,\bar d_a
 +Y_e^{ib}(\widetilde{s})\,\phi_d\,\bar e_i\,\ell_b\;,
\end{equation}
where $a\in\{1,2,4,5\}$ and $b\in\{2,3,7,8\}$. The Yukawa matrices 
at order 10 are
\begin{eqnarray}
 Y_u^{ij}(\widetilde{s}) & = &
 \left(\begin{array}{ccc}
 	g & \widetilde{s}^6 & \widetilde{s}^4 \\
 	\widetilde{s}^3 & 0 & \widetilde{s}^7 \\
 	\widetilde{s}^7 & \widetilde{s}^7 & 0
 \end{array}\right)\;,\\
 Y_d^{ia}(\widetilde{s}) & = &
 \left(\begin{array}{cccc}
  	0 & \widetilde{s}^6 & \widetilde{s}^2 & \widetilde{s}^2 \\
 	\widetilde{s}^5 & 0 & \widetilde{s}^5 & \widetilde{s}^5 \\
 	0 & \widetilde{s}^5 & \widetilde{s} & \widetilde{s}
 \end{array}\right)\;,\\
 Y_e^{ib}(\widetilde{s}) & = &
 \left(\begin{array}{cccc}
 	0 & \widetilde{s}^6 & 0 & 0 \\
 	\widetilde{s}^5 & 0 & 0 & 0 \\
 	0 & \widetilde{s}^5 & 0 & 0
 \end{array}\right)\;.
\end{eqnarray}
The low-energy $3\times3$  $Y_d$ and $Y_e$ Yukawa matrices\footnote{Here we
neglect corrections to the K\"ahler potential.}
   are obtained by
integrating out one linear combination of $\bar d_a$ ($a=1,2,4,5$) which pairs
up with $d_4$ and one linear combination of $\ell_b$ ($b=2,3,7,8$) which pairs
up with $\bar \ell_4$. Their precise form depends on various coefficients, so
let us only discuss their main features.

The quark Yukawa matrices have  the full rank such that, in general, 
there are no massless
eigenstates. The lepton Yukawa matrix has rank 2 implying that
the electron is massless to order 8 in the superpotential.
Also, there appears one massless pair of $f$--plets in the hidden
sector, which somewhat lowers the gaugino condensation scale. 

Assuming that the relevant $\widetilde{s}_i$ have VEVs below the
string scale, the Yukawa couplings are hierarchical and resemble the
Froggatt--Nielsen structure \cite{Froggatt:1978nt}.  However, the hierarchy
appears due to the string selection rules rather than the \U1 charge assignment
only as in the original Froggatt--Nielsen mechanism.

It is remarkable that the up--type quarks  tend to be  heavier than  the
down--type quarks, which in turn are heavier than the leptons.  We also note
that the Yukawa matrices generally contain non--trivial CP phases due to complex
$\tilde s_i$ VEVs.

\subsection{Proton stability and $\boldsymbol{\BmL}$ breakdown}

The \BmL symmetry enforces absence of renormalizable operators  leading to
proton decay. However, non--renormalizable \BmL conserving operators such as 
\begin{equation}
 W ~=~
 \kappa_{ijkl}^{(1)}~q_i\,q_j\,q_k\,\ell_l
 +
 \kappa_{ijkl}^{(2)}~\bar u_i\,\bar u_j\,\bar d_k\,\bar e_l 
 \label{eq:Wprotondecay}
\end{equation}
also induce proton decay.  The resulting constraint on $\kappa_{ijkl}^{(1)}$ and
$\kappa_{ijkl}^{(2)}$ involving  the first two generations is very tight
\cite{Hinchliffe:1992ad,Murayama:1994tc},
\begin{equation}
 \kappa_{ijkl}^{(1,2)}
 ~\lesssim~\frac{10^{-8}}{M_\mathrm{P}}\;.
\end{equation}

The operators (\ref{eq:Wprotondecay}) are induced both directly and by
integrating out the vector--like matter. For example, integrating out
a heavy pair $d_2 \bar d_6$ from the superpotential $ W = q_2 q_2
d_2 + q_3 \ell_3 \bar d_6 $ yields $W\sim q_2 q_2 q_3 \ell_3 $.  These
operators can be suppressed either by tuning the $\tilde s_i$ VEVs or
by an additional, perhaps discrete, symmetry
\cite{Dreiner:2005rd,Tatar:2006dc}.  This issue will be discussed
elsewhere.

\label{sec:BreakBmL}

Breaking \BmL is a difficult issue.  It has to occur at an energy
scale $M_{\BmL}$ well below $M_\mathrm{string}$. In the following, we
assume that the scale $M_{\BmL}$ is generated dynamically, without
breaking SUSY.

\BmL breaking VEVs   fill in the zeros of the mass matrices, in particular,
(\ref{eq:Mlstilde}). This generates the $\mu$--term, $W = \mu ~\phi_u \phi_d$. 
Assuming that apart from $\tilde s_i$, only  states charged under \BmL get
non--zero VEVs, its magnitude is ($n\geq 1$)
\begin{equation}
 \mu~\sim~\frac{M_{\BmL}^{n+1}}{M_\mathrm{string}^n}\;.
\end{equation}
For an intermediate scale $M_{\BmL}$, this can give a phenomenologically viable
$\mu$--term.

\BmL breakdown  generates masses for the right--handed neutrinos. Our
construction has the necessary ingredients for the seesaw, i.e., neutrino
Yukawa couplings and large Majorana neutrino masses. A detailed analysis of
this issue will be presented elsewhere.

Finally, small $R$--parity violating couplings are generated. Their
magnitude is given by $(M_{\BmL}/M_\mathrm{string})^m$ with $m$
depending on the type of the coupling. For $m\geq 2$, an intermediate
scale $M_{\BmL}$ suppresses proton decay sufficiently
\cite{Dreiner:1997uz}.

To conclude, in this section we have studied a vacuum configuration
with conserved \BmL at the string scale. This suppresses
renormalizable $R$--parity violating couplings as well as the
$\mu$--term. Furthermore, flavour structures \`a la Froggatt--Nielsen
arise as a consequence of the string selection rules.


\section{Outlook}
\label{sec:Outlook}

Guided by the idea of local grand unification we have constructed an orbifold
compactification of the heterotic string which leads to the supersymmetric
standard model gauge group and particle content. 
The model has large vacuum degeneracy. For certain vacua
with unbroken \BmL symmetry, the resulting phenomenology is particularly
attractive.  In this case, one pair of Higgs doublets is massless
automatically, with the  subsequently generated    $\mu$--term
being due to \BmL breaking.
The top quark Yukawa  coupling is of the order of the gauge coupling 
and the arising  pattern of Yukawa couplings
is  reminiscent of the Froggatt-Nielsen textures.

These results can be the first steps towards a fully realistic theory. They
immediately lead, however, to further questions which concern detailed properties 
of SUSY vacua,
$B\!-\!L$  breakdown, incorporation of the seesaw mechanism,   identification of
$R$--parity and proton decay. Furthermore, effects of string threshold corrections
and other contributions on gauge coupling unification  have to be studied.
Eventually, one would like to determine quantitatively the
 Yukawa couplings for specific supersymmetric vacua.

On the conceptual side, a  deeper understanding of the decoupling of  exotic
states is particularly desirable. Orbifolds often represent special  points in
the moduli space of more general Calabi--Yau compactifications. Non--zero vacuum
expectation values of specific standard model singlets  correspond to other
points in the moduli space where the orbifold singularities  have been blown up. Since
these vacuum expectation values also generate mass terms, at least some of the
unwanted exotic states should be absent in compactifications on smooth
manifolds. The orbifold limit of Calabi-Yau compactifications is well understood
for the standard embedding \cite{Polchinski:1998rr} but remains to be studied in
detail for non--standard embeddings which are relevant to the models
presented in this paper.

Finally, it is important to search for other models in the framework  of the
$\E8\times\E8$ and the $\SO{32}$ heterotic string with localized
$\boldsymbol{16}$--plets of $\SO{10}$ \cite{Choi:2004wn,Nilles:2006np}.  It
would also be very interesting to understand the connection between orbifold
compactifications  and  compactifications on Calabi--Yau manifolds endowed with
vector bundles \cite{Bouchard:2005ag,Braun:2005nv,Braun:2006me,
Bouchard:2006dn,Blumenhagen:2006ux}, which have many phenomenologically
appealing features.

\subsection*{Acknowledgments}

We would like to thank  R.~Blumenhagen, M.~Cveti{\u{c}}, A.~Hebecker,
T.~Kobayashi, W.~Lerche, M.~Lindner, J.~Louis, F.~Pl\"oger, S.~Raby,
S.~Ramos-S\'anchez, S.~Stieberger, S.~Theisen, P.~K.~S.~Vaudrevange  and in
particular H.~P.~Nilles  for valuable discussions.
We are indebted to  P.~K.~S.~Vaudrevange for pointing out an error 
in the string selection rules
presented in an earlier version of this paper.

This work was partially supported by the EU 6th Framework Program
MRTN-CT-2004-503369 ``Quest for Unification'' and MRTN-CT-2004-005104
``ForcesUniverse'' as well as the virtual institute VIPAC of the Helmholtz
society.

\clearpage
\appendix
\section{Sample calculations}
\label{sec:SampleCalculation}

In this appendix, we present details of the spectrum calculation for our model. 
These calculations are straightforward but tedious. For
practical purposes, it is convenient to automatize them by means of a computer
algebra system.

\subsection{Gauge group}

The 4D gauge group is obtained by subjecting the \E8 roots
$p$ ($p^2=2$) to the projection conditions
\begin{subequations}
\begin{eqnarray}
 p\cdot V_6 & \in & \mathbbm{Z}\;,\label{eq:pV6}\\
 p\cdot W_2 & \in & \mathbbm{Z}\;,\label{eq:pW2}\\
 p\cdot W_3 & \in & \mathbbm{Z}\;.\label{eq:pW3}
\end{eqnarray}
\end{subequations}
Consider condition (\ref{eq:pV6}). The roots of the first \E8 surviving the twist
are
\begin{equation}
 p~\in~\left\{\pm(1,1,0^6),\pm(1,-1,0^6),\pm(0^3,\underline{1,1,0,0,0}),
 (0^3,\underline{1,-1,0,0,0})\right\}\;,
\end{equation}
where the underline denotes permutations and the superscripts indicate repeated
entries. The simple roots are by definition the smallest linearly independent
positive roots (cf.\ \cite{Cahn:1985wk}).
For a suitable choice of positivity, they read
\begin{eqnarray}
\lefteqn{ \{p_\mathrm{sr}\}
 ~= ~\left\{ (1, 1, 0^6), (1, -1, 0^6),\right.}\nonumber\\
 & & \phantom{\left\{\right.}\left.
 (0^3, 1, -1,0^3), (0^4, 1, -1,0^2),
 (0^5, 1, -1,0),(0^6, 1, -1),(0^6, 1, 1)
 \right\}\;.
 \label{eq:simpleRootsSO10}
\end{eqnarray}
Calculating the Cartan matrix $A_{ij}=p_\mathrm{sr}^i\cdot p_\mathrm{sr}^j$,
one finds that
the simple roots in the first line correspond to the raising operators of two \SU2
factors whereas those in the second line correspond to \SO{10}.
Thus, the gauge group after twisting is $\SO{10}\times\SU2^2$, which also corresponds to the
local gauge symmetry at the origin (cf.\
Tab.~\ref{tab:LocalGUTs}). 

The Wilson line projections \eqref{eq:pW2} and \eqref{eq:pW3} 
lead to the simple roots
\begin{equation}\label{eq:SimpleRootsSM}
 \left\{(0^5, 1, -1, 0), (0^6, 1, -1), 
 (0^3, 1, -1, 0^3)\right\} \;,
\end{equation}
which correspond to the gauge groups \SU3 and \SU2 in 4D. 
All \E8 Cartan generators
survive the projection. They give rise to the Cartan generators of \SU3 and \SU2
and to five \U1 generators. The latter can be represented by vectors
perpendicular to the simple roots (Eq.~\eqref{eq:U1generators}). 
The surviving subgroup of the second \E8 is obtained
analogously.

\subsection{Untwisted sector}

The untwisted sector states are obtained from the projection
\begin{subequations}
\begin{eqnarray}
 p\cdot V_6-q\cdot v_6& \in & \mathbbm{Z}\;,\quad
 p\cdot V_6~\not\in~\mathbbm{Z}\;, \label{eq:pV62}\\
 p\cdot W_2 & \in & \mathbbm{Z}\;,\label{eq:pW22}\\
 p\cdot W_3 & \in & \mathbbm{Z}\;,\label{eq:pW32}
\end{eqnarray}
\end{subequations}
with $p^2=2$.
There are 118 weights transforming in the first \E8 which survive the first
projection \eqref{eq:pV62}. They include
\begin{equation}
 \left\{
 \left(\tfrac{1}{2},-\tfrac{1}{2},\tfrac{1}{2},\text{odd}\:(\pm\tfrac{1}{2})^5
 \right),
 \left(-\tfrac{1}{2},\tfrac{1}{2},\tfrac{1}{2},
 \text{odd}\:(\pm\tfrac{1}{2})^5\right)
 \right\}\;,\label{eq:SampleWeights}
\end{equation}
where ``$\text{odd}\:(\pm\frac{1}{2})^5$'' denotes 5 entries
$\pm\tfrac{1}{2}$ with an odd number of `$-$' signs.
The Dynkin labels of these representations 
are obtained by multiplying the above weights by the simple
roots \eqref{eq:simpleRootsSO10}. 
One finds that \eqref{eq:SampleWeights} is $(\boldsymbol{16},\boldsymbol{1},\boldsymbol{2})$
of $\SO{10}\times\SU2^2$ (cf.\ \cite{Slansky:1981yr}).

The Wilson
line projections eliminate some of the states such that the 4D result is 
 $(\overline{\boldsymbol{3}},\boldsymbol{1})$ of 
$\SU3\times\SU2$ plus non--Abelian singlets. The second \E8 states are 
determined analogously.

\subsection{$\boldsymbol{T_1}$}

\begin{table}[t]
\centerline{\begin{tabular}{|l|l|l|l|}
\hline
$k$ & $\omega^{(k)}$ & $\overline{\omega}^{(k)}$ & $\delta c^{(k)}$\\
\hline
1 & $\frac{1}{6}(5,4,3)$ & $\frac{1}{6}(1,2,3)$ & $11/36$ \\
2 & $\frac{1}{3}(2,1,3)$ & $\frac{1}{3}(1,2,3)$ & $2/9$ \\
3 & $\frac{1}{2}(1,2,1)$ & $\frac{1}{2}(1,2,1)$ & $1/4$ \\
4 & $\frac{1}{3}(1,2,3)$ & $\frac{1}{3}(2,1,3)$ & $2/9$ \\
\hline
\end{tabular}}
\caption{$\omega^{(k)}$, $\overline{\omega}^{(k)}$ and $\delta c^{(k)}$ in
$\mathbbm{Z}_{6-\mathrm{II}}$ orbifolds with $v_6=\frac{1}{6}(-1,-2,3;0)$.}
\label{tab:Example4omegas}
\end{table}

The first step is to solve the mass equations \eqref{eq:MassEquations}. For
convenience, the quantities $\omega^{(k)}$, $\overline{\omega}^{(k)}$ and 
$\delta c^{(k)}$ appearing in \eqref{eq:MassEquations} are listed in
Tab.~\ref{tab:Example4omegas}. Consider now the $(\theta,0)$ sector, i.e.\
$V_{(\theta,0)}=V_6$. For $\widetilde{N}=0$, the shifted $\E8\times\E8$ momenta
$p_\mathrm{sh}\equiv p+V_6$ with $p\in \Lambda_{\E8\times\E8}$ are\footnote{An 
efficient way to solve automatically the mass equations is presented in
\cite{Vaudrevange:2004dt}.}
\begin{equation}
 \{p_\mathrm{sh}\} ~ = ~
 \bigl\{
 \left(0, 0, -\tfrac{1}{6}, 
 \text{odd}\:(\pm\tfrac{1}{2})^5\right)\, (\tfrac{1}{3}, 0^7)
 \bigr.\}\;.
\end{equation}
Using the Dynkin labels, it is straightforward to show that these weights
transform as $\boldsymbol{16}$ of the local \SO{10}. The
corresponding \SO8 lattice shifted momenta are given by
\begin{equation}
 q_\mathrm{sh}~=~ \{(\tfrac{1}{3}, \tfrac{1}{6}, 0; -\tfrac{1}{2}),
 (-\tfrac{1}{6}, -\tfrac{1}{3}, -\tfrac{1}{2}; 0)\}\;.
\end{equation}
They describe the fermion and the boson of an $N=1$ left--chiral
superfield. 
As stated in Sec.~\ref{sec:StringsOnOrbifold}, 
solutions to the mass equation in the $T_1$ sector are 
twist invariant and all appear in the 4D spectrum.
The above $\boldsymbol{16}$--plet thus produces one complete generation
of the SM matter.

Apart from the
$\boldsymbol{16}$--plet, the massless spectrum contains one $(\boldsymbol{2},\boldsymbol{1})$ and
two $(\boldsymbol{1},\boldsymbol{2})$ representations under the local $\SU2^2$. 
Other $T_1$ states are obtained
by solving the mass equations for $V_f=V_6+n_2\,W_2+n_3\,W_3$
with $0\le n_2\le1$ and $0\le n_3\le 2$.

\subsection{$\boldsymbol{T_2}$}

Consider the $(\theta^2,0)$ sector. The local gauge group is given by
the \E8 roots satisfying 
$V_{(\theta^2,0)}\cdot p=0$ mod 1, where $V_{(\theta^2,0)}=2V_6$. 
 This yields $\SO{14}\times[\SO{14}]$. 
The corresponding massless matter at the origin is 
\begin{equation}
 (\boldsymbol{14};\boldsymbol{1})\oplus(\boldsymbol{1};\boldsymbol{14})\oplus
 (\boldsymbol{1};\boldsymbol{1})\oplus(\boldsymbol{1};\boldsymbol{1})\;,
\label{T2example}
\end{equation}
where the non--Abelian singlets have non--zero oscillator numbers. 
On the right--moving side, one has 4 solutions to the mass equations with $v_3=2v_6$
which combine into an $N=2$ multiplet.
As explained in Secs.~\ref{sec:StringsOnOrbifold} and
\ref{sec:Z3xZ2geometry}, 
the next step is to form linear combinations of the massless states
which produce \Z6 eigenstates. These are then subject to the conditions
\eqref{eq:ProjectionT2T4} with $q_\gamma\in\{0,\tfrac{1}{2},1\}$ and
$p_\mathrm{sh}\cdot W_2\in\mathbbm{Z}$.
The resulting spectrum is chiral.

\subsection{$\boldsymbol{T_3}$}

The local gauge shifts are $V_f=3(V_6+n_2\,W_2)$ with $0\le n_2\le1$. 
The corresponding local
gauge groups and matter are shown in
Tab.~\ref{tab:LocalGUTs}. Again, one must impose projection conditions
\eqref{eq:ProjectionT3}, now with $q_\gamma\in\{0,\pm\tfrac{1}{3},1\}$ and
$p_\mathrm{sh}\cdot W_3\in\mathbbm{Z}$.

\subsection{$\boldsymbol{T_4}$}

The $T_4$ states are obtained analogously to the $T_2$ states, with the only
difference being the local shift $V_f=4(V_6+n_3\,W_3)$ and $v=4v_6$.

\subsection{$\boldsymbol{T_5}$}

The fermionic component of the massless right mover has $q_4= +1/2$. 
The massless states are CP--conjugates of the $T_1$ sector and 
no left--chiral superfields arise in $T_5$.

\section{Additional material for the selection rules}

\label{sec:AESR}

This appendix contains additional information on the string selection rules of
Sec.~\ref{sec:CouplingsAndSelectionRules} and outlines of proofs of some
statements.

\subsection{Sublattices}

The space group rule states that $\ell_i$ of the space group elements have to
add up to zero up to shifts in the corresponding  sublattices.  For
concreteness, these sublattices are listed in Tab.~\ref{tab:sublattices}.

\begin{table}[h]
\centerline{
\begin{tabular}{|l|c|c|c|}
\hline
 sublattice & \G2 plane & \SU3 plane & \SO4 plane\\
\hline
 $(\mathbbm{1}-\theta^1)\,\Lambda$
 & $n\,e_1 + m\, e_2$ 
 & $n\,e_3 + (-n-3m)\, e_4$ 
 & $2n\,e_5 + 2m\, e_6$ \\
\hline
 $(\mathbbm{1}-\theta^2)\,\Lambda$
 & $3n\,e_1 + m\, e_2$ 
 & $n\,e_3 + (-n-3m)\, e_4$ 
 & $-$ \\
\hline
 $(\mathbbm{1}-\theta^3)\,\Lambda$
 & $2n\,e_1 +2 m\, e_2$ 
 & $-$ 
 & $2n\,e_5 + 2m\, e_6$ \\
\hline
\end{tabular}
}
\caption{Sublattices $(\mathbbm{1}-\theta^k)\Lambda$. The integers $n$, $m$ are
varied  independently   in  each plane. 
Note that 
$(\mathbbm{1}-\theta^{6-k})\Lambda=(\mathbbm{1}-\theta^k)\Lambda$.}
\label{tab:sublattices}
\end{table}

\subsection{On Eq.~(\ref{eq:SpaceGroupSelectionRules})}

The coupling among $n$ states must satisfy Eq.~\eqref{sgr}. 
Using the multiplication law for the space group, one has
\begin{eqnarray}
 (\theta^{k_1},\ell_1)\,(\theta^{k_2},\ell_2)\dots(\theta^{k_n},\ell_n)
 ~=~(\theta^{k_1}\,\theta^{k_2}\cdots\theta^{k_n},\ell_1+\theta^{k_1}\ell_2+\dots+
 \theta^{k_1}\cdots\theta^{k_{n-1}}\,\ell_n). \nonumber
\end{eqnarray}
The rule $\sum\limits_{r=1}^n k_r  = 0\mod 6$ is then obvious.  Further, by
shifting the $\ell_i$,
\begin{equation}
 \ell_i~\to~\ell_i+(1-\theta^{k_i})\,\lambda_i\;,
\end{equation}
one can always achieve
\begin{equation}
 \ell_1+\theta^{k_1}\ell_2+\dots+
 \theta^{k_1}\cdots\theta^{k_{n-1}}\,\ell_n
 ~\to~\ell_1+\ell_2+\dots+\ell_n\;.
\end{equation}
Thus, $\sum \ell_i =0$ up to the sublattice $\sum \Lambda_{k_i}$.

\subsection{On the selection rules in the $\boldsymbol{\mathrm{G}_2}$ plane}

Consider the \G2 plane selection rule for  a coupling of string states.  If
$T_1$ states are involved, the sublattice  $\sum \Lambda_{k_i} = \sum (1-
\theta^{k_i}) \Lambda$ is the entire lattice and all fixed points can couple.
Similarly, there is no restriction  when $T_3$ and $T_{2}$ (or $T_{4}$) sectors
are present simultaneously. 

Suppose now that the coupling involves only the $T_2$ states. The corresponding
fixed points are x,y,z (Sec.~\ref{sec:Z3xZ2geometry}). 
x is at the origin
and is \Z6 invariant, while y and z  are interchanged under
$\theta$--twisting. The couplings consistent with the space group selection rule
for the \G2 plane are ${\rm x}^n$, ${\rm x\,y\,z}$, ${\rm y^3}$, ${\rm z^3}$ 
and higher couplings built out
of these blocks.   In terms of $\gamma$--eigenstates, this
means that the coupling of any number of $q_\gamma=0$ states to a single
$q_\gamma \not= 0$ state is prohibited, while the others are allowed, i.e.
$\{q_{\gamma(1)},\dots q_{\gamma(n)}\}
~\not\in~\text{permutations}\{x,0,\dots0\}$ with $x\not=0$.
Similar considerations apply to
couplings of the type $T_4\dots T_4$, $T_2 T_2\dots T_4\, T_4$ and $T_3\dots
T_3$.

\section{Models with 3 local $\boldsymbol{16}$--plets}
\label{app:NoGo}

In this appendix, we discuss the obstacles to obtaining 3 equivalent
families of $\boldsymbol{16}$--plets in $\Z{N}$ orbifold models with $N\le 6$.
Triplication of families could in principle be a result of the presence of 3
equivalent fixed points, which support SM matter in the first twisted sector
$T_1$.\footnote{One could also entertain the possibility of obtaining 3
equivalent families from higher twisted sectors. However, such states are
subject to additional projection conditions which usually destroy either the
equivalence of families or their GUT sctructure.} We however find that this
simple possibility cannot be realized, at least in $\Z{N \le 6 }$.

First of all, in the $\Z{3}$ orbifold one does not have a local
$\boldsymbol{16}$--plet because one cannot break $\E{8}$ to $\SO{10}$ by a
$\Z{3}$ twist. Then, in $\Z{4}$ orbifolds there is no triplication due to
geometry, i.e.\ the number of equivalent fixed points is under no circumstances
divisible by 3. The next simplest possibility is the $\Z{6}$ which we examine in
detail below.

For $\Z{6}$ orbifolds, all possible local shifts $V_6$ are listed in
\cite{Katsuki:1989qz}, together with the corresponding local groups and local
$T_1$ states. Among them, there are only 5 local shifts $V_6$ which have a local
$\SO{10}$ and a $\boldsymbol{16}$--plet. There are 3 of them in
$\Z{6-\mathrm{I}}$ models [$v_6 = {1\over 6}(-1,-1,2)$],
\begin{eqnarray}
 V_6 = {1\over 6}(2,2,2,0^5)\,(2,1,1,0^5)\,,\;
 {1\over 6}(3,3,2,0^5)\,(2,2,0^6)\,,\;
 {1\over 6}(4,1,1,0^5)\,(0^8)\,,
\end{eqnarray}
and 2 in $\Z{6-\mathrm{II}}$ models [$v_6 = {1\over 6}(-1,-2,3)$],
\begin{eqnarray}
 V_6 = {1\over 6}(2,2,2,0^5)\,(1,1,0^6)\,,\;
 {1\over 6}(3,3,2,0^5)\,(2,0^7)\,.
\end{eqnarray}
These local shifts can be accompanied by Wilson lines $W_2$, $W_2'$, and $W_3$,
depending on the geometry of the orbifold~\cite{Kobayashi:1990mi}.

We demand that the $\SO{10}$ be broken to $\SU{3}\times\SU{2}\times\U1^2$ by the
orbifold action. This requires at least two different Wilson lines. The
$\Z{6-\mathrm{I}}$ models allow for only one Wilson line $W_3$, which destroys
the triplication, and hence we discard them. The $\Z{6-\mathrm{II}}$ models
allow for combinations of $(W_2, W_2')$, $(W_2,W_3)$, and $(W_2, W_2', W_3)$. 
Among them, only the first one can produce three equivalent fixed points with
local $\SO{10}$ symmetry and a $\boldsymbol{16}$--plet (cf.
Fig.~\ref{fig:SequentialFamilies}). We therefore concentrate on these models,
namely,
\begin{eqnarray}
 V_6 & = & {1\over 6}(2,2,2,0^5)(1,1,0^6)\;,\quad
 W_2 ~=~ \mathrm{any}\,,\; W'_2~=~\mathrm{any}\,,
 \label{eq:V6W2W2-1}
 \\
 V_6 & = & {1\over 6}(3,3,2,0^5)(2,0^7)\;,\quad
 W_2 ~=~ \mathrm{any}\,,\; W'_2~=~\mathrm{any}\,.
 \label{eq:V6W2W2-2}
\end{eqnarray}

Naively, one may think that the number of models to be studied is enormous.
However, employing symmetry transformations of the local shifts and Wilson
lines which produce equivalent models, one can show that most of the models are
redundant. These symmetries, which include lattice translations and 
Weyl reflections, have been used in Ref.~\cite{Giedt:2000bi} for a systematic
classification of inequivalent models in $\Z{3}$ orbifolds. We have performed a
similar classification of the $\Z{6}$ models and found that there are at most
69 inequivalent models of type \eqref{eq:V6W2W2-1} and at most 129 inequivalent
models of type \eqref{eq:V6W2W2-2}. At this stage, we have only required modular
invariance and $\SO{10}$ breakdown to $\SU{3}\times\SU{2}\times\U1^2$.

As the next step, we have studied the massless spectrum of these models and
identified quantum numbers of exotic states.  Remarkably, we found that all of
these models contain exotic states which are chiral with respect to $\SU{3}_c
\times \SU{2}_\mathrm{L} \times \U1_Y$.\footnote{In $\SO{10}$, there are two
distinct choices of $\U1_Y$ which exchange the definitions of up--type and
down--type right--handed quarks. We have checked both possibilities.} Such
states cannot be decoupled and, therefore, the low energy spectrum contains
exotic particles beyond the MSSM.  We thus conclude that geometric triplication
of $\boldsymbol{16}$--plets is not possible in $\Z{N\le 6}$ orbifolds.

\clearpage
\setcounter{equation}{0}
\addtolength{\evensidemargin}{-1.25cm}
\addtolength{\oddsidemargin}{-1.25cm}
\section{Tables}
\label{sec:Tables}

\subsection{States of the model of Sec.~\ref{sec:Model}}

\subsubsection{Survey of local GUTs}

\begin{longtable}{|c|c|l|l|}
\hline
 $k$ & $n_3$ & $n_2=0$ & $n_2=1$\\
\hline
\endhead
\hline
\endfoot
\hline
\caption{Local GUT groups and representations. Non--Abelian singlets and \U1
factors are omitted. The brackets $[\dots]$ indicate subgroups of the second
\E8. For different $k$, $n_3$ and $n_2$ the groups are in general embedded
differently into \E8. The local GUTs can be inferred from  the tables of
Ref.~\cite{Katsuki:1989qz}.
}
\label{tab:LocalGUTs}
\endlastfoot
 $1$ & $0$ & 
 $\SO{10}\times\SO{4}\times \left[\SO{14}\right]$
 & 
 $\SO{8}\times\SU{4} \times \left[\SU{7}\right]$\\
 & & 
 $\left(\boldsymbol{16},\boldsymbol{1},\boldsymbol{1};\boldsymbol{1}\right)
 \oplus
 2\times\left(\boldsymbol{1},\boldsymbol{2},\boldsymbol{1};\boldsymbol{1}\right)
 \oplus
 \left(\boldsymbol{1},\boldsymbol{1},\boldsymbol{2};\boldsymbol{1}\right)$
 & 
 $ \left(\boldsymbol{1},\boldsymbol{4};\boldsymbol{1}\right)$\\
\hline
 $1$ & $1$ &
 $\SO{12} \times \left[\SO{8}\times\SU{4}\right]$ &
 $\SO{8}\times\SU{4} \times \left[\SU{7}\right]$\\
 & & 
 $\left(\boldsymbol{1};\overline{\boldsymbol{8}},\boldsymbol{1}\right)
 \oplus
 \left(\boldsymbol{1};\boldsymbol{1},\overline{\boldsymbol{4}}\right)$ 
 &
 $\left(\boldsymbol{1},\boldsymbol{4};\boldsymbol{1}\right)$\\ 
\hline
 $1$ & $2$ &
 $\SU{7} \times \left[\SO{8}\times\SU{4}\right]$
 &
 $\SO{8}\times\SU{4} \times \left[\SO{10}\times\SO4\right]$\\
 & & 
 $\left(\boldsymbol{1};\boldsymbol{1},\overline{\boldsymbol{4}}\right)$ &
 $\left(\boldsymbol{1},\overline{\boldsymbol{4}};\boldsymbol{1},\boldsymbol{1},\boldsymbol{2}\right)$\\
\hline
\hline
 $2$ & $0$ & \multicolumn{2}{|c|}{$\SO{14} \times \left[\SO{14}\right]$}\\
 & & \multicolumn{2}{|c|}{$\left(\boldsymbol{14};\boldsymbol{1}\right)\oplus
  \left(\boldsymbol{1};\boldsymbol{14}\right)$}\\
\end{longtable}

\subsubsection{Spectrum of the model of Sec.~\ref{sec:Model}}

\begin{longtable}{|c|c|c|c|c|c|ccc|l|c|c|c|c|c|c|c|c|c|}
\hline
 & $k$ & $n_3$ & $n_2$ & $n_2'$ & $q_\gamma$ & 
 	$R_1$ & $R_2$ & $R_3$ & $\text{irrep}$ 
	& $q_Y$ & $q_2$ & $q_3$ & $q_4$ & $q_5$ & $q_6$ & $q_7$ & $q_8$ & $q_9$\\
\hline
\endhead
\hline
\endfoot
\hline
\caption{All SM non-singlet representations in terms of \emph{left-chiral}
states. The \U1 charges refer to the basis of generators
\eqref{eq:U1generatorsSaturdayAfternoon}. The $H$--momenta $R_i$ are listed for
the bosonic components.
}
\label{tab:LongTableNonSingletsSaturdayAfternoonModel}
\endlastfoot
 $\bar e_{1}$ & $0$ & $*$ & $*$ & $*$ & $0$ & $-1$ & $0$ & $0$ & $\left(\boldsymbol{1},\boldsymbol{1};\boldsymbol{1},\boldsymbol{1}\right)$ & $1$ & $\frac{1}{2}$ & $-\frac{1}{2}$ & $\frac{1}{2}$ & $-\frac{1}{2}$ & $0$ & $0$ & $0$ & $0$\\
 $\bar u_{1}$ & $0$ & $*$ & $*$ & $*$ & $0$ & $-1$ & $0$ & $0$ & $\left(\overline{\boldsymbol{3}},\boldsymbol{1};\boldsymbol{1},\boldsymbol{1}\right)$ & $-\frac{2}{3}$ & $\frac{1}{2}$ & $-\frac{1}{2}$ & $\frac{1}{2}$ & $-\frac{1}{2}$ & $0$ & $0$ & $0$ & $0$\\
\hline
 $q_{1}$ & $0$ & $*$ & $*$ & $*$ & $0$ & $0$ & $-1$ & $0$ & $\left(\boldsymbol{3},\boldsymbol{2};\boldsymbol{1},\boldsymbol{1}\right)$ & $\frac{1}{6}$ & $\frac{1}{2}$ & $\frac{1}{2}$ & $-\frac{1}{2}$ & $-\frac{1}{2}$ & $0$ & $0$ & $0$ & $0$\\
\hline
 $\ell_{1}$ & $0$ & $*$ & $*$ & $*$ & $0$ & $0$ & $0$ & $-1$ & $\left(\boldsymbol{1},\boldsymbol{2};\boldsymbol{1},\boldsymbol{1}\right)$ & $-\frac{1}{2}$ & $1$ & $0$ & $0$ & $-1$ & $0$ & $0$ & $0$ & $0$\\
 $\bar \ell_{1}$ & $0$ & $*$ & $*$ & $*$ & $0$ & $0$ & $0$ & $-1$ & $\left(\boldsymbol{1},\boldsymbol{2};\boldsymbol{1},\boldsymbol{1}\right)$ & $\frac{1}{2}$ & $-1$ & $0$ & $0$ & $1$ & $0$ & $0$ & $0$ & $0$\\
\hline
\hline
 $\bar e_{2}$ & $1$ & $0$ & $0$ & $0$ & $0$ & $-\frac{1}{6}$ & $-\frac{1}{3}$ & $-\frac{1}{2}$ & $\left(\boldsymbol{1},\boldsymbol{1};\boldsymbol{1},\boldsymbol{1}\right)$ & $1$ & $0$ & $0$ & $-\frac{1}{6}$ & $-\frac{1}{2}$ & $\frac{1}{3}$ & $0$ & $0$ & $0$\\
 $\ell_{2}$ & $1$ & $0$ & $0$ & $0$ & $0$ & $-\frac{1}{6}$ & $-\frac{1}{3}$ & $-\frac{1}{2}$ & $\left(\boldsymbol{1},\boldsymbol{2};\boldsymbol{1},\boldsymbol{1}\right)$ & $-\frac{1}{2}$ & $0$ & $0$ & $-\frac{1}{6}$ & $\frac{3}{2}$ & $\frac{1}{3}$ & $0$ & $0$ & $0$\\
 $\bar u_{2}$ & $1$ & $0$ & $0$ & $0$ & $0$ & $-\frac{1}{6}$ & $-\frac{1}{3}$ & $-\frac{1}{2}$ & $\left(\overline{\boldsymbol{3}},\boldsymbol{1};\boldsymbol{1},\boldsymbol{1}\right)$ & $-\frac{2}{3}$ & $0$ & $0$ & $-\frac{1}{6}$ & $-\frac{1}{2}$ & $\frac{1}{3}$ & $0$ & $0$ & $0$\\
\end{longtable}

\begin{longtable}{|c|c|c|c|c|c|ccc|l|c|c|c|c|c|c|c|c|c|}
\hline
 & $k$ & $n_3$ & $n_2$ & $n_2'$ & $q_\gamma$ & 
 	$R_1$ & $R_2$ & $R_3$ & $\text{irrep}$ 
	& $q_Y$ & $q_2$ & $q_3$ & $q_4$ & $q_5$ & $q_6$ & $q_7$ & $q_8$ & $q_9$\\
\hline
\endhead
\hline
\endfoot
\hline
\caption{Same as Tab~\ref{tab:LongTableNonSingletsSaturdayAfternoonModel} 
for SM singlets.}
\label{tab:LongTableSingletsSaturdayAfternoonModel}
\endlastfoot
 $s_{1}$ & $0$ & $*$ & $*$ & $*$ & $0$ & $-1$ & $0$ & $0$ & $\left(\boldsymbol{1},\boldsymbol{1};\boldsymbol{1},\boldsymbol{1}\right)$ & $0$ & $0$ & $0$ & $0$ & $0$ & $\frac{1}{2}$ & $-1$ & $-\frac{1}{2}$ & $-2$\\[0.6mm]
 $s_{2}$ & $0$ & $*$ & $*$ & $*$ & $0$ & $-1$ & $0$ & $0$ & $\left(\boldsymbol{1},\boldsymbol{1};\boldsymbol{1},\boldsymbol{1}\right)$ & $0$ & $0$ & $0$ & $0$ & $0$ & $\frac{1}{2}$ & $-1$ & $-\frac{1}{2}$ & $2$\\[0.6mm]
 $\bar f_{1}$ & $0$ & $*$ & $*$ & $*$ & $0$ & $-1$ & $0$ & $0$ & $\left(\boldsymbol{1},\boldsymbol{1};\overline{\boldsymbol{4}},\boldsymbol{1}\right)$ & $0$ & $0$ & $0$ & $0$ & $0$ & $\frac{1}{2}$ & $1$ & $\frac{1}{2}$ & $-1$\\[0.6mm]
\hline
 $s_{3}$ & $0$ & $*$ & $*$ & $*$ & $0$ & $0$ & $-1$ & $0$ & $\left(\boldsymbol{1},\boldsymbol{1};\boldsymbol{1},\boldsymbol{1}\right)$ & $0$ & $-\frac{1}{2}$ & $-\frac{1}{2}$ & $-\frac{1}{2}$ & $-\frac{5}{2}$ & $0$ & $0$ & $0$ & $0$\\[0.6mm]
 $\bar f_{2}$ & $0$ & $*$ & $*$ & $*$ & $0$ & $0$ & $-1$ & $0$ & $\left(\boldsymbol{1},\boldsymbol{1};\overline{\boldsymbol{4}},\boldsymbol{1}\right)$ & $0$ & $0$ & $0$ & $0$ & $0$ & $1$ & $0$ & $0$ & $1$ \\[0.6mm]
\hline
\hline
 $s_{4}$ & $1$ & $0$ & $0$ & $0$ & $0$ & $-\frac{1}{6}$ & $-\frac{1}{3}$ & $-\frac{1}{2}$ & $\left(\boldsymbol{1},\boldsymbol{1};\boldsymbol{1},\boldsymbol{1}\right)$ & $0$ & $0$ & $0$ & $-\frac{1}{6}$ & $-\frac{5}{2}$ & $\frac{1}{3}$ & $0$ & $0$ & $0$\\[0.6mm]
 $s_{5}$ & $1$ & $0$ & $0$ & $0$ & $0$ & $-\frac{1}{6}$ & $\frac{2}{3}$ & $-\frac{1}{2}$ & $\left(\boldsymbol{1},\boldsymbol{1};\boldsymbol{1},\boldsymbol{1}\right)$ & $0$ & $-\frac{1}{2}$ & $-\frac{1}{2}$ & $\frac{1}{3}$ & $0$ & $\frac{1}{3}$ & $0$ & $0$ & $0$\\[0.6mm]
 $s_{6}$ & $1$ & $0$ & $0$ & $0$ & $0$ & $-\frac{1}{6}$ & $\frac{2}{3}$ & $-\frac{1}{2}$ & $\left(\boldsymbol{1},\boldsymbol{1};\boldsymbol{1},\boldsymbol{1}\right)$ & $0$ & $\frac{1}{2}$ & $\frac{1}{2}$ & $\frac{1}{3}$ & $0$ & $\frac{1}{3}$ & $0$ & $0$ & $0$\\
\end{longtable}

\begin{table}[!h]
\centerline{%
\begin{tabular}{|l|c||l|c||l|c|}
\hline
field & charge & field & charge & field & charge\\
\hline
 $s_{1}$ & $0$  &  $s_{2}$ & $0$  &  $s_{3}$ & $0$  \\
 $s_{4}$ & $1$  &  $s_{5}$ & $-\frac{4}{3}$  &  $s_{6}$ & $1$  \\
 $s_{7}$ & $\frac{2}{3}$  &  $s_{8}$ & $2$  &  $s_{9}$ & $-\frac{4}{3}$  \\
 $s_{10}$ & $1$  &  $s_{11}$ & $1$  &  $s_{12}$ & $-\frac{4}{3}$  \\
 $s_{13}$ & $1$  &  $s_{14}$ & $\frac{2}{3}$  &  $s_{15}$ & $2$  \\
 $s_{16}$ & $-\frac{4}{3}$  &  $s_{17}$ & $1$  &  $s_{18}$ & $\frac{2}{3}$  \\
 $s_{19}$ & $\frac{2}{3}$  &  $s_{20}$ & $\frac{2}{3}$  &  $s_{21}$ & $\frac{4}{3}$  \\
\hline
\end{tabular}
}
\caption{Anomalous charges of the SM singlets $s_i$ and $h_i$.}
\label{tab:AnomalousChargesSinglets}
\end{table}
\clearpage

\subsection{Monomials}

\begin{longtable}{|l|l|l|}
\hline
\endfirsthead
\hline
\endhead
\hline
\endfoot
\hline
\caption{Examples of gauge invariant monomials.}
\label{tab:GaugeInvariantMonomials}
\endlastfoot
$s_{1}\,s_{19 (23)}\,s_{20 (24)}\,s_{57 (60)}$ &
$s_{2}\,s_{18 (22)}\,s_{20 (24)}\,s_{57 (60)}$ &
$s_{3}\,s_{6 (10,13,17)}\,s_{26 (28)}\,s_{57 (60)}$\\[0.6mm]
$s_{4 (11)}\,s_{26 (28)}\,s_{57 (60)}$ &
$s_{4 (11)}\,s_{27 (29)}\,s_{39}\,s_{40}$ &
$s_{5 (9,12,16)}\,s_{8 (15)}\,s_{56}$\\[0.6mm]
$s_{5 (9,12,16)}\,s_{21 (25)}\,s_{68}$ &
$s_{6 (10,13,17)}\,s_{7 (14)}\,s_{55}$ &
$s_{30 (32)}\,s_{56}$\\[0.6mm]
$s_{31 (33)}\,s_{55}$ &
$s_{34}\,s_{53 (54)}$ &
$s_{35 (37)}\,s_{64}$\\[0.6mm]
$s_{36 (38)}\,s_{63}$ &
$s_{39}\,s_{59 (62)}$ &
$s_{40}\,s_{58 (61)}$\\[0.6mm]
$s_{41}\,s_{57 (60)}$ &
$s_{42 (45)}\,s_{69}$ &
$s_{43 (46)}\,s_{68}$\\[0.6mm]
$s_{44 (47)}\,s_{67}$ &
$s_{48}\,s_{65 (66)}$ &
$s_{49 (51)}\,s_{50 (52)}$\\[0.6mm]
\end{longtable}

\begin{longtable}{|c|c|c|}
\hline
\endfirsthead
\hline
\endhead
\hline
\endfoot
\hline
\caption{Examples of gauge invariant monomials carrying negative net anomalous
charge.}
\label{tab:GaugeInvariantMonomialsGen}
\endlastfoot
 $s_{5 (9,12,16)}\,s_{5 (9,12,16)}\,s_{39}\,s_{55}\,s_{56}$  &
 $s_{5 (9,12,16)}\,s_{34}\,s_{39}\,s_{49 (51)}\,s_{55}$ &
 $s_{34}\,s_{35 (37)}\,s_{39}\,s_{40}\,s_{67}$  \\[0.6mm]
 $s_{34}\,s_{39}\,s_{39}\,s_{43 (46)}\,s_{55}$  &
 $s_{35 (37)}\,s_{39}\,s_{40}\,s_{42 (45)}\,s_{57 (60)}$  &
 $s_{35 (37)}\,s_{55}\,s_{56}\,s_{57 (60)}\,s_{67}$  \\[0.6mm]
 $s_{39}\,s_{39}\,s_{40}\,s_{42 (45)}\,s_{63}$  &
 $s_{39}\,s_{55}\,s_{56}\,s_{63}\,s_{67}$  &
\end{longtable}

\begin{longtable}{|c|c|c|}
\hline
\endfirsthead
\hline
\endhead
\hline
\endfoot
\hline
\caption{Examples of gauge invariant monomials carrying negative net anomalous
charge for $s_i$ from $U,T_2,T_4$.}
\label{tab:GaugeInvariantMonomialsFflat}
\endlastfoot
  $s_{34}\,s_{35 (37)}\,s_{39}\,s_{40}\,s_{67}$  &
  $s_{34}\,s_{39}\,s_{39}\,s_{43 (46)}\,s_{55}$  &
  $s_{35 (37)}\,s_{39}\,s_{40}\,s_{42 (45)}\,s_{57 (60)}$  \\[0.6mm]
  $s_{35 (37)}\,s_{55}\,s_{56}\,s_{57 (60)}\,s_{67}$  &
  $s_{39}\,s_{39}\,s_{40}\,s_{42 (45)}\,s_{63}$  &
  $s_{39}\,s_{55}\,s_{56}\,s_{63}\,s_{67}$  \\[0.6mm]
\end{longtable}

\begin{longtable}{|l|}
\hline
\endfirsthead
\endhead
\hline
\endfoot
\hline
\caption{Examples of gauge invariant monomials involving only the singlets of
Eq.~\eqref{eq:stilde}.}
\label{tab:GaugeInvariantMonomials2}
\endlastfoot
 $s_{12}\,s_{40}\,s_{40}\,s_{61}\,h_{2}\,h_{4}\,h_{7}\,h_{14}$\\
 $s_{16}\,s_{40}\,s_{40}\,s_{61}\,h_{2}\,h_{4}\,h_{7}\,h_{14}$\\
 $s_{1}\,s_{16}\,s_{19}\,s_{40}\,s_{57}\,h_{2}\,h_{4}\,h_{14}\,h_{14}$\\
 $s_{1}\,s_{16}\,s_{19}\,s_{40}\,s_{60}\,h_{2}\,h_{4}\,h_{14}\,h_{14}$\\
 $s_{14}\,s_{16}\,s_{40}\,s_{40}\,s_{61}\,h_{2}\,h_{4}\,h_{5}\,h_{14}$\\
 $s_{16}\,s_{39}\,s_{40}\,s_{40}\,s_{61}\,h_{2}\,h_{2}\,h_{4}\,h_{14}$\\
 $s_{40}\,s_{40}\,s_{53}\,s_{60}\,s_{61}\,h_{2}\,h_{2}\,h_{5}\,h_{14}$\\
 $s_{40}\,s_{40}\,s_{54}\,s_{60}\,s_{61}\,h_{2}\,h_{2}\,h_{5}\,h_{14}$\\
 $s_{16}\,s_{24}\,s_{39}\,s_{40}\,s_{40}\,s_{61}\,h_{2}\,h_{2}\,h_{4}\,h_{7}$\\
 $s_{3}\,s_{24}\,s_{40}\,s_{60}\,h_{2}\,h_{4}\,h_{4}\,h_{5}\,h_{7}\,h_{7}$\\
 $s_{16}\,s_{18}\,s_{19}\,s_{24}\,s_{24}\,s_{39}\,s_{54}\,s_{60}\,s_{60}\,h_{2}\,h_{2}\,h_{4}\,h_{5}$\\
 $s_{16}\,s_{19}\,s_{22}\,s_{24}\,s_{24}\,s_{39}\,s_{54}\,s_{60}\,s_{60}\,h_{2}\,h_{2}\,h_{4}\,h_{5}$
\end{longtable}

\begin{longtable}{|c|}
\hline
\endfirsthead
\hline
\endhead
\hline
\endfoot
\hline
\caption{Examples of gauge invariant monomials carrying negative net anomalous 
charge and involving only the singlets of Eq.~\eqref{eq:stilde}.}
\label{tab:GaugeInvariantMonomialsNC}
\endlastfoot
 $s_{12}\,s_{12}\,s_{39}\,s_{40}\,h_{5}\,h_{14}$\\
 $s_{12}\,s_{16}\,s_{39}\,s_{40}\,h_{5}\,h_{14}$\\
 $s_{12}\,s_{12}\,s_{24}\,s_{39}\,s_{40}\,h_{5}\,h_{7}$\\
 $s_{12}\,s_{12}\,s_{24}\,s_{39}\,s_{39}\,s_{40}\,h_{2}\,h_{5}$\\
 $s_{12}\,s_{12}\,s_{39}\,s_{40}\,s_{40}\,s_{61}\,h_{5}\,h_{14}$\\
 $s_{3}\,s_{40}\,s_{40}\,s_{57}\,h_{5}\,h_{5}\,h_{7}\,h_{7}$\\
 $s_{3}\,s_{40}\,s_{40}\,s_{60}\,h_{5}\,h_{5}\,h_{7}\,h_{7}$\\
 $s_{12}\,s_{12}\,s_{14}\,s_{40}\,s_{40}\,h_{5}\,h_{5}\,h_{7}\,h_{14}$\\
 $s_{12}\,s_{12}\,s_{40}\,s_{53}\,s_{57}\,h_{5}\,h_{5}\,h_{14}\,h_{14}$\\
 $s_{12}\,s_{12}\,s_{40}\,s_{54}\,s_{57}\,h_{5}\,h_{5}\,h_{14}\,h_{14}$\\
 $s_{3}\,s_{12}\,s_{39}\,s_{40}\,s_{57}\,h_{4}\,h_{5}\,h_{5}\,h_{14}$
\end{longtable}

\subsection{$\boldsymbol{B\!-\!L}$ charges}
\hspace*{1cm}
\begin{sideways}
\begin{minipage}{\textheight}%
\refstepcounter{table}
\label{tab:BmLs}
\centerline{Table~\ref{tab:BmLs}: $B-L$ charges of the $s_i$.}
\centerline{\begin{tabular}{|r|ccccccccccccccccccccccc|}
\hline
$i$ & 1 & 2 & 3 & 4 & 5 & 6 & 7 & 8 & 9 & 10 & 11 & 12 & 13 & 14 & 15 & 16 & 17 & 
 18 & 19 & 20 & 21 & 22 & 23 \\
$q_{B-L}$ & 0 & 0 & 0 & 1 & 0 & 1 & 0 & -1 & 0 & 1 & 1 & 0 & 1 & 0 & -1 & 0 & 1 & 0 & 0 
 & 0 & 1 & 0 & 0  \\
\hline
$i$ & 24 & 25 & 26 & 27 & 28 & 29 & 30 & 31 & 32 & 33 & 34 & 35 & 36 & 37 & 38 & 
39 & 40 & 41 & 42 & 43 & 44 & 45 & 46 \\
$q_{B-L}$ & 0 & 1 & -1 & -1 & -1 & -1 & -1 & 1 & -1 & 1 & 0 & 1 & -1 & 
1 & -1 & 0 & 0 & 0 & -1 & 1 & 1 & -1 & 1  \\
\hline
$i$ &47 & 48 & 49 & 50 & 51 & 52 & 53 & 54 & 55 & 56 & 57 & 58 & 59 & 60 & 61 & 62 & \
63 & 64 & 65 & 66 & 67 & 68 & 69 \\
 $q_{B-L}$ & 1 & 0 & 1 & -1 & 1 & -1 & 0 & 0 & \
 -1 & 1 & 0 & 0 & 0 & 0 & 0 & 0 & 1 & -1 & 0 & 0 & -1 & -1 & 1\\
\hline
\end{tabular}
}
\end{minipage}
\end{sideways}
\hspace*{1cm}
\begin{sideways}
\begin{minipage}{\textheight}
\refstepcounter{table}
\label{tab:BmLspsm}
\centerline{Table~\ref{tab:BmLspsm}: $B-L$ charges of the $s^\pm_i$.}
\centerline{
\begin{tabular}{|r|cccccccccccccccc|}
\hline
$i$ & 1 & 2 & 3 & 4 & 5 & 6 & 7 & 8 & 9 & 10 & 11 & 12 & 13 & 14 & 15 & 16 \\
\hline
$q_{B-L}(s^+_i)$ & $\frac{1}{2}$ & $\frac{1}{2}$ & $-\frac{1}{2}$ & $-\frac{1}{2}$ 
& $\frac{1}{2}$ & $\frac{3}{2}$ & $\frac{1}{2}$ & $\frac{3}{2}$ & $-\frac{3}{2}$ 
& $\frac{1}{2}$ & $\frac{1}{2}$ & $\frac{1}{2}$ & $-\frac{3}{2}$ & $\frac{1}{2}$ 
& $\frac{1}{2}$ & $\frac{1}{2}$ \\
$q_{B-L}(s^-_i)$ & $\frac{1}{2}$ & $\frac{1}{2}$ & $-\frac{1}{2}$ & $-\frac{1}{2}$ 
& $-\frac{3}{2}$ & $-\frac{1}{2}$ & $-\frac{3}{2}$ & $-\frac{1}{2}$ & $\frac{3}{2}$ 
& $-\frac{1}{2}$ & $-\frac{1}{2}$ & $-\frac{1}{2}$ & $\frac{3}{2}$ & $-\frac{1}{2}$ 
& $-\frac{1}{2}$ & $-\frac{1}{2}$\\
\hline
\end{tabular}
}
\end{minipage}
\end{sideways}
\hspace*{1cm}
\begin{sideways}
\begin{minipage}{\textheight}
\refstepcounter{table}
\label{tab:BmLm}
\centerline{Table~\ref{tab:BmLm}: $B-L$ charges of the $m_i$.}
\centerline{
\begin{tabular}{|r|cccccccc|}
\hline
$i$ & 1 & 2 & 3 & 4 & 5 & 6 & 7 & 8 \\
\hline
$q_{B-L}(m_i)$ & $-\frac{1}{2}$ & $-\frac{1}{2}$ & $\frac{1}{2}$ & $\frac{1}{2}$ 
& $-\frac{1}{2}$ & $\frac{1}{2}$ & $-\frac{1}{2}$ & $\frac{1}{2}$ \\
\hline
\end{tabular}
}
\end{minipage}
\end{sideways}
\clearpage
\subsection{Mass matrices}
\label{sec:MassMatrices}

\subsubsection{Mass matrices for generic singlet vevs}
\begin{eqnarray}
 \mathcal{M}_m^{ij}(s) & = &
 \left(\begin{array}{cccccccc}
 0   & s^6 & s^5 & s^5 & s^6 & s^1 & s^6 & s^5\\
 s^6 & 0   & s^5 & s^5 & s^6 & s^5 & s^6 & s^1\\
 s^5 & s^5 & 0   & 0   & s^1 & s^5 & s^4 & s^5\\
 s^5 & s^5 & 0   & 0   & s^4 & s^5 & s^1 & s^5\\
 s^6 & s^6 & s^1 & s^4 & 0   & s^5 & s^5 & s^5\\
 s^1 & s^5 & s^5 & s^5 & s^5 & 0   & s^5 & s^6\\
 s^6 & s^6 & s^4 & s^1 & s^5 & s^5 & 0   & s^5\\
 s^5 & s^1 & s^5 & s^5 & s^5 & s^6 & s^5 & 0  \\
 \end{array}\right)\;,\label{eq:Mm(s)}\\
 \mathcal{M}_s^{ij}(s) & = &
 \left(\begin{array}{cccccccccccccccc}
s^5 & s^5 & s^5 & s^5 & s^6 & s^6 & s^6 & s^6 & s^5 & s^5 & s^5 & s^6 & s^5 & s^5 & s^5 & s^6\\
s^5 & s^5 & s^5 & s^5 & s^6 & s^6 & s^6 & s^6 & s^5 & s^5 & s^5 & s^6 & s^5 & s^5 & s^5 & s^6\\
s^1 & s^5 & s^5 & s^5 & s^6 & s^6 & s^6 & s^6 & s^5 & s^1 & s^1 & s^6 & s^5 & s^5 & s^5 & s^6\\
s^5 & s^1 & s^5 & s^5 & s^6 & s^6 & s^6 & s^6 & s^5 & s^5 & s^5 & s^6 & s^5 & s^1 & s^1 & s^6\\
s^6 & s^6 & s^5 & s^5 & s^6 & s^1 & s^6 & s^5 & s^5 & s^5 & s^5 & 0   & s^5 & s^5 & s^5 & 0  \\
s^5 & s^5 & s^6 & s^6 & s^1 & s^6 & s^5 & s^6 & s^5 & s^1 & s^1 & s^6 & s^5 & s^5 & s^5 & s^6\\
s^6 & s^6 & s^5 & s^5 & s^6 & s^5 & s^6 & s^1 & s^5 & s^5 & s^5 & 0   & s^5 & s^5 & s^5 & 0  \\
s^5 & s^5 & s^6 & s^6 & s^5 & s^6 & s^1 & s^6 & s^5 & s^5 & s^5 & s^6 & s^5 & s^1 & s^1 & s^6\\
s^5 & s^5 & s^5 & s^5 & s^5 & s^6 & s^5 & s^6 & s^5 & s^5 & s^5 & s^5 & s^5 & s^5 & s^5 & s^5\\
s^1 & s^5 & s^5 & s^5 & s^1 & s^5 & s^5 & s^5 & s^5 & s^1 & s^5 & s^5 & s^5 & s^5 & s^5 & s^5\\
s^1 & s^5 & s^5 & s^5 & s^1 & s^5 & s^5 & s^5 & s^5 & s^5 & s^1 & s^5 & s^5 & s^5 & s^5 & s^5\\
s^5 & s^5 & s^5 & s^5 & s^6 & 0   & s^6 & 0   & s^5 & s^5 & s^5 & s^5 & s^5 & s^5 & s^5 & s^5\\
s^5 & s^5 & s^5 & s^5 & s^5 & s^6 & s^5 & s^6 & s^5 & s^5 & s^5 & s^5 & s^5 & s^5 & s^5 & s^5\\
s^5 & s^1 & s^5 & s^5 & s^5 & s^5 & s^1 & s^5 & s^5 & s^5 & s^5 & s^5 & s^5 & s^1 & s^5 & s^5\\
s^5 & s^1 & s^5 & s^5 & s^5 & s^5 & s^1 & s^5 & s^5 & s^5 & s^5 & s^5 & s^5 & s^5 & s^1 & s^5\\
s^5 & s^5 & s^5 & s^5 & s^6 & 0   & s^6 & 0   & s^5 & s^5 & s^5 & s^5 & s^5 & s^5 & s^5 & s^5
 \end{array}\right)\;,\nonumber\\ 
 & & \label{eq:Ms(s)}\\
 \mathcal{M}_f(s)& = &
 \left(\begin{array}{cccc}
 s^6 & s^5 & s^5 & s^3\\
 s^6 & s^5 & s^5 & s^3\\
 0 & s^5 & s^3 & s^6\\
 0 & s^5 & s^3 & s^6
 \end{array}\right)\;,\label{eq:Mf(s)}\\
 \mathcal{M}_w(s) & = &
 \left(\begin{array}{ccccc}
 s^1 & s^5 & s^5 & s^5 & s^5\\
 s^5 & s^1 & s^5 & s^5 & s^5\\
 s^5 & s^5 & s^5 & s^3 & s^3\\
 s^5 & s^5 & s^3 & s^3 & s^3\\
 s^5 & s^5 & s^3 & s^3 & s^3\\
 \end{array}\right)\;.\label{eq:Mw(s)}
\end{eqnarray}

\subsubsection{Mass matrices for the $\boldsymbol{B\!-\!L}$ preserving vacuum of Sec.~\ref{sec:Pheno}}

\begin{eqnarray}
 \mathcal{M}_m^{ij}(\widetilde{s}) & = &
 \left(\begin{array}{cccccccc}
 0 & 0 & 0 & \widetilde{s}^8 & 0 & \widetilde{s} & 0 & \widetilde{s}^5 \\
 0 & 0 & \widetilde{s}^8 & 0 & 0 & \widetilde{s}^5 & 0 & \widetilde{s} \\
 0 & \widetilde{s}^8 & 0 & 0 & \widetilde{s} & 0 & \widetilde{s}^5 & 0 \\
 \widetilde{s}^8 & 0 & 0 & 0 & \widetilde{s}^5 & 0 & \widetilde{s} & 0 \\
 0 & 0 & \widetilde{s} & \widetilde{s}^5 & 0 & \widetilde{s}^7 & 0 & \widetilde{s}^7 \\
 \widetilde{s} & \widetilde{s}^5 & 0 & 0 & \widetilde{s}^7 & 0 & \widetilde{s}^7 & 0 \\
 0 & 0 & \widetilde{s}^5 & \widetilde{s} & 0 & \widetilde{s}^7 & 0 & \widetilde{s}^7 \\
 \widetilde{s}^5 & \widetilde{s} & 0 & 0 & \widetilde{s}^7 & 0 & \widetilde{s}^7 & 0
 \end{array}\right)\;,\label{eq:Mmstilde}\\
 \mathcal{M}_s^{ij}(\widetilde{s}) & = &
 \left(\begin{array}{cccccccccccccccc}
 0 & 0 & 0 & 0 & 0 & \widetilde{s}^6 & 0 & 0 & 0 & \widetilde{s}^8 & \widetilde{s}^8 & 0 & 0 & 0 & 0 & 0 \\
 0 & 0 & 0 & 0 & 0 & 0 & 0 & \widetilde{s}^6 & 0 & 0 & 0 & 0 & 0 & \widetilde{s}^8 & \widetilde{s}^8 & 0 \\
 \widetilde{s}^8 & \widetilde{s}^8 & 0 & 0 & 0 & 0 & 0 & 0 & 0 & 0 & 0 & 0 & 0 & 0 & 0 & 0 \\
 \widetilde{s}^8 & \widetilde{s}^8 & 0 & 0 & 0 & 0 & 0 & 0 & 0 & 0 & 0 & 0 & 0 & 0 & 0 & 0 \\
 0 & 0 & \widetilde{s}^6 & 0 & 0 & \widetilde{s}^8 & 0 & \widetilde{s}^8 & 0 & \widetilde{s}^5 & \widetilde{s}^5 & 0 & 0 & \widetilde{s}^8 & \widetilde{s}^8 & \widetilde{s}^8 \\
 0 & 0 & 0 & 0 & \widetilde{s}^8 & 0 & \widetilde{s}^8 & 0 & 0 & 0 & 0 & 0 & 0 & 0 & 0 & 0 \\
 0 & 0 & 0 & \widetilde{s}^6 & 0 & \widetilde{s}^8 & 0 & \widetilde{s}^8 & 0 & \widetilde{s}^8 & \widetilde{s}^8 & \widetilde{s}^8 & 0 & \widetilde{s}^5 & \widetilde{s}^5 & 0 \\
 0 & 0 & 0 & 0 & \widetilde{s}^8 & 0 & \widetilde{s}^8 & 0 & 0 & 0 & 0 & 0 & 0 & 0 & 0 & 0 \\
\end{array}\right)\;,\nonumber\\
 & &\label{eq:Msstilde} \\
 \mathcal{M}_f^{ij}(\widetilde{s}) & = &
  \left(
 \begin{array}{cccc}
 0 & \widetilde{s}^5 & \widetilde{s}^8 & \widetilde{s}^6 \\
 0 & \widetilde{s}^5 & \widetilde{s}^8 & \widetilde{s}^6 \\
 0 & \widetilde{s}^6 & \widetilde{s}^6 & 0 \\
 0 & \widetilde{s}^6 & \widetilde{s}^6 & 0
 \end{array}
 \right)\;,\label{eq:Mfstilde}\\
 \mathcal{M}_w^{ij}(\widetilde{s}) & = &
 \left(
 \begin{array}{ccccc}
 \widetilde{s} & \widetilde{s}^5 & \widetilde{s}^8 & 0 & 0 \\
 \widetilde{s}^5 & \widetilde{s} & \widetilde{s}^8 & \widetilde{s}^7 & \widetilde{s}^7 \\
 \widetilde{s}^8 & \widetilde{s}^8 & 0 & \widetilde{s}^6 & \widetilde{s}^6 \\
 0 & \widetilde{s}^7 & \widetilde{s}^6 & \widetilde{s}^6 & \widetilde{s}^6 \\
 0 & \widetilde{s}^7 & \widetilde{s}^6 & \widetilde{s}^6 & \widetilde{s}^6
 \end{array}
 \right)\;.\label{eq:Mwstilde}
\end{eqnarray}

\clearpage
\subsection{Survey of orbifold GUT limits}

\begin{table}[!ht]
\begin{center}
\begin{tabular}{|r|ccc|l|l|}
\hline
 & \multicolumn{3}{|c|}{plane} & 
& \\
dim. & $\mathrm{G}_2$ & $\SU3$ & $\SO4$ & conditions
& SUSY, bulk groups\\
\hline
10&
\CenterObject{\includegraphics[scale=0.3]{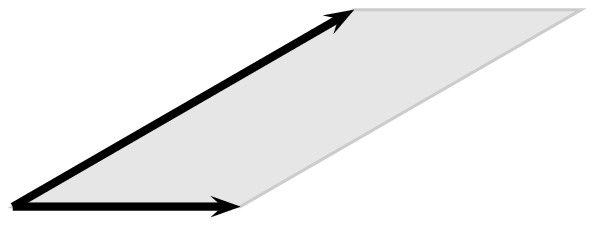}}
& 
\CenterObject{\includegraphics[scale=0.3]{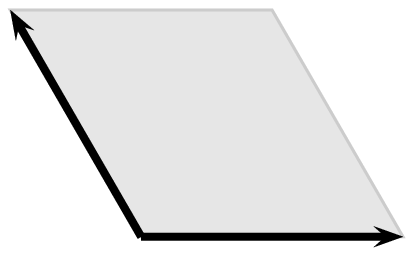}}
& 
\CenterObject{\includegraphics[scale=0.3]{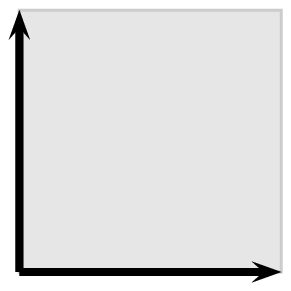}}
& -- & $N=4$, $\mathrm{E}_8$ \\[2mm]
\hline
9&
\CenterObject{\includegraphics[scale=0.3]{G2Torus.eps}}
& 
\CenterObject{\includegraphics[scale=0.3]{SU3Torus.eps}}
& 
\CenterObject{\includegraphics[scale=0.3]{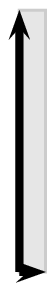}}
& $p\cdot W_2\in\mathbbm{Z}$ & $N=4$, $\SO{16}$ \\[2mm]
9&
\CenterObject{\includegraphics[scale=0.3]{G2Torus.eps}}
& 
\CenterObject{\includegraphics[scale=0.3]{SU3Torus.eps}}
& 
\CenterObject{\includegraphics[scale=0.3]{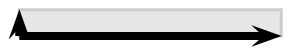}}
& $-$ & $N=4$, $\E{8}$ \\[2mm]
\hline
8&
\CenterObject{\Huge$\bullet$}
& 
\CenterObject{\includegraphics[scale=0.3]{SU3Torus.eps}}
& 
\CenterObject{\includegraphics[scale=0.3]{SO4Torus.eps}}
& -- &  $N=4$, $\mathrm{E}_8$\\[2mm]
8&
\CenterObject{\includegraphics[scale=0.3]{G2Torus.eps}}
& 
\CenterObject{\Huge$\bullet$}
& 
\CenterObject{\includegraphics[scale=0.3]{SO4Torus.eps}}
& $p\cdot W_3\in\mathbbm{Z}$  &  $N=4$, $\mathrm{E}_6\!\times\!\SU3$ \\[2mm]
8&
\CenterObject{\includegraphics[scale=0.3]{G2Torus.eps}}
& 
\CenterObject{\includegraphics[scale=0.3]{SU3Torus.eps}}
& 
\CenterObject{\Huge$\bullet$}
&
$p\cdot W_2\in\mathbbm{Z}$ 
& $N=4$, $\SO{16}$ 
\\
\hline
\end{tabular}
\end{center}
\caption{Survey of the orbifold GUTs in different dimensions. The bullet
indicates small compact dimensions.  $\U1$ factors and subgroups of the second
\E8 are omitted.}
\label{tab:OGL}
\end{table}

\clearpage
\renewcommand{\nomname}{List of frequently used symbols}
\addcontentsline{toc}{section}{\nomname}

\begin{theglossary} 

 \nomgroup{E}

  \item [{$e_a$}]\begingroup lattice vectors\refeqpage\nomeqref {2.1}
		\nompageref{6}

 \nomgroup{F}

  \item [{$f$}]\begingroup fixed point\refeqpage\nomeqref {2.7}
		\nompageref{7}

 \nomgroup{L}

  \item [{$\lambda^I$}]\begingroup left--moving fermions\refeqpage\nomeqref {2.10}
		\nompageref{7}

 \nomgroup{N}

  \item [{$n_2$}]\begingroup localization quantum number in the \SO4 plane\refpage\nomeqref {3.18}
		\nompageref{22}
  \item [{$n_2'$}]\begingroup localization quantum number in the \SO4 plane\refpage\nomeqref {3.18}
		\nompageref{22}
  \item [{$n_3$}]\begingroup localization quantum number in the \SU3 plane\refpage\nomeqref {3.16}
		\nompageref{21}

 \nomgroup{P}

  \item [{$p$}]\begingroup $p\in\Lambda_{\E8\times\E8}$: $\E8\times\E8$ root lattice vector (`momentum')\refeqpage\nomeqref {2.37}
		\nompageref{11}
  \item [{$\psi^i$}]\begingroup right--moving fermions\refeqpage\nomeqref {2.9}
		\nompageref{7}
  \item [{$\widetilde{\psi}^i$}]\begingroup complex NSR fermions\refeqpage\nomeqref {2.11}
		\nompageref{7}

 \nomgroup{Q}

  \item [{$q$}]\begingroup $q\in\Lambda_{\SO8}^*$: \SO8 weight (`momentum') \refeqpage\nomeqref {2.37}
		\nompageref{11}
  \item [{$q_\gamma$}]\begingroup additional quantum number in $T_{k>1}$ twisted sectors of non--prime orbifolds\refeqpage\nomeqref {2.58}
		\nompageref{15}

 \nomgroup{R}

  \item [{$R_i$}]\begingroup invariant $H$--momenta\refeqpage\nomeqref {4.7}
		\nompageref{24}

 \nomgroup{T}

  \item [{$\theta$}]\begingroup twist\refeqpage\nomeqref {2.3}
		\nompageref{6}

 \nomgroup{V}

  \item [{$V_f$}]\begingroup local gauge shift\refeqpage\nomeqref {2.32}
		\nompageref{10}
  \item [{$V_N$}]\begingroup gauge shift vector\refeqpage\nomeqref {2.18}
		\nompageref{8}
  \item [{$v_N$}]\begingroup twist vector\refeqpage\nomeqref {2.4}
		\nompageref{6}

 \nomgroup{X}

  \item [{$X_\mathrm{L,R}^i$}]\begingroup string coordinates\refeqpage\nomeqref {2.9}
		\nompageref{7}

 \nomgroup{Z}

  \item [{$Z^i$}]\begingroup complex string coordinates\refeqpage\nomeqref {2.11}
		\nompageref{7}
  \item [{$z^i$}]\begingroup complex coordinates of the torus\refpage\nomeqref {2.0}
		\nompageref{6}

\end{theglossary}

\clearpage

\addcontentsline{toc}{section}{References}

\providecommand{\bysame}{\leavevmode\hbox to3em{\hrulefill}\thinspace}

\end{document}